\documentclass[fleqn,usenatbib]{mnras}

\usepackage{mathptmx}

\usepackage[T1]{fontenc}

\DeclareRobustCommand{\VAN}[3]{#2}
\let\VANthebibliography\thebibliography
\def\thebibliography{\DeclareRobustCommand{\VAN}[3]{##3}\VANthebibliography}

\usepackage{graphicx}	
\usepackage{amsmath}	
\usepackage{amssymb}	
\usepackage{caption}
\usepackage{subcaption}
\usepackage{tablefootnote}
\usepackage{lineno}     
\usepackage{threeparttable} 
\usepackage{orcidlink}  
\usepackage{longtable} 
\usepackage{booktabs} 
\usepackage[font=small,labelfont=bf]{caption} 
\usepackage{xcolor,colortbl} 
\usepackage[symbol]{footmisc} 

\title[Energy dependence of QPOs in accreting XRPs]{Energy dependence of Quasi-periodic oscillations in accreting X-ray pulsars}

\author[H. Manikantan et al.]{
Hemanth Manikantan,$^{1,2}$\thanks{E-mail: hemanth.manikantan@inaf.it}$^{\orcidlink{0000-0001-9404-1601}}$
Biswajit Paul,$^{1}$
Rahul Sharma$^{1}$$^{\orcidlink{0000-0003-0366-047X}}$
Pragati Pradhan$^{3}$$^{\orcidlink{0000-0002-113$1-3$059}}$
and Vikram Rana$^{1}$$^{\orcidlink{0000-0003-1703-8796}}$
\\
$^{1}$Astronomy \& Astrophysics Department, Raman Research Institute, CV Raman Avenue, Sadashivanagar, Bangalore 560080, India\\
$^{2}$INAF Istituto di Astrofisica e Planetologia Spaziali,
Via del Fosso del Cavaliere 100, 00133 Roma, Italy\\
$^{3}$Department of Physics, Embry Riddle Aeronautical University, Prescott Campus, 3700 Willow Creek Road, Prescott, AZ 86301, USA\\
}

\date{Accepted XXX. Received YYY; in original form ZZZ}

\pubyear{2024}

\begin{document}
\label{firstpage}
\pagerange{\pageref{firstpage}--\pageref{lastpage}}
\maketitle

\begin{abstract}
We present the results from an investigation of the energy dependence of Quasi-Periodic Oscillations (QPOs) exhibited by accreting X-ray pulsars using data from archival \textit{XMM-Newton}, \textit{NuSTAR}, \textit{RXTE}, and \textit{NICER} observations. In a search for the presence of QPOs in 99 \textit{XMM-Newton} and \textit{NuSTAR} observations, we detected QPOs in eleven observations of five sources, viz., 4U 1626--67 (48 mHz), IGR J19294+1816 (30 mHz), V 0332+53 (2, 18 and 40 mHz), Cen X--3 (30 mHz), and XTE J1858+034 (180 mHz). A positive correlation of the QPO rms amplitude with energy is exhibited by 4U 1626--67, IGR J19294+1816, Cen X--3 and XTE J1858+034, while no energy dependence is observed in V 0332+53. We also analysed the energy spectrum to decouple thermal (soft-excess) from non-thermal emission and determine if the soft-excess has different QPO properties. We found no evidence for different QPO characteristics of the soft excess. The \textit{NuSTAR} observations of V 0332+53 during the Type-I outburst in 2016 show the presence of twin QPOs at 2.5 mHz and 18 mHz, while the \textit{XMM-Newton} and \textit{NuSTAR} observations during the Type-II outburst in 2015 show a QPO at 40 mHz. We review the observed QPO properties in the context of QPOs found in other types of accreting sources and the models usually used to explain the QPOs in accreting X-ray pulsars.
\end{abstract}

\begin{keywords}
pulsars: general -- accretion discs -- X-rays: binaries -- methods: data analysis {-- X-rays: individual: 4U 1626--67, IGR J19294+1816, V 0332+53, Cen X--3, XTE J1858+034}
\end{keywords}



\section{Introduction}
Accreting X-ray pulsars (XRPs) are strongly magnetized rotating neutron stars, having surface magnetic fields of the order of 10$^{12}$ Gauss, that accretes matter from a binary stellar companion either via Roche lobe overflow or from its strong stellar wind. Quasi-periodic oscillations (QPOs), exhibited as a concentration of Fourier power at frequencies of a few ten mHz in the power spectral density (PSD), are a transient phenomenon in XRPs. These mHz QPO have been reported in about twenty XRPs (See the list of sources in \citealt{james2010discovery} and \citealt{graman2021astrosat}) in both transient and persistent sources. 

The strong magnetic field of the neutron star impedes the formation of an accretion disk inside the magnetospheric radius ($r$\textsubscript{M}) at about a few 1000 km from the neutron star. Therefore, the radius of the inner accretion disk is expected to be of the order of $r$\textsubscript{M}. QPOs are usually exhibited by XRPs at frequencies of a few ten mHz, and since the Keplerian orbital frequency of matter around a canonical 1.4 M$_\odot$ neutron star at $r$\textsubscript{M}$\sim5000$ km is $\frac{1}{2\pi}\sqrt{\frac{GM_{\textrm{NS}}}{r_\textrm{M}^3}}$ $\sim200$ mHz, the QPOs are qualitatively associated with phenomena related to the inner accretion disk. {The QPO is XRPs are usually observed from a few keV up to $\sim$40 keV \citep{qu2005discovery_v0332,harsha_cenx3_2008ApJ...685.1109R,nespoli2010discovery_1a1118,devasia2011timing,graman2021astrosat,2023rahul_lmcx4}, but 1A 0535+262 is an exception where QPO is only present in hard X-rays above $\sim$25 keV \citep{finger1996quasi,ma_1a0535_mHzqpo_Erelation-distinct}.} The accretion disk is considered to be always present in persistent sources. On the contrary, transient sources are believed to have phases of formation of a temporary accretion disk around the neutron star during their luminous phases.

The two most commonly used models to explain QPOs in XRPs are the Magnetospheric beat-frequency model (BFM; \citealt{alpar_shazam1985gx5}) and the Keplerian frequency model (KFM; \citealt{klis1997kilohertz}). BFM models the QPO as a modulation in the mass accretion rate on to the NS poles at the beat frequency between the spin frequency of NS ($\nu_{\textrm{NS}}$) and the orbital frequency of the inner accretion disk. Matter is channelled onto the neutron star from inhomogeneity in the inner accretion disk along the spinning magnetic field lines. According to BFM, $\nu_{\textrm{QPO}}=\nu_{\textrm{k}}-\nu_{\textrm{NS}}$, where $\nu_{\textrm{k}}$ is the Keplerian orbital frequency of the inner accretion disk and $\nu_{\textrm{NS}}$ is the spin period of the neutron star. KFM models the QPO as an effect due to the blobs of matter in the inner accretion disk intercepting the emission from the neutron star, because of which the observer perceives a modulation in the X-ray flux. According to KFM, $\nu_{\textrm{QPO}}=\nu_{\textrm{k}}$. KFM places a constraint on the observed QPO frequency based on the neutron star spin frequency $\nu_{\textrm{NS}}<\nu_{\textrm{QPO}}$, based on the argument of centrifugal inhibition of accreted matter beyond this limit. BFM, on the other hand, provides constraints on the minimum X-ray flux to be maintained by accretion based on the centrifugal inhibition limit \citep{FINGER1998_qpo-review}. Assuming the radius of the inner accretion disk also scales by the accretion rate $\dot{M}^{-\frac{2}{7}}$ (and hence the X-ray luminosity), the Keplerian frequency of the inner accretion disk, and thereby the QPO frequency, is expected to increase with luminosity in both KFM and BFM (\citealt{FINGER1998_qpo-review}).

QPOs exhibit variability as a function of photon energy, which puts constraints on the applicable models that explain the origin of QPO. So far no general characteristic trend has been recognised in XRPs with respect to the QPO rms {amplitude} as a function of energy. Correlation (IGR J19294+1816; \citealt{graman2021astrosat}, KS 1947+300; \citealt{james2010discovery}, XTE J1858+034; \citealt{mukherjee2006variablextej1854}, LMC X--4; \citealt{2023rahul_lmcx4}), anti-correlation (A 1118-615; \citealt{nespoli2010discovery_1a1118}, V 0332+53; \citealt{qu2005discovery_v0332}, Cen X--3 \citealt{liu_Cenx3_40mhZqpo_insight}, {GX 304--1; \citealt{devasia2011timing})} and no correlation (Cen X--3; \citealt{harsha_cenx3_2008ApJ...685.1109R}) have all been reported. Most of the studies mentioned above were performed using \textit{RXTE}/PCA data, which was sensitive to photons in the 2--30 keV energy band. There are also cases where unusual dependency is seen, for instance, the detection of QPO at 80 keV and the convex-shaped relation between QPO rms and photon energy in 1A 0535+262 peaking at around 60 keV from the \textit{Insight}/HXMT data which offers 1--250 keV spectral coverage \citep{ma_1a0535_mHzqpo_Erelation-distinct}.

We performed a comprehensive search for QPOs in the archival \textit{XMM-Newton} and \textit{NuSTAR} observations of XRPs. {Additional data from \textit{RXTE}/PCA and \textit{NICER} were analysed for 4U 1626--67.} In observations where QPO was detected, we constructed the variation of QPO rms as a function of photon energy and analysed the energy spectrum. 

{The paper is arranged as follows. Section~\ref{sec:ODR} describes the instruments utilized, data reduction and the methods employed for timing and spectral analyses. The results of the timing and spectral analysis for each source are provided in Section~\ref{sec:SAR}. The discussion of findings and interpretation of the results is given in Section~\ref{sec:DAI}. The log of all observations utilised in this work and fits on individual power spectral densities are given in Appendices~\ref{app:qpofit} and \ref{app:obslog}.}

\section{Observations and Data Reduction}\label{sec:ODR}

\subsection{Instruments and Data reduction}

\textit{XMM-Newton}:
The PN-type European Photon Imaging Camera (EPIC-PN) onboard the \textit{X-ray Multi-Mirror Mission} (\textit{XMM-Newton}) is an array of twelve pn-CCDs coupled to focussing optics, sensitive to photons in 0.15--15 keV \citep{struder2001spic-pn}. EPIC-PN has an effective area of around $1000$ cm$^2$ at 1.5 keV. We followed the standard data reduction steps from \textit{XMM-Newton} data analysis threads\footnote{\url{https://www.cosmos.esa.int/web/xmm-newton/sas-threads}}. First, the event list for the EPIC-PN instrument was generated from the Observation Data Files (ODF) with the tool {\sffamily epproc}, using the calibration files generated with the tool {\sffamily cifbuild}. The generated event time stamps were then corrected for the motion of the earth around the barycenter of the solar system using the tool {\sffamily barycen} and subsequently filtered for time intervals of high background particle flaring. We also checked the observations for pile-up, following the steps mentioned in SAS Thread \textit{epatplot}\footnote{\url{https://www.cosmos.esa.int/web/XMM-Newton/sas-thread-epatplot}}. If found, the piled-up data were removed by excluding the central core of the PSF in Imaging mode observations using an annular source region, and by removing the boresight columns in Timing mode observations\footnote{\url{http://xmm-tools.cosmos.esa.int/external/xmm_user_support/documentation/sas_usg/USG/epicpileuptiming.html}}.

The source and background light curves and spectra were extracted from the resulting events file with the tool {\sffamily evselect}. A circular source and annular background region were used for imaging mode observations, while rectangular strips were used for timing mode observations.

\textit{NuSTAR}:
\textit{NuSTAR} operating in 3--79 keV, consists of two CdZnTe detectors paired to separate hard X-ray focussing optics and have a total effective area of about 1000 cm$^2$ at 10 keV \citep{harrison2013nustar}. The two detectors are called Focal Plane Modules (FPM) A and B. We followed the standard data reduction steps from The NuSTAR Data Analysis Software Guide\footnote{\url{https://heasarc.gsfc.nasa.gov/docs/nustar/analysis/nustar_swguide.pdf}}, using the  NuSTAR Data Analysis Software package {\sffamily NuSTARDAS v2.1.1}. The filtered and calibrated events files were generated with the tool {\sffamily nupipeline} using the NuSTAR calibration database (CALDB) version 20210315. Using {\sffamily nuproducts}, the source and background light curves were generated from circular regions of FPMA and FPMB modules. The light curves from FPMA and FPMB were summed together using task \texttt{lcmath}.

\textit{RXTE}/PCA:
The Proportional Counter Array (PCA) on the \textit{Rossi X-ray Timing Explorer} (\textit{RXTE}) consists of five collimated large-area Xenon filled proportional counter units with a total effective area of $\sim6500$ cm$^2$, sensitive to photons in 2--60 keV \citep{2006jahoda}. PCA light curves in different energy bands were extracted from the GoodXenon Event mode data files using the tool {\sffamily seextrct} using the photon energy to channel conversion table given here\footnote{\url{https://heasarc.gsfc.nasa.gov/docs/xte/e-c_table.html}}. We have used only one \textit{RXTE}/PCA observation (of 4U 1626--67{, OID P10101}) in this work.

{\textit{NICER}: The \textit{Neutron star Interior Composition Explorer} (\textit{NICER}) is an X-ray observatory installed on the International Space Station (ISS). The X-ray Timing Instrument (XTI) onboard \textit{NICER} comprises 56 X-ray concentrator (XRC) optics coupled to SDD detectors \citep{nicer}. The XTI operates in the 0.2$-$10 keV energy range and has an effective area of $\sim1900$ cm$^2$ at 1.5 keV. XTI lightcurves in different energy bands were extracted using the tool \texttt{nicerl3-lc} and the spectra were extracted using the tool \texttt{nicerl3-spect} after screening the data from the noisy detectors (\texttt{DET\_ID} 14 and 34). The background lightcurves and spectra were generated using the Space Weather background model. The lightcurves were corrected for the solar-system barycenter using the tool \texttt{barycorr}. We have only used the \textit{NICER} observations of 4U 1626$-$67 (Table~\ref{tab:qpoobscatalogue}) in this work.}

\subsection{Method of analysis}

For each observation listed in Table~\ref{tab:obscat1}, {the background subtracted} light curves were generated with a bin size of 1 s. Setting the bin size to 1 s enables the construction of PSD up to 500 mHz, facilitating the ability to check for mHz QPOs. The PSD of each light curve was generated using the {\small XRONOS} tool \texttt{powspec}\footnote{\url{https://heasarc.gsfc.nasa.gov/lheasoft/ftools/fhelp/powspec.txt}}. The light curves were divided into segments of length 4096 s {(1024 s for \textit{NICER})} and the power spectra obtained from the lightcurve segments were averaged to improve the signal-to-noise ratio of PSD \citep{vdk_xraytiming_1989ASIC..262...27V}. The resulting PSD was normalised so that it has units of  (rms/mean)$^2$ Hz$^{-1}$ so that integrating the PSD over frequency gives the fractional rms squared variability. An expected flat noise level of {$\sim2$/mean} was also subtracted from the PSD (\citealt{vdk_xraytiming_1989ASIC..262...27V}, \citealt{belloni_1990A&A...230..103B}). Normalisation of PDS and subtraction of the expected noise level was achieved by setting norm $=-2$ in \texttt{powspec}. The fractional rms amplitude from QPO in \textit{NICER} observations were estimated from non-background-corrected lightcurves, and hence the estimated QPO fractional rms were corrected for background by scaling it with a factor of {$\sqrt{(S+B)}/{S}$ \citep{belloni_1990A&A...230..103B}}, where $S$ and $B$ are the source and background count rates, respectively.

QPOs appeared as a relatively wide asymmetric bump {in the PSD} (See Appendix.\ref{app:qpofit}) {and were identified through visual inspection}. Apart from QPOs some of the PSDs also showed sharp narrow features (See, for example, Figs.~\ref{fig:4u1626_qpofit_xmmn}, \ref{fig:4u1626_qpofit_rxtepca}, \ref{fig:igrj19294_qpofit} and \ref{fig:cenx3_qpofit}) corresponding to the spin period of the pulsar and its harmonics. PSD to a factor of four frequency range on either side of the QPO vicinity {($0.25\nu$\textsubscript{QPO}--$4\nu$\textsubscript{QPO}) or a factor of 8 for the low \textit{Q}-factor cases (V 0332$+$53 and Cen X$-$3)} was fitted with the combination of a \texttt{powerlaw} or \texttt{lorentzian} (for the continuum), and a \texttt{lorentzian} (for the QPO). {The sharp spikes in PSD corresponding to the pulsar spin period and its harmonics were removed before performing the fit.} The centre ($\nu$\textsubscript{QPO}), width ({width}\textsubscript{QPO}) of \texttt{lorentzian} and the integrated fractional rms-squared power under the \texttt{lorentzian} ($\rm{P}_{rms}\pm\Delta\rm{P}_{rms}$) was then estimated. Fractional rms {amplitude} variability of the QPO was estimated as $\sqrt{\rm{P}_{rms}}$  $\pm$ {(${\Delta\rm{P}_{rms}}/2\sqrt{\rm{P}_{rms}}$)}. To assess the variation of QPO fractional rms in different energy ranges, the procedure was repeated on PSDs derived from light curves in different energy bands. The errors assigned to the QPO fractional rms values are their 68\% (1$\sigma$) confidence intervals and all other parameters are their 90\% {(2.7$\sigma$)} confidence intervals, unless otherwise stated. {The fractional rms amplitude for the high-frequency (200 mHz) QPO in XTE J1858+034 was assessed after constructing the PSD from lightcurve with 0.5 s bin size so that the Nyquist frequency is 1 Hz.}

We analysed energy-resolved lightcurves for a total of 99 observations of 29 X-ray pulsars and QPOs were identified in {eleven} of them {(Table~\ref{tab:qpoobscatalogue}). The very fact that QPOs were not detected in all the observations of a particular source, and in sources which have previous reports of QPO, indicates the transient nature of QPOs in XRPs.}

{In the observations with a QPO detection, we also performed the spectral analysis. We used the package \texttt{XSPEC} v12.13.1 \citep{arnaud1999xspec} for performing the spectral analysis. Unless stated otherwise, we have binned the spectrum to a minimum of 25 counts per bin.}

\section{Sources and Results}\label{sec:SAR}

\begin{table*}
	\centering
	\caption{Observations catalogue for timing and spectral analysis, of sources with QPO detection.}
	\label{tab:qpoobscatalogue}
        \scalebox{0.86}{
	\begin{tabular}{lllclccc} 
		\hline
		Source & Observatory/Instrument & Obs. ID & Observing mode &Start date (Duration in ks)& Avg. count rate$^\dagger$ {(cts s$^{-1}$)} &Piled-up &Avg. count rate$^\ddagger$ {(cts s$^{-1}$)} \\
		\hline
		4U 1626--67 & \textit{XMM-Newton}/PN &0111070201 &PrimeSmallWindow &24-08-2001 (16) &$33.20\pm0.06$ &- &-\\
		& &0152620101 &PrimeSmallWindow &20-08-2003 (84) &$27.63\pm0.02$ &XRL$^\S$ &-\\
		& \textit{RXTE}/PCA &P10101 &Good Xenon &10-02-1996 (395) &$306.90\pm0.05^\mathparagraph$ &- &-\\
        & {\textit{NICER}/XTI} & 62038001XX$^\S$ &{N/A} &{28-04-2023 (40)$^\S$} &{$22.12\pm0.04^\S$} &- &-\\
		IGR J19294+1816 & \textit{XMM-Newton}/PN &0841190101 & PrimeFullWindow &13-10-2019 (67) &20.57$\pm$0.03 &PU &$2.74\pm0.09$\\
		V 0332+53 & \textit{XMM-Newton}/PN & 0763470301 & FastTiming &10-09-2015 (32) &423.68$\pm$0.19 &PU &$194.7\pm0.1$\\
		  & & 0763470401 & FastTiming &16-09-2015 (31) &280.12$\pm$0.12 &PU &$162.0\pm0.1$\\
        &\textit{NuSTAR}/FPM &{80102002004} &{N/A} &{10-09-2015 (15)} &{$209.40\pm0.10$} &{-} &{-}\\
        & &{80102002006} &{N/A} &{16-09-2015 (17)} &{$136.60\pm0.10$} &{-} &{-}\\
        & &90202031002 &N/A &30-07-2016 (44) &$24.52\pm0.02^{||}$ &- &-\\
        & &90202031004 &N/A &31-07-2016 (44) &$19.64\pm0.02^{||}$ &- &-\\
		Cen X--3 &\textit{XMM-Newton}/PN &0400550201 &FastTiming &{12-06-2006} (80) &$452.10\pm0.07$ &No &{-}\\
            XTE J1858+034 &\textit{NuSTAR}/FPM &90501348002 &N/A &03-11-2019 (90) &$17.27\pm0.02^{||}$ &- &-\\
		\hline
	\end{tabular}
 }
 \begin{tablenotes}
    \item   $^\dagger$  Before pile-up correction.
    \item   $^\ddagger$ After pile-up correction.
    \item   $^\S$ This \textit{XMM-Newton} observation is affected by X-ray loading.
    \item   $^\mathparagraph$ Across all five PCUs of \textit{RXTE} that were ON during this observation.
    \item   $^{||}$ In FPMA module.
    \item  {$^{\S}$ XX=03, 04, 06, 11, 12, 15--17, 22, 32, 33, 35--42, 45--47, 49. The start date of OID 6203800103 and total on-source exposure across all the observations (out of a total elapsed time duration of 5933 ks) are given. The average count rate in 0.5--10 keV of the combined lightcurve from the observations is given.}
 \end{tablenotes}
\end{table*}

\subsection{4U 1626--67}\label{sec:4u1626}
4U 1626--67 is a persistent Low mass Ultra-compact X-ray binary in which a 130 mHz spinning strongly magnetized ($\sim3\times10^{12}$ G) neutron star (\citealt{coburn_rxte_review}, and references therein) is accreting Oxygen and Neon rich matter from a companion by Roche lobe overflow, that is assessed from the presence of Oxygen/Neon emission complex at 1 keV in its energy spectrum \citep{schulz_1626_chandrahetg_ONeaccdisk}. The X-ray {spectrum} of the persistent accreting pulsar 4U 1626--67 has been extensively studied, and it usually exhibits a soft blackbody component along with the power-law \citep{4u1626_camero20124u}. A QPO at 48 mHz is well established in the source at multiple wavelengths, in the Optical band with 3\% rms amplitude \citep{deepto_1626_opricalQPO}, UV band with 3\% rms amplitude in the near-UV to 15\% in the far-UV \citep{deepto_1626_uvQPO} and X-ray band with 15\% rms (\citealt{shinoda_1626_qpo_discovery}, \citealt{kaur_1626_rxte}). The 48 mHz QPO is observed in 4U 1626--67 when the pulsar is spinning down, and lower frequency QPOs (36 and 40 mHz) are observed when the source is spinning up \citep[See][and references therein]{jain2010}.

\begin{figure}
    \centering
    \includegraphics[width=\columnwidth]{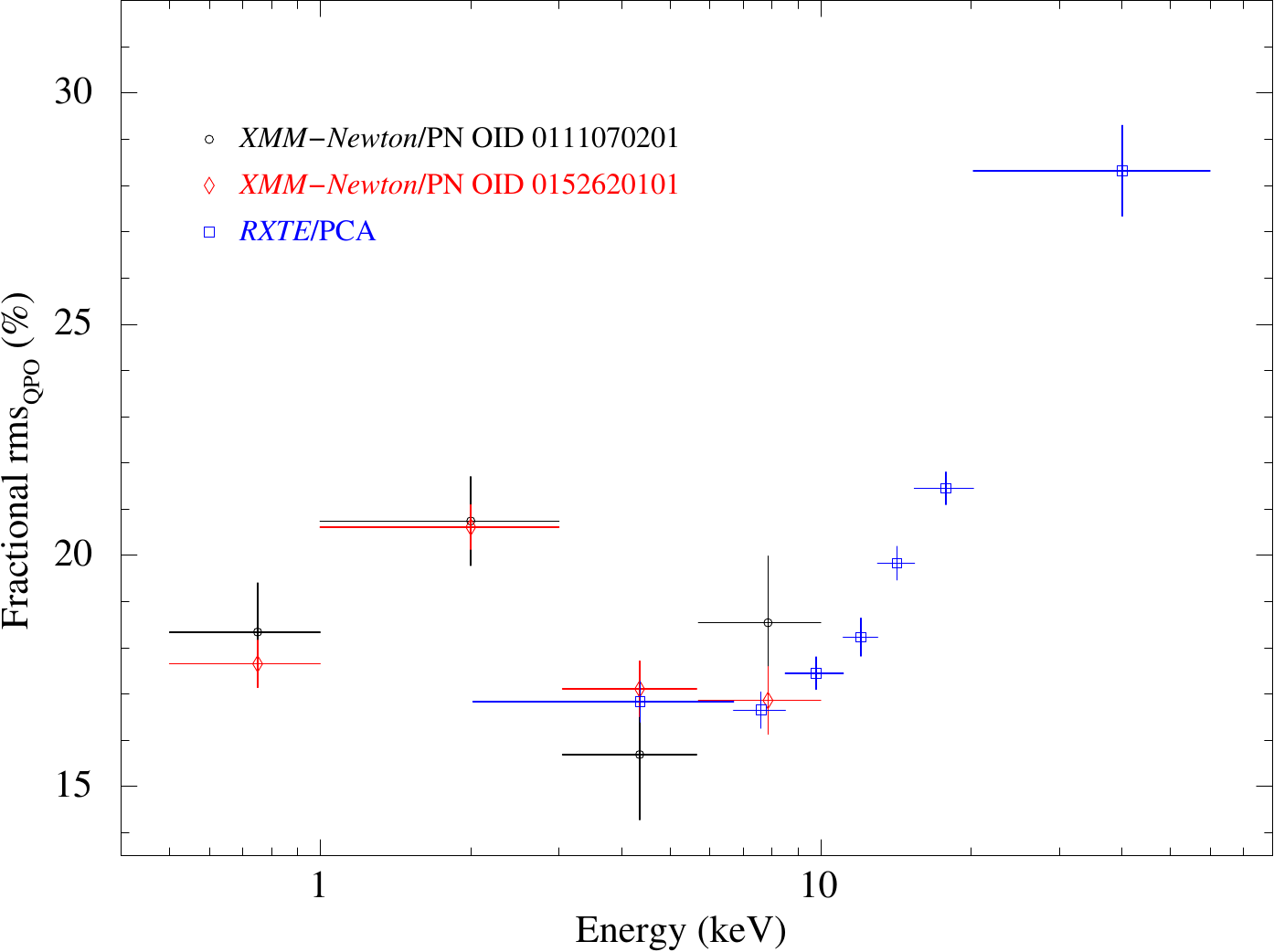}
    \includegraphics[width=\columnwidth]{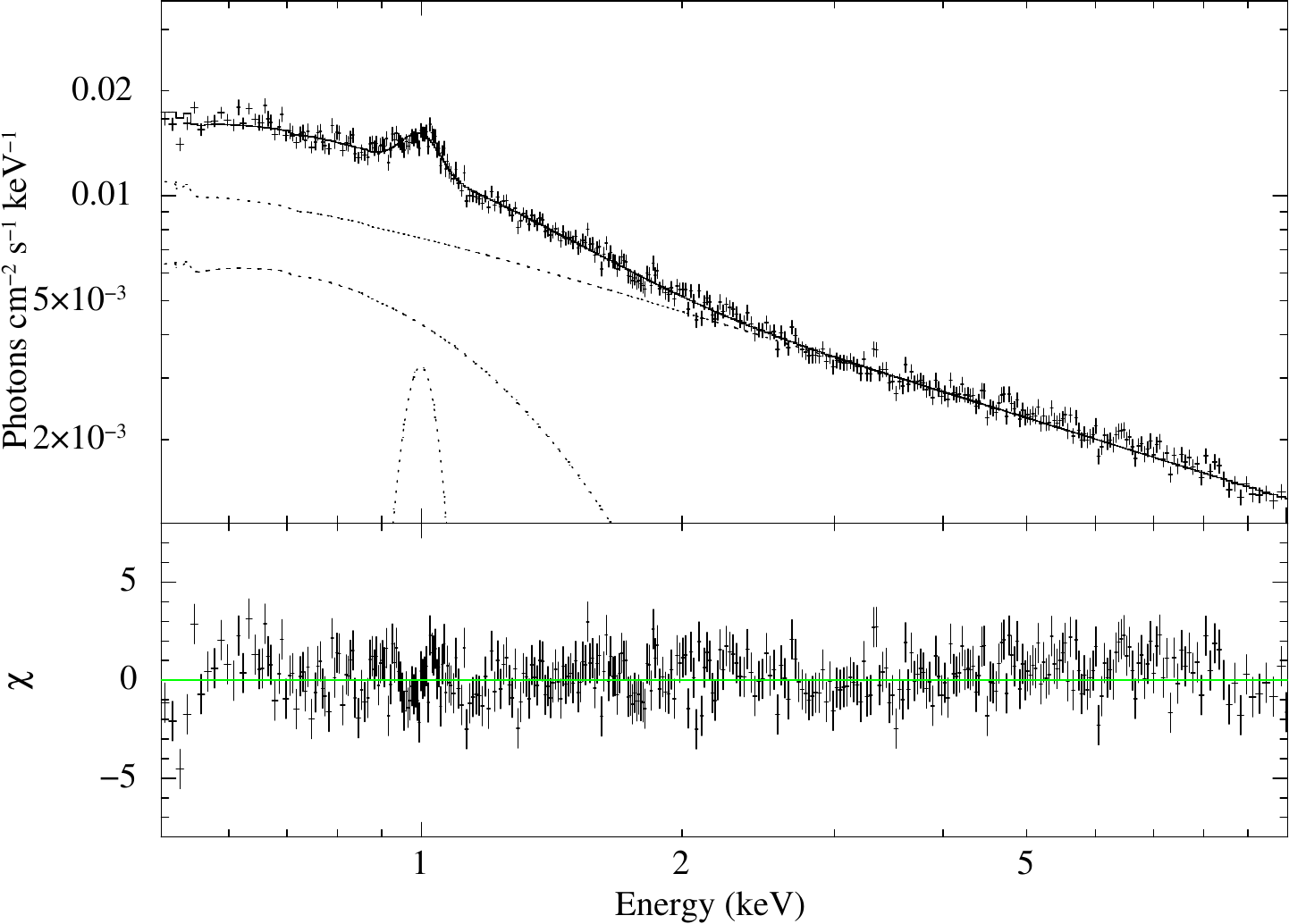}
    \caption{Top: The energy dependent variation of QPO fractional rms amplitude in 4U 1626--67 from \textit{XMM-Newton}/PN (red, black) and \textit{RXTE}/PCA (blue) observations. Bottom: The 0.5--10 keV unfolded spectrum and residuals to the best-fit model \texttt{tbabs * (powerlaw + gaussian + bbody)} on the \textit{XMM-Newton} Obs. ID 0111070201. Soft excess was modelled with a black body component of $kT\sim$ 0.3 keV. }
    \label{fig:4u1626-results}
\end{figure}

\begin{figure}
    \centering
    \includegraphics[width=\columnwidth]{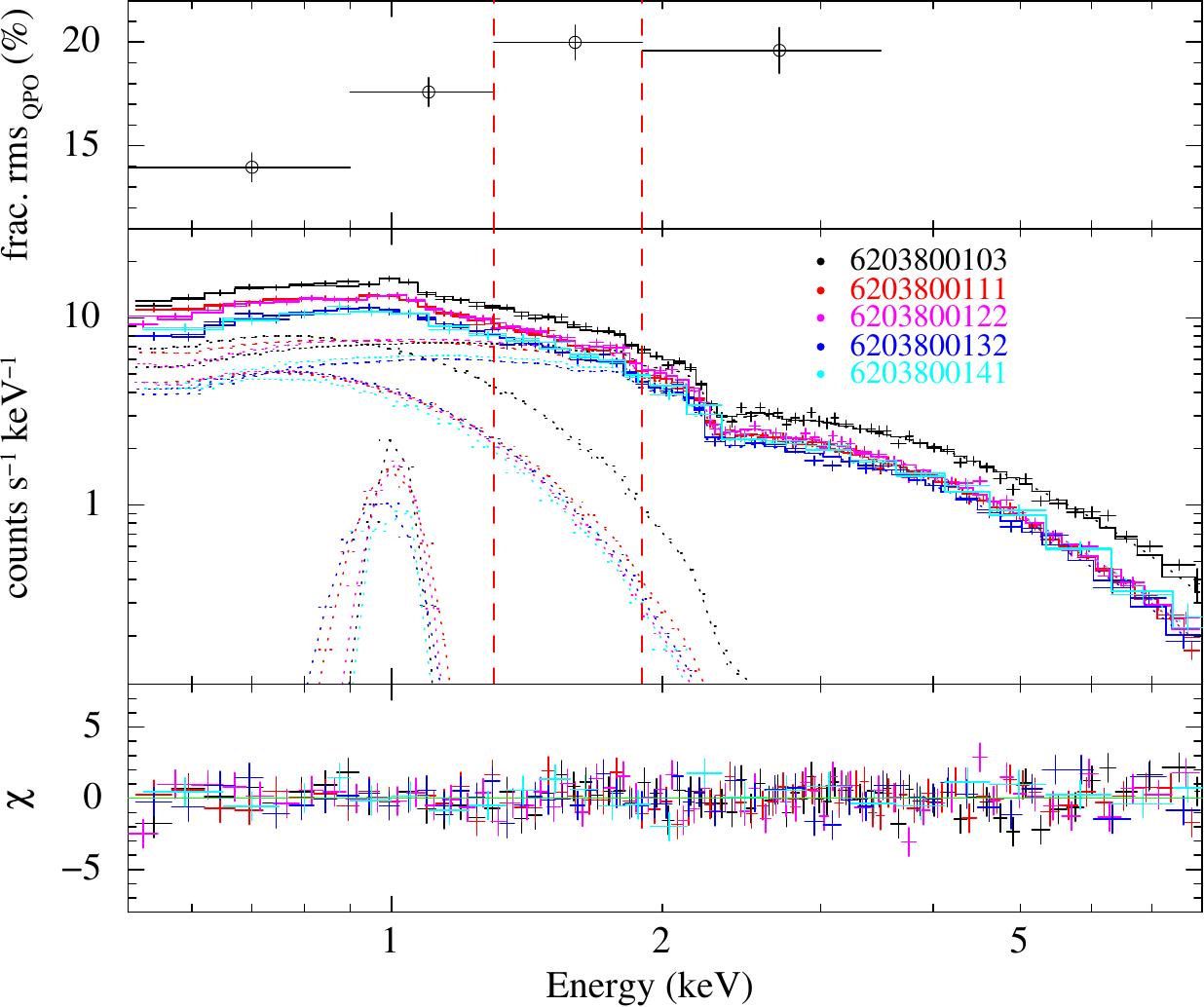}
    \caption{The top panel shows QPO fractional rms amplitude in 4U 1626$-$67 in different energy bands from combined PDS of all the \textit{NICER} observations. The middle panel shows the \textit{NICER} photon energy spectrum and the best fit composite spectral model \texttt{tbabs * (powerlaw + bbody + gaussian)} and the contribution from individual model components, especially the soft excess and the Neon line at 1 keV, on selected five observations. The vertical dashed lines indicate the energy range 1.3--1.9 keV, where the QPO rms peaks (See Section~\ref{disc:1626_qpo_softexcess}). The bottom panel shows residuals to the best fit spectral model.}
    \label{fig:1626-nicer-results}
\end{figure}

The 48 mHz QPO is present in two \textit{XMM-Newton} observations (Fig.~\ref{fig:4u1626_qpofit_xmmn}) \citep{1626_aru_xmm_2014_qpo}. To extend the {study of} QPO {variability} to higher energy bands,
 we also selected an \textit{RXTE}/PCA observation having $\sim147$ ks on-source exposure in which 48 mHz QPO was reported by \cite{kaur_1626_rxte}. {The band-limited noise between 12 to 192 mHz in the PDS of both the \textit{XMM-Newton}/PN observations were fitted with a \texttt{powerlaw} and the QPO at $48$ mHz was fitted with a \texttt{Lorentzian} (Fig.~\ref{fig:4u1626_qpofit_xmmn}). The band-limited noise between 12 to 192 mHz in the \textit{RXTE}/PCA observation was fitted with a \texttt{powerlaw}, a broad low frequency \texttt{Gaussian} at $\sim12$ mHz, a narrow \texttt{Lorentzian} at $\sim80$ mHz for the QPO-pulsar-sideband \citep{kommers1998sidebands_4u1626}, and the QPO at $48$ mHz was fitted with a \texttt{Lorentzian} (Fig.~\ref{fig:4u1626_qpofit_rxtepca}).} We determined the energy-resolved variation of QPO in the 0.5--60 keV energy band ({Fig.~\ref{fig:4u1626-results}}). {The fractional rms amplitude of QPO exhibits a steadily rising trend in 3--60 keV, increasing from about 15\% to 28\%. However, it deviates from this trend below 3 keV, where the QPO rms is high at around 20\% in 1--3 keV and 18\% in 0.5--1 keV. To study the QPO characteristics below 3 keV in finer energy segments, we analysed the combined PSD from multiple \textit{NICER} observations (Table~\ref{tab:qpoobscatalogue}) of 4U 1626$-$67 having the 48 mHz QPO. The band-limited noise between 12 to 192 mHz in the PDS of \textit{NICER} observation was fitted with a \texttt{powerlaw} and the QPO at $48$ mHz was fitted with a \texttt{Lorentzian} (Fig.~\ref{fig:4u1626_qpofit_nicer}). We estimated the fractional rms amplitude of the QPO in five energy bands; 0.5--0.9 (14\%), 0.9--1.3 (17\%), 1.3--1.9 (20\%), and 1.9--3.5 keV (19\%) (Fig.~\ref{fig:1626-nicer-results}). A detailed discussion on the energy dependence of QPO rms below 3 keV, especially from \textit{NICER} observations, is given in Section~\ref{disc:1626_qpo_softexcess}.}

We analysed {the 0.5--10 keV} spectra from both \textit{XMM-Newton} observations {and the 0.5--8 keV spectra from all the \textit{NICER} observations. We adopted the spectral model from \cite{1626_aru_xmm_2018}.} A {\texttt{powerlaw}} could fit the continuum, and the strong Neon emission complex at $\sim$ 1 keV {was} fitted with a \texttt{Gaussian} profile. The spectra also showed the presence of soft excess, which was modeled with a black body component of $kT_\textrm{BB}\sim$ 0.3 keV. {The finally used spectral model is \texttt{tbabs*(powerlaw+bbody+gaussian)} } (Figs.~\ref{fig:4u1626-results},~\ref{fig:1626-nicer-results}). {The best fitting spectral model parameters for all the observations are given in Table.~\ref{tab:4u1626_spectra}.}

\begin{table*}
    \centering
    \caption{The results of spectral fit performed on two observations of 4U 1626--67 having QPO. {All observations were fitted with the composite spectral model \texttt{tbabs*(powerlaw+bbody+gaussian)}.} The best fit model parameter values and their errors are given. The errors quoted on all parameters are their 90\% confidence ranges.}
    \scalebox{1.0}{
    \begin{tabular}{l|cccc}
        \hline
         &\multicolumn{2}{c}{\textit{XMM-Newton}/PN}& \textit{NICER}\\
         Obs. ID& 0111070201 &0152620101 &{62038001XX$^{\ell}$}\\
         \hline
         nH$^\dagger$ &$0.05\pm0.01$ &$0.07\pm0.01$ &{$0.07\pm0.03$} &\\
         PhoIndex ($\Gamma$) &$0.81\pm0.02$ &$0.80\pm0.01$ &{$0.65\pm0.11$}\\
         N\textsubscript{PL}$^\ddagger$ &($8.1\pm0.2)\times10^{-3}$ &$(6.8\pm0.1)\times10^{-3}$ &{$(5.1\pm0.6)\times10^{-3}$}\\
         kT\textsubscript{bbody} (keV) &$0.27\pm0.01$ &$0.24\pm0.01$ &{$0.23\pm0.02$}\\
         N\textsubscript{bbody}$^\S$ &$(1.2\pm0.1)\times10^{-4}$ &$(1.10\pm0.04)\times10^{-4}$ &{$(7.8\pm2.3) \times 10^{-5}$}\\
         E\textsubscript{Gauss} &$1.00\pm0.01$ &$1.01\pm0.01$ &{$1.00\pm0.02$}\\
             $\sigma$\textsubscript{Gauss} &$0.04\pm0.01$ &$0.02^{+0.01}_{-0.02}$ &{$0.01^\star$}\\
         N\textsubscript{Gauss}$^\P$ &$(3.9\pm0.5)\times10^{-4}$ &$(2.1\pm0.2)\times10^{-4}$ &{$(1.2\pm0.3) \times 10^{-4}$}\\
         Flux\textsubscript{2--20 keV} ($10^{-10}$ erg s\textsuperscript{-1} cm\textsuperscript{-2})& $2.90\pm0.01$ &$2.31\pm0.01$ &{$1.70\pm0.17$ to $6.05\pm0.17$}\\
         
         \\
         $\chi^2$ (dof) &1823 (1642) &2153 (1896) &{74(94) to 122(95)}\\
         $\chi^2$\textsubscript{red} &1.11 &1.14 &{0.79 to 1.29}\\
         \hline
    \end{tabular}
    }
    \begin{tablenotes}
        \item $^\star$ Frozen.
        \item $^\dagger$ in units of 10$^{22}$ atoms cm$^{-2}$.
        \item $^\ddagger$ Normalization in units of photons s$^{-1}$ cm$^{-2}$ keV$^{-1}$ at 1 keV.
        \item $^\S$ Normalization in units of $10^{37}$ ergs s$^{-1}$ kpc$^{-2}$.
        \item $^\P$ Total photons s$^{-1}$ cm$^{-2}$ in the gaussian line.
        \item  {$^{\ell}$ XX = 03, 04, 06, 11, 12, 15--17, 22, 32, 33, 35--42, 45--47, and 49. {The quoted best-fitting parameter values are the error-weighted mean and standard error of the parameter estimates from individual observations. The ranges of variation of flux and fit-statistic are given. The best fitting model on observations XX=38 and 45 having relatively short exposure duration did not require a blackbody component.}}
    \end{tablenotes}
    \label{tab:4u1626_spectra}
\end{table*}

\subsection{IGR J19294+1816}
IGR J19294+1816 is a transient High mass X-ray binary in which a spinning ($\nu$\textsubscript{spin}$\sim$83 mHz) strongly magnetized ($\sim4\times10^{12}$ G) neutron star \citep{1916_tsygankov_crsf} accretes matter from a Be-type companion star \citep{1916_be_discovery_IR} in a 117 d orbit \citep{1916_orbitalP}. QPO was reported in the \textit{AstroSat}/LAXPC observation at 32 mHz during {the luminosity decline phase of a Type-I} outburst of the source during the periastron passage in 2019 \citep{19294_xmm_outburst_Atel,graman2021astrosat}.

\begin{figure}
    \centering
    \includegraphics[width=\columnwidth]{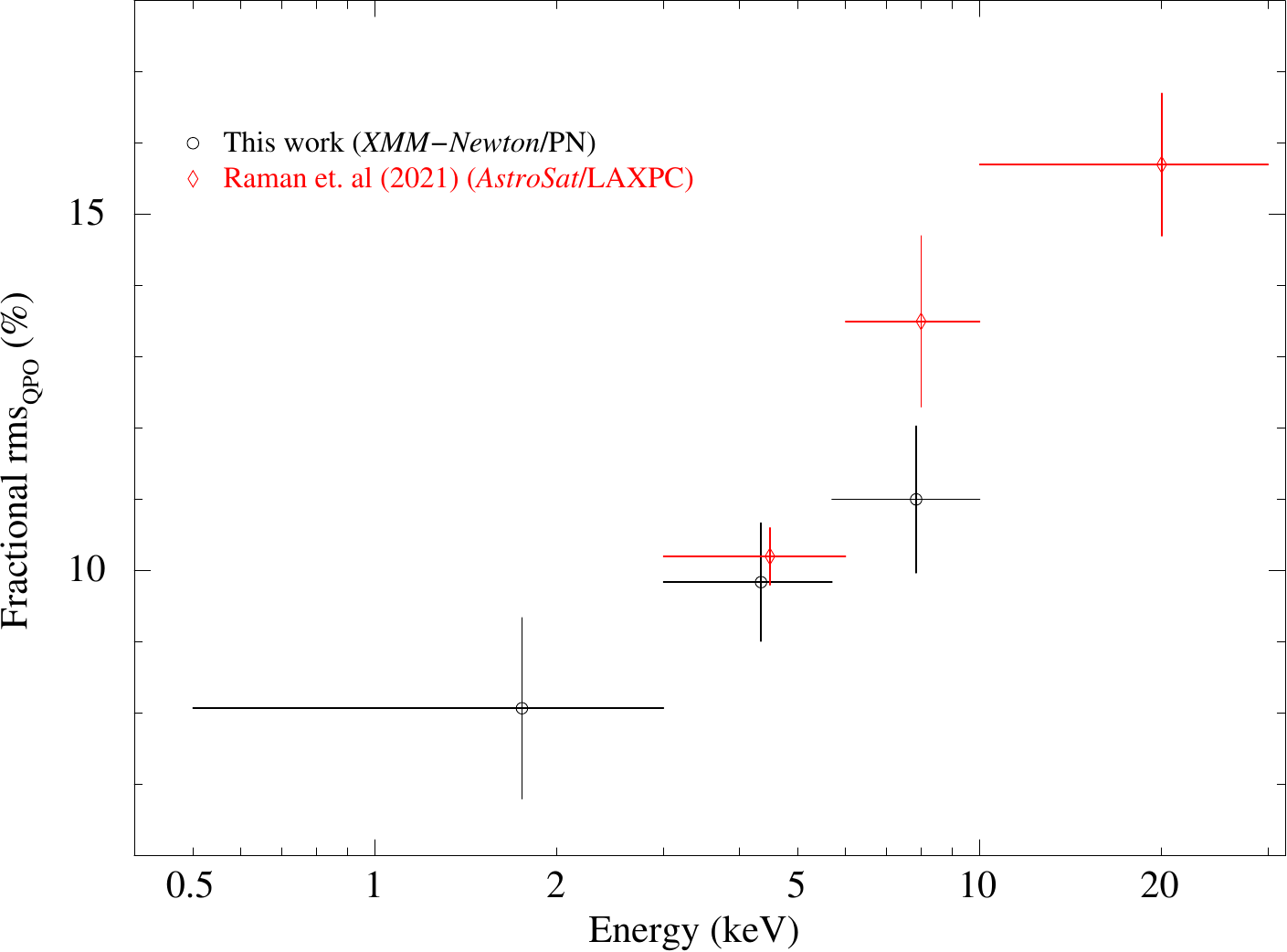}
    \includegraphics[width=\columnwidth]{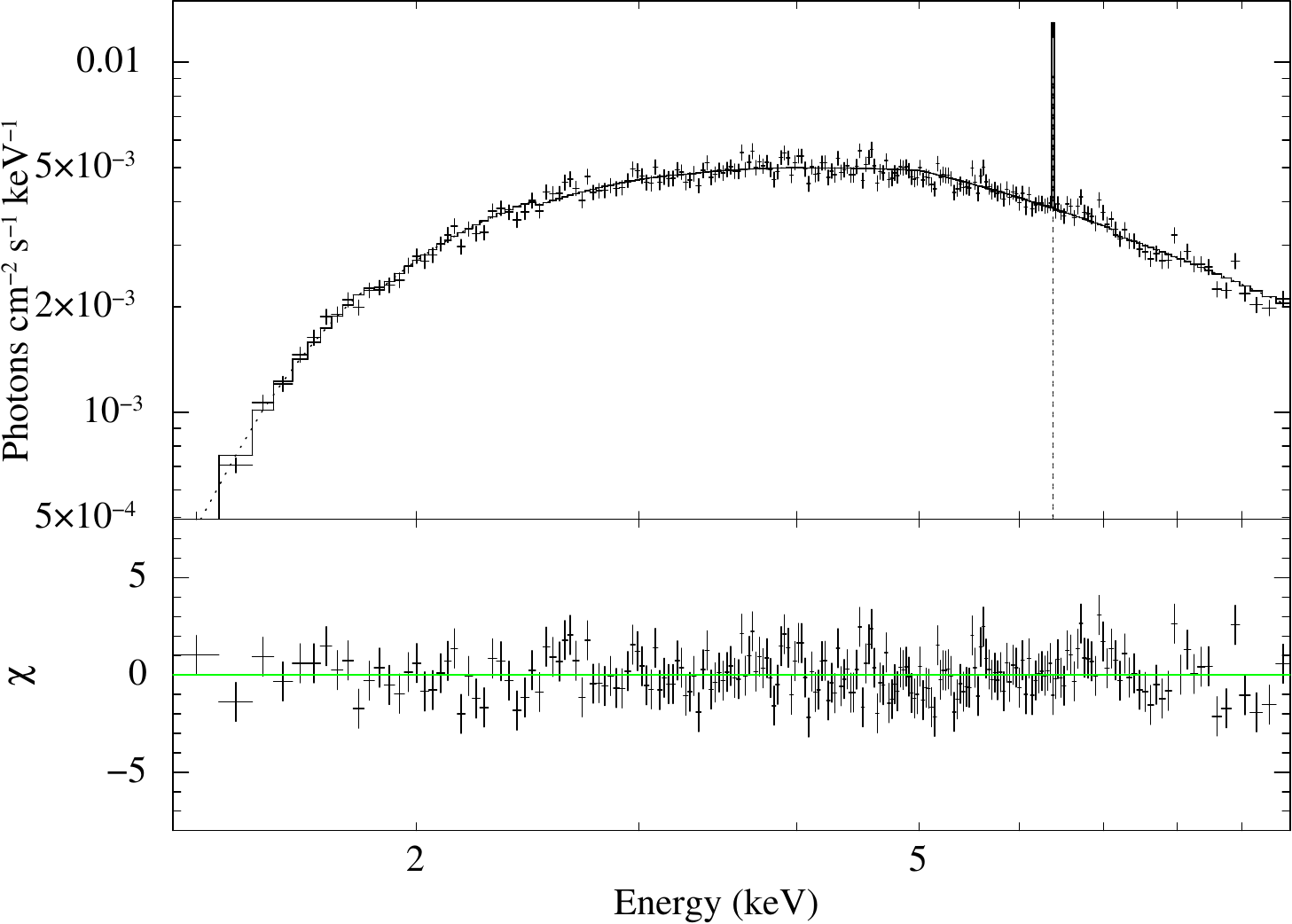}
    \caption{Top: The energy dependent variation of QPO fractional rms amplitude in IGR J19294+1816 from \textit{XMM-Newton} (\textit{this work}; black) and \textit{Astrosat}/LAXPC (red; \citealt{graman2021astrosat}). Bottom: The 0.5--10 keV unfolded \textit{XMM-Newton}/PN spectrum and residuals to the best fit model \texttt{tbabs * (powerlaw * highecut + gaussian)}.}
    \label{fig:igrj19294}
\end{figure}

We detected QPO in an \textit{XMM-Newton} observation at 30 mHz (Fig.~\ref{fig:igrj19294_qpofit}) {during the rising phase of the same 2019 outburst}. {The band-limited noise in PDS between 8 mHz and 128 mHz was fitted with a \texttt{powerlaw} and the QPO at $30$ mHz was fitted with a \texttt{Lorentzian} (Fig.~\ref{fig:igrj19294_qpofit}).} The fractional rms amplitude of QPO shows an increasing trend in the 0.5--10 keV range {and the trend continues till 30 keV (Fig.~\ref{fig:igrj19294}), as evident from} the results of \cite{graman2021astrosat}.

{We tried to fit different continuum models to the 0.5--10 keV \textit{XMM-Newton}/PN spectrum and found that \texttt{powerlaw*highecut} and \texttt{powerlaw*FDcut} gave good fits to the continuum with acceptable values for all the spectral parameters.} Residuals left by the iron fluorescence line could be modelled with a \texttt{Gaussian} emission profile. Since the width of the \texttt{Gaussian} was not constrained by the fit, we fixed it to 10 eV{ (the spectral resolution of EPIC-PN at 6.4 keV is 150 eV). The finally used spectral model is \texttt{tbabs*(powerlaw*highecut+gaussian)} (Fig.~\ref{tab:igrj19294_spectra}). The best fitting spectral model parameters are given in Table.~\ref{tab:igrj19294_spectra}.}

\begin{table}
    \centering
    \caption{The results of spectral fit performed on one observation of IGR J19294+1816 having QPO. The best fit model parameter values and their errors are given. The errors quoted on all the parameters are their 90\% confidence ranges.}
    \begin{tabular}{l|c}
        \hline
         Obs. ID& 0841190101\\
         \hline
         nH$^\dagger$ &$3.62\pm0.17$\\
         PhoIndex ($\Gamma$) &$0.41\pm0.07$\\
         N\textsubscript{PL}$^\ddagger$ &$0.010\pm0.001$\\
         cutoffE (keV) &$4.96\pm0.30$\\
         foldE (keV) &$7.03\pm0.50$\\
         E\textsubscript{Gauss,Fe} &$6.39\pm0.01$\\
         $\sigma$\textsubscript{Gauss,Fe} &0.01$^\star$\\
         N\textsubscript{Gauss,Fe}$^\P$ &$(2.9\pm0.4)\times10^{-4}$\\
         Flux\textsubscript{2--20 keV} ($10^{-10}$ erg s\textsuperscript{-1} cm\textsuperscript{-2}) &$4.04\pm0.02$\\
         \\
         $\chi^2$ (dof) &679 (669)\\
         $\chi^2$\textsubscript{red} &1.02\\
         \hline
    \end{tabular}
    \begin{tablenotes}
        \item $^\star$ Frozen.
        \item $^\dagger$ in units of 10$^{22}$ atoms cm$^{-2}$.
        \item $^\ddagger$ Normalization in units of photons s$^{-1}$ cm$^{-2}$ keV$^{-1}$ at 1 keV.
        \item $^\P$ Total photons s$^{-1}$ cm$^{-2}$ in the gaussian line.
    \end{tablenotes}
    \label{tab:igrj19294_spectra}
\end{table}

\subsection{V 0332+53}
V 0332+53 is a transient High mass X-ray binary in which a 227 mHz pulsar accretes matter from a Be-type companion star in a 34 d eccentric ($e\sim0.3$) orbit \citep{v0332_stella_intro}. QPO was discovered in the source during a {Type-II} outburst {in 1989} from \textit{Ginga}/LAC observations by \cite{V0332_Ginga_takeshima_QPO-dicovery} at 51 mHz with about 5\% rms in the 2.3--37.2 keV energy band. The 51 mHz QPO and another 220 mHz QPO (centred at the NS spin frequency) were found in later observations during {the 2004/2005 Type-II} outburst decay phase {with} \textit{RXTE}/PCA \citep{qu2005discovery_v0332}, and \textit{INTEGRAL}/JEM-X and IBIS \citep{v0332_mowlavi_integral_twinQPO} observations. \cite{qu2005discovery_v0332} also observed that the QPO central frequency does not vary with flux and that the QPO rms stays constant as a function of photon energy till 10 keV and drops beyond 10 keV.

\begin{figure}
    \centering
    \includegraphics[width=\columnwidth]{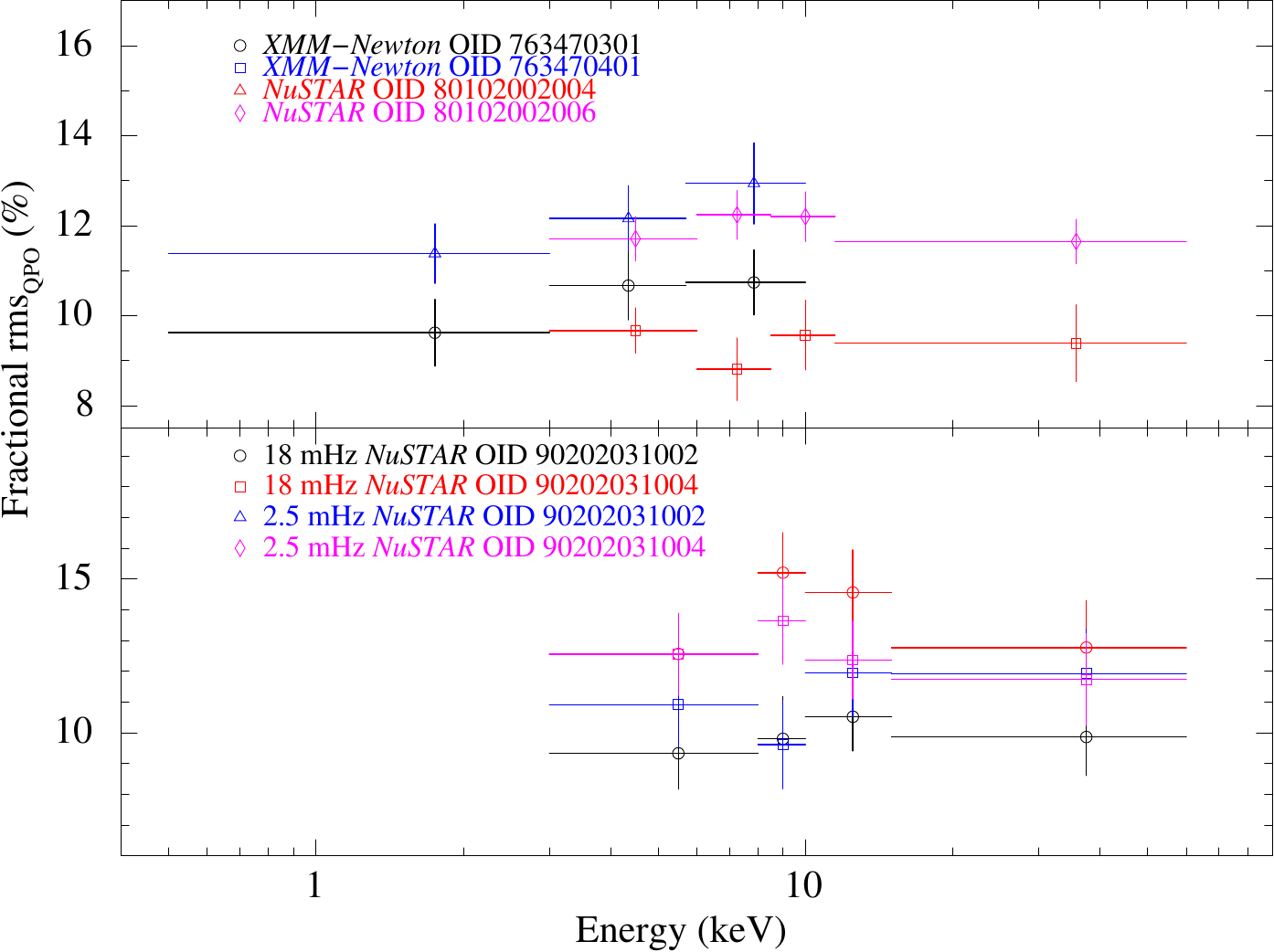}
    \includegraphics[width=\columnwidth]{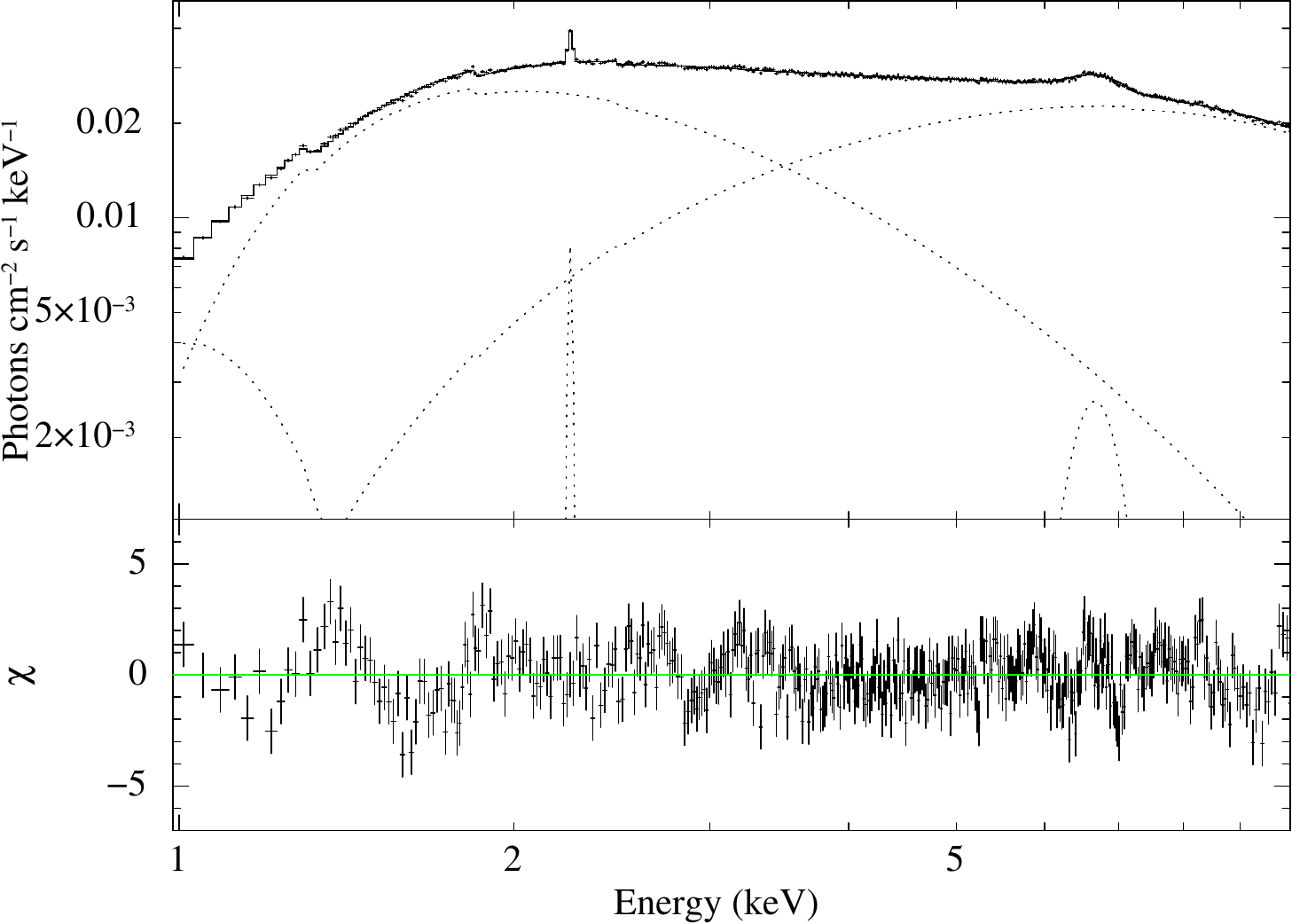}
    \includegraphics[width=\columnwidth]{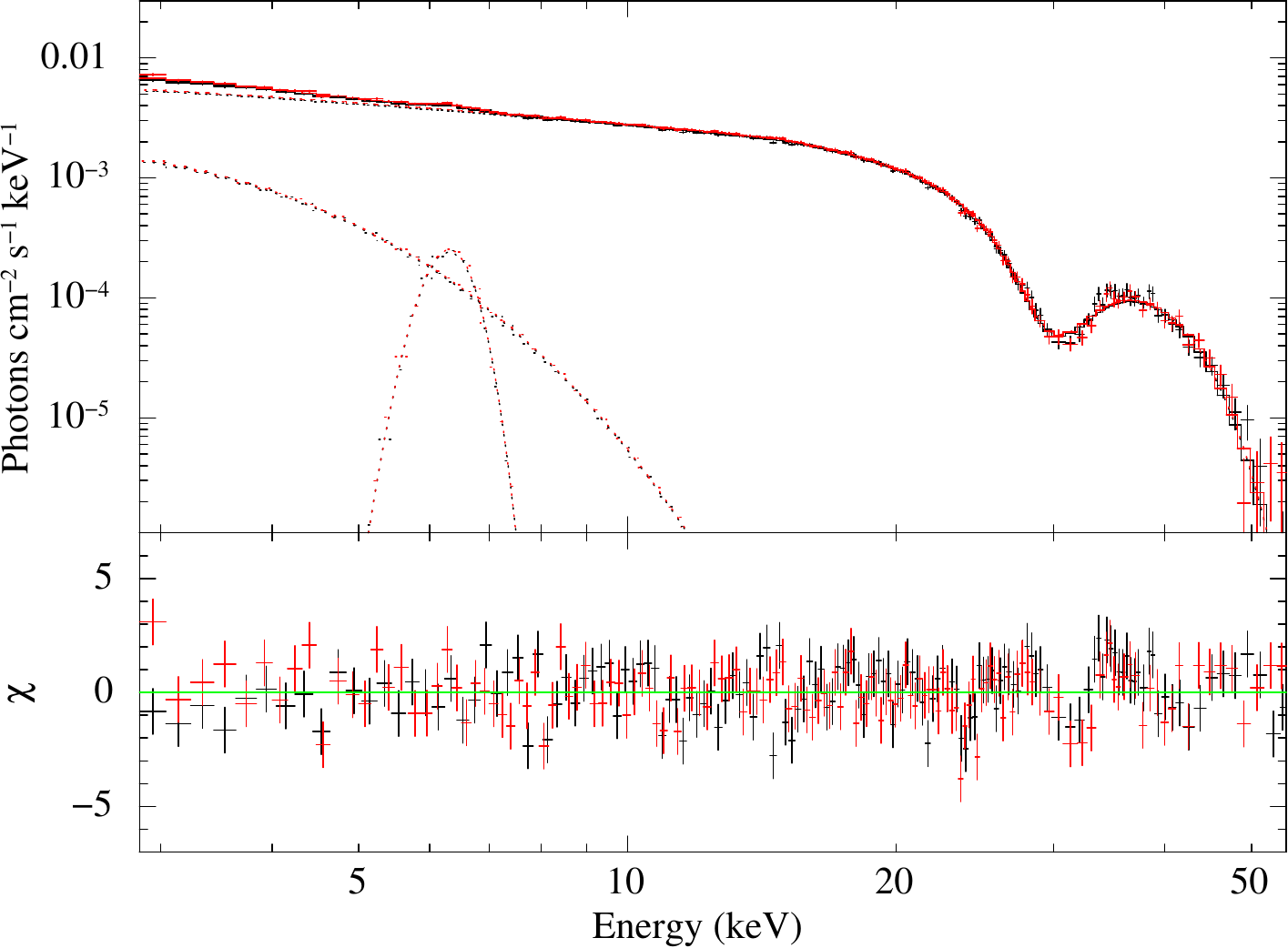}
    \caption{ Top: The top and bottom panels show the energy dependence of the $\sim$40 mHz, and $\sim$2.5, $\sim$18 mHz QPOs in V 0332+53. Middle: The 1$-$10 keV unfolded spectrum and residuals to the best fit spectral model \texttt{tbabs * (NPEX + gaussian + bbody)} on the \textit{XMM-Newton} observation OID 0763470401. Bottom: The 3$-$55 keV unfolded spectrum and residuals to the best fit spectral model \texttt{tbabs * (powerlaw * mplcut + gaussian + bbody)} on the \textit{NuSTAR} observation OID 90202031002.}
    \label{fig:v0332_qpormsvE}
\end{figure}

Two {quasi-simultaneous} observations from \textit{XMM-Newton} {and \textit{NuSTAR} during the 2015 Type-II outburst} show the presence of a QPO at about 40 mHz (Figs.~\ref{fig:v0332_qpofit-XMM_2015},~\ref{fig:v0332_qpofit-NuSTAR_2015}), and two observations from \textit{NuSTAR} { during a Type-I outburst in 2016} show twin QPOs at 2.5 mHz and 18 mHz (Fig.~\ref{fig:v0332_qpofit-NuSTAR_2016}). { The \textit{XMM-Newton}/PN and \textit{NuSTAR} PSDs of the 2015 observations in the $0.005-0.24$ Hz range were fitted with a wide \texttt{Lorentzian} for the band-limited noise and another narrow \texttt{Lorentzian} at $40$ mHz for the QPO (Figs.~\ref{fig:v0332_qpofit-XMM_2015},~\ref{fig:v0332_qpofit-NuSTAR_2015}). The \textit{NuSTAR} PSDs of 2016 observations in the $0.005-0.12$ Hz range were fitted with \texttt{powerlaw * highecut} for the broadband noise and two \texttt{Lorentzian} profiles at $\sim2.5$ mHz and $\sim18$ mHz for the twin QPOs (Fig.~\ref{fig:v0332_qpofit-NuSTAR_2016}). The rms amplitude of 40 mHz QPO in the 2015 observations, and the 2.5 mHz and 18 mHz QPO in the 2016 observations does not exhibit any secular trend in energy (Fig.~\ref{fig:v0332_qpormsvE}). The quality factor of the 2.5 mHz QPO is higher than the 18 mHz and 40 mHz QPOs. The \textit{NuSTAR} PSDs of both the 2016 observations also show the presence of a faint QPO-like structure at about 100 mHz (Fig.~\ref{fig:v0332_qpofit-NuSTAR_2016}). {We would like to point out that an additional sixth \textit{NuSTAR} observation also exists during the rising phase, before the peak of the 2015 outburst. However, the PDS of this \textit{NuSTAR} observation has a complex shape with four QPO-like features at 2, 9, 57 and 225 mHz frequencies. However, the analysis and interpretation of that observation are beyond the scope of this work.} 

We analysed the spectra from the two \textit{XMM-Newton} and four \textit{NuSTAR} observations. The \textit{XMM-Newton}/PN spectra were re-binned with the {\small HEASOFT} tool {\small FTGROUPPHA} such that each spectral bin has a minimum number of counts equivalent to a signal-to-noise ratio (SNR) of $\geq50$ per bin. We tried to fit different continuum models to the 1--10 keV \textit{XMM-Newton}/PN spectra of both observations and found that absorbed \texttt{powerlaw}, \texttt{compTT} and \texttt{NPEX} models could fit the continuum well. Residuals were left at around 6.7 keV, and 2.2 keV in both observations and we used two \texttt{Gaussian}s centred at 6.7 keV and 2.2 keV to model them. The width of the 2.2 keV line was not constrained by the fit in both observations and therefore we froze it to 10 eV. We finally used the \texttt{NPEX} model as other continuum models left residuals around the iron fluorescence region. Soft-excess-like residuals were visible at the lowest energies only in OID 0763470301 and we used a \texttt{bbody} with $kT$$\sim$0.1 keV to model it. The finally used model for the two \textit{XMM-Newton} observations are \texttt{tbabs*(NPEX+gaussian$_1$+gaussian$_2$+bbody)}, and \texttt{tbabs*(NPEX+gaussian$_1$+gaussian$_2$)}.

The \textit{NuSTAR}/FPM spectra were re-binned with the same tool as per the optimal binning scheme developed by \cite{kaastra2016optimal}. We found that the continuum of the 3$-$55 keV \textit{NuSTAR} spectra could be well-modelled with an absorbed \texttt{powerlaw} with a high energy cutoff (\texttt{highecut}, \texttt{newhcut}) or \texttt{compTT}. We used the simplest high energy cutoff model \texttt{highecut}. An additional \texttt{gabs} centred at the cutoff energy was included to flatten the kink at E$_\mathrm{cut}$ (\texttt{mplcut}; \citealt{coburn_rxte_review}).  The absorption column density could not be constrained by the fit, so we froze it to the Galactic value of $0.7\times10^{22}$ atoms cm$^{-2}$. Significant residuals were left at $\sim$30 keV (the fundamental CRSF; \citealt{vybornov_v0332_nustar_crsf}) and they could be modelled with a \texttt{gabs} or a \texttt{cyclabs} model, and we adopted \texttt{cyclabs} owing to its better fit-statistic. However, to flatten the residuals completely in 2015 observations, we added one more narrow \texttt{gabs} at $\sim$30 keV. This could be due to poor modelling of the underlying continuum or a complex CRSF profile \citep{v0332_nustar_qpo_noqpo_doroshenko2017luminosity}. Residuals were also left at $\sim$55 keV due to the harmonic CRSF \citep{vybornov_v0332_nustar_crsf}, which could be eliminated by the harmonic CRSF parameters of the \texttt{cyclabs}. Residuals were left at 6.4 keV (iron fluorescence), and we fitted it with a \texttt{Gaussian}. Soft-excess-like residuals were visible and we modelled it with a \texttt{bbody} of $kT$$\sim$0.3 keV in 2015 observations and $kT$$\sim$1.0 keV in 2016 observations. The final used model for the 2015 and 2016 \textit{NuSTAR} observations are \texttt{tbabs*(powerlaw*mplcut*cyclabs*gabs+bbody+gaussian)}, and \texttt{tbabs*(powerlaw*mplcut*cyclabs+bbody+gaussian)}, respectively. The best fitting spectral model parameters are given in Table~\ref{tab:v0332_spectra}.}

\begin{table*}
    \centering
    \caption{ The results of spectral fit performed on four observations of V 0332+53 having a QPO. The best fit model parameter values and their errors are given. The errors quoted on all the parameters are their 90\% confidence ranges.}
    \scalebox{0.8}{
    \begin{tabular}{l|cccccccc}
        \hline
         &\multicolumn{2}{c}{\textit{XMM-Newton}/PN}& \multicolumn{4}{c}{\textit{NuSTAR}$^\top$}\\
         Obs. ID& 0763470401 &0763470301 &80102002004 &80102002006 &90202031002 &90202031004\\
         Continuum &\texttt{tbabs*NPEX} &\texttt{tbabs*NPEX} &\texttt{tbabs*powerlaw*mplcut} &\texttt{tbabs*powerlaw*mplcut} &\texttt{tbabs*powerlaw*mplcut} &\texttt{tbabs*powerlaw*mplcut}\\
         Energy range &1--10 keV &1--10 keV &3--55 keV &3--55 keV &3--55 keV &3--55 keV\\
         \hline
         nH$^\dagger$ &$2.19\pm0.07$ &$2.17\pm0.06$ &$2.07_{-0.90}^{+1.04}$ &$0.43_{-0.44}^{+0.94}$ &$0.7^\star$ &$0.7^\star$\\
         PhoIndex ($\Gamma$) &$1.06\pm0.05$ &$1.04\pm0.05$ &$0.53\pm0.02$ &$0.49\pm0.02$ &$0.57_{-0.04}^{+0.03}$ &$0.57_{-0.02}^{+0.04}$\\
         N\textsubscript{CPL/PL-1}$^\ddagger$ &$0.15\pm0.01$ &$0.18\pm0.01$ &$0.18\pm0.01$ &$0.106_{-0.003}^{+0.006}$ &$0.011\pm0.001$ &$0.008\pm0.001$\\
         E\textsubscript{cut} &$3.41\pm0.03$ &$3.27\pm0.03$ &$10.90_{-0.21}^{+0.18}$ &$11.95_{-0.21}^{+0.18}$ &$15.17_{-0.39}^{+0.84}$ &$16.31_{-0.36}^{+0.39}$\\
         E\textsubscript{fold} &$-$ &$-$ &$17.23_{-1.32}^{+1.65}$ &$16.79_{-0.97}^{+0.96}$ &$26.21_{-9.24}^{+4.30}$ &$16.58_{-2.88}^{+4.34}$\\
         N\textsubscript{CPL/PL-2}$^\ddagger$ &$(2.93\pm0.08)\times10^{-3}$ &$(4.00\pm0.10)\times10^{-3}$ &$-$ &$-$ &$-$ &$-$\\
         E\textsubscript{Gauss,Fe} &$6.64\pm0.02$ &$6.66\pm0.02$ &$6.44\pm0.02$ &$6.46\pm0.03$ &$6.3\pm0.10$ &$6.37\pm0.07$\\
         $\sigma$\textsubscript{Gauss,Fe} &$0.25\pm0.02$ &$0.35_{-0.02}^{+0.03}$ &$0.34\pm0.04$  &$0.38\pm0.04$ &$0.36_{-0.15}^{+0.17}$ &$0.15_{-0.15}^{+0.13}$\\
         N\textsubscript{Gauss,Fe}$^\P$ &$(1.35\pm0.08)\times10^{-3}$ &$(2.34\pm0.15)\times10^{-3}$ &$0.005\pm0.001$ &$(3.77\pm0.44)\times10^{-3}$ &$(2.25)\times10^{-4}$ &$(1.23_{-0.37}^{+0.04})\times10^{-4}$\\
         E\textsubscript{Gauss,2} &$2.25\pm0.01$ &$2.25\pm0.01$ &$-$ &$-$ &$-$ &$-$\\
         $\sigma$\textsubscript{Gauss,2} &$0.01^\star$ &$0.01^\star$ &$-$ &$-$ &$-$ &$-$\\
         N\textsubscript{Gauss,2}$^\P$ &$(2.65\pm0.52)\times10^{-4}$ &$(3.53\pm0.60)\times10^{-4}$ &$-$ &$-$ &$-$ &$-$\\
         kT\textsubscript{bbody} (keV) &$-$ &$0.089\pm0.003$ &$0.33_{-0.06}^{+0.04}$ &$0.34\pm0.10$ &$0.88_{-0.07}^{+0.09}$ &$0.97_{-0.09}^{+0.08}$\\
         N\textsubscript{bbody}$^\S$ &$-$ &$0.10\pm0.02$ &$0.03_{-0.01}^{+0.05}$ &$0.01_{-0.01}^{+0.05}$ &$(3.38^{+0.69}_{-0.57})\times10^{-4}$ &$(3.17_{-0.58}^{+0.68})\times10^{-4}$\\
         Flux\textsubscript{2--20 keV} ($10^{-10}$ erg s\textsuperscript{-1} cm\textsuperscript{-2}) &$29.05\pm0.02$ &$35.01\pm0.03$ &$124.80\pm0.08$ &$83.07\pm0.07$ &$7.51\pm0.02$ &$6.08\pm0.01$\\
         const &$-$ &$-$ &$0.980\pm0.001$ &$0.978\pm0.0.002$ &$1.025\pm0.004$ &$1.04\pm0.01$\\
         \\
         $\chi^2$ (dof) &1502.83 (1256) &1600.27 (1417) &538.75 (401) &537.59 (377) &418.66 (331) &364.12 (323)\\
         $\chi^2$\textsubscript{red} &1.20 &1.13 &1.34 &1.43 &1.27 &1.13\\
         \hline
    \end{tabular}
    }
    \begin{tablenotes}
        \item $^\star$ Frozen.
        \item $^\dagger$ in units of 10$^{22}$ atoms cm$^{-2}$.
        \item $^\ddagger$ Normalization in units of photons s$^{-1}$ cm$^{-2}$ keV$^{-1}$ at 1 keV.
        \item $^\P$ Total photons s$^{-1}$ cm$^{-2}$ in the gaussian line.
        \item {$^\S$ Normalization in units of $10^{37}$ ergs s$^{-1}$ kpc$^{-2}$.}
        \item $^\top$ A CRSF at 29 keV and its harmonic is modelled with \texttt{cyclabs} and \texttt{gabs} models, the parameters of which are not given in this table.
    \end{tablenotes}
    \label{tab:v0332_spectra}
\end{table*}

\subsection{Cen X--3}
Cen X--3 is a persistent 208 mHz X-ray pulsar accreting matter from a massive O-type 20 M{$_\odot$} supergiant companion star in a compact, nearly circular 2.2 d orbit \citep[][and references therein]{suchY-cenx3_intro}. The accretion is expected to be partly from stellar wind and partly from the accretion disk due to the overall spin-up trend of the pulsar \citep{petterson_cenx3_wind_plus_disk_accretion}. Cen X--3 shows QPO at different frequencies ranging from 30 mHz (this work) to 90 mHz \citep{harsha_cenx3_2008ApJ...685.1109R}. \cite{harsha_cenx3_2008ApJ...685.1109R} and \cite{liu_Cenx3_40mhZqpo_insight} have shown that the QPO frequency or rms showed no dependence on the X-ray luminosity. Even though {\cite{harsha_cenx3_2008ApJ...685.1109R} reported that the 40 mHz QPO rms does not show any energy dependence in the 1996--1998 \textit{RXTE} observations}, recently \cite{liu_Cenx3_40mhZqpo_insight} showed that the 40 mHz QPO rms decreases from 13\% at 2 keV to about 9\% at 17 keV {in the 2020 \textit{Insight-HXMT} observations}. They further argued that the QPO frequency and rms have an orbital dependence and that the QPO photons show an overall soft lag.

\begin{figure}
    \centering
    \includegraphics[width=\columnwidth]{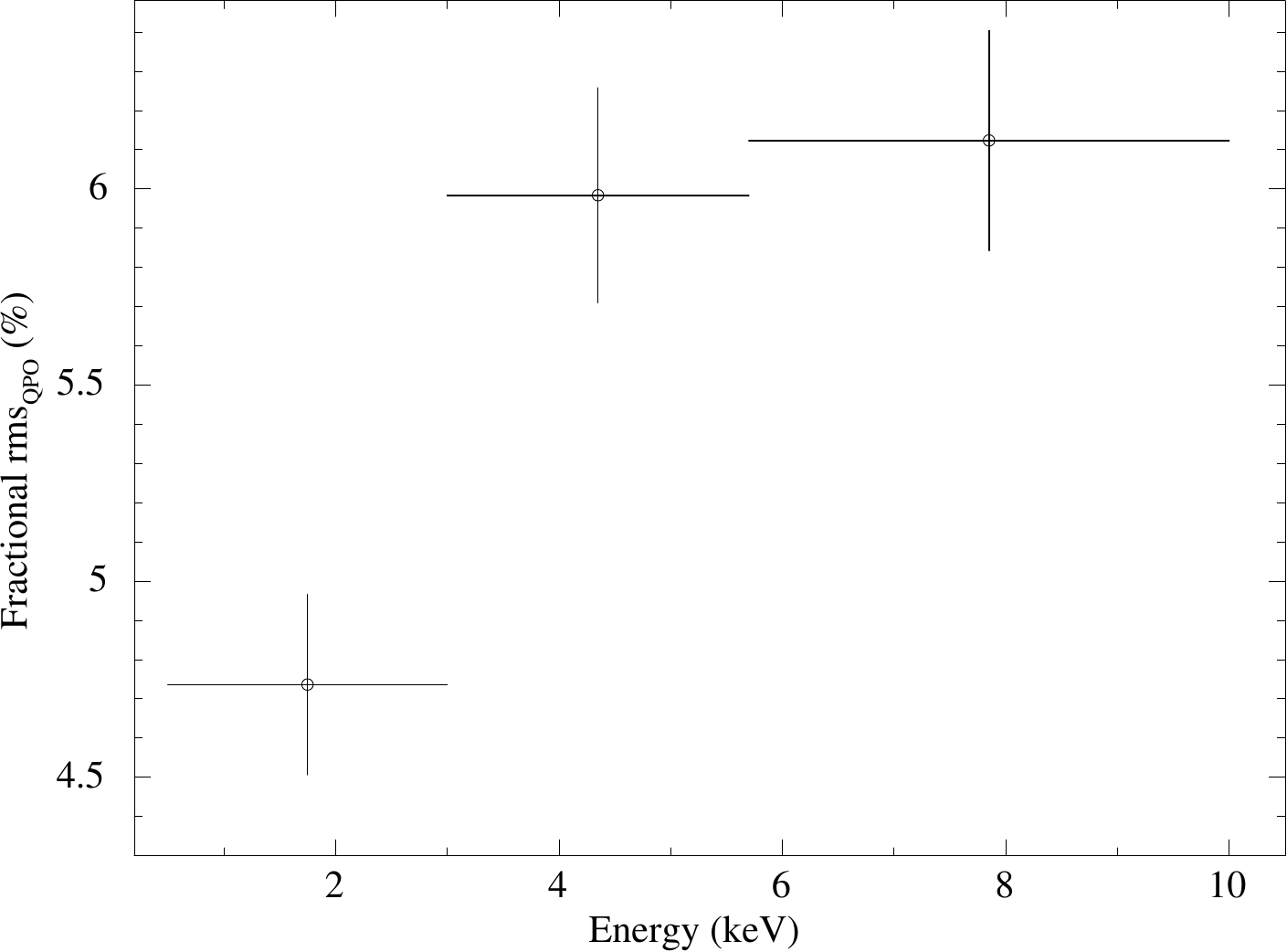}
    \includegraphics[width=\columnwidth]{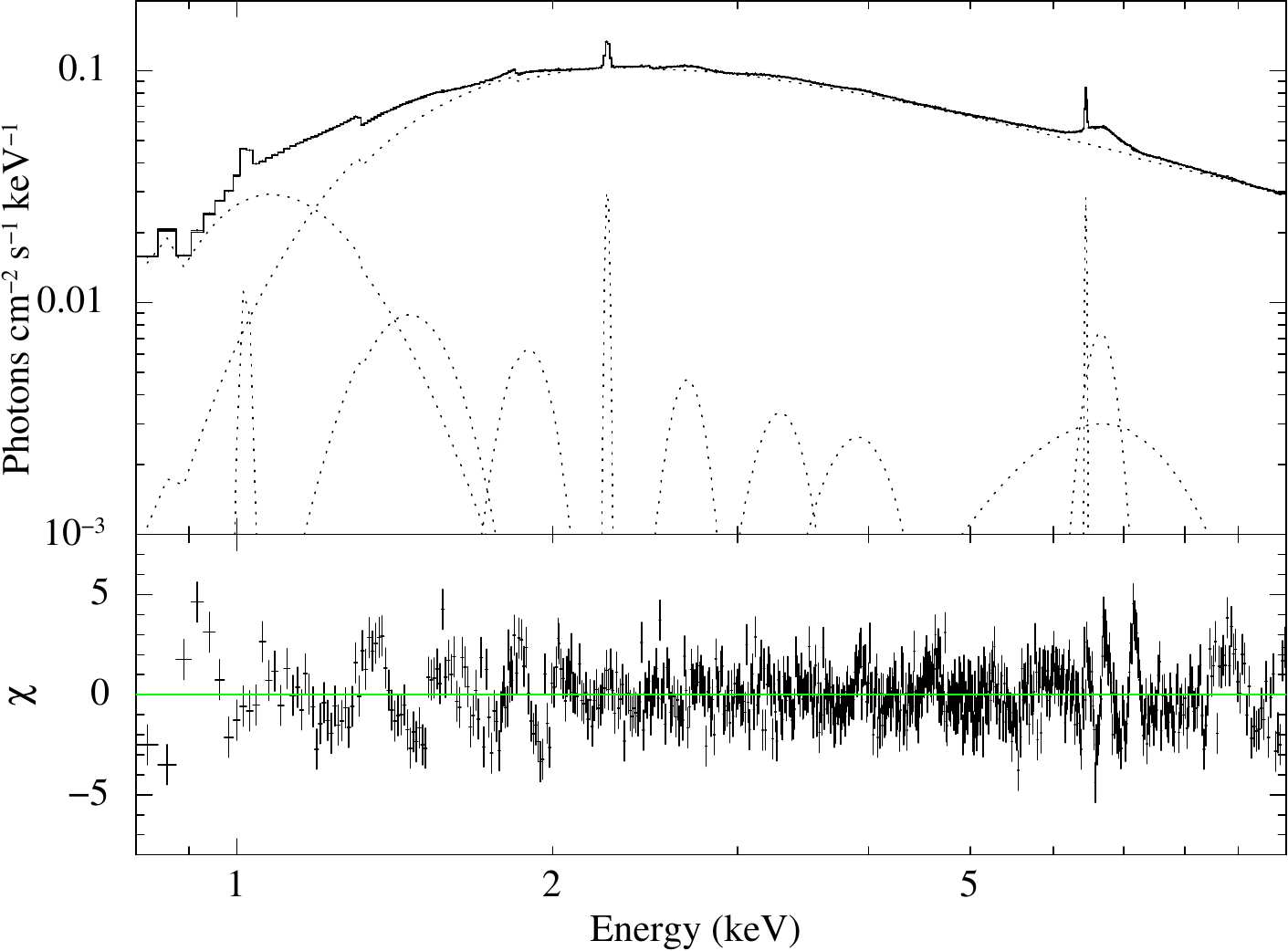}
    \caption{Top: The energy dependence of QPO rms in Cen X--3 from \textit{XMM-Newton}. Bottom: The 0.8--10 keV unfolded \textit{XMM-Newton}/PN spectrum and residuals to the best fit model \texttt{tbabs * (cutoffpl + bbody + (10)gaussians)}.}
    \label{fig:cenx3_results}
\end{figure}

We detected the presence of a QPO at around 30 mHz in one \textit{XMM-Newton} observation {that was conducted in the out-of-eclipse orbital phase. The PDS in 7.5$-$120 mHz range was modelled with the combination of a \texttt{powerlaw} for the band-limited noise and a \texttt{Lorentzian} at $30$ mHz for the QPO (Fig.~\ref{fig:cenx3_qpofit}).} The QPO rms shows a weak increasing trend with photon energy, {increasing} from about {5}\% in 0.5--3 keV to about 6\% in {3--}10 keV (Fig.~\ref{fig:cenx3_results}).

{The \textit{XMM-Newton}/PN spectrum was re-binned such that each bin has counts equivalent to a minimum SNR of 50. We adopted the spectral model of the source used in \cite{aftab2019x}. The spectrum was modelled with an absorbed \texttt{powerlaw} for the continuum, a \texttt{bbody} with kT$\sim0.1$ keV, and eleven \texttt{Gaussian} emission lines. The iron line emission region is complex, and it was modelled with a combination of two narrow \texttt{Gaussian}s at 6.44 keV and 6.65 keV, and a broad \texttt{Gaussian} at 6.65 keV. The complex iron region is possibly due to three distinct iron lines at 6.4, 6.7 and 7 keV \citep{naik2012investigation-cenx3_ironlines}. The finally used best fitting spectral model is \texttt{tbabs*(powerlaw+bbody+11$\times$Gaussians}) (Fig.~\ref{fig:cenx3_results}). The best fitting spectral model parameters are given in Table~\ref{tab:cenx3_spectra}.}

\begin{table}
    \centering
    \caption{The results of spectral fit performed on the \textit{XMM-Newton} observation of Cen X--3 having a QPO. The best fit model parameter values and their errors are given. The errors quoted on all the parameters are their 90\% confidence ranges.}
    \begin{tabular}{l|c}
        \hline
         Obs. ID& 040550201\\
         \hline
         Continuum -- --\\
         nH\textsubscript{1} &$2.43\pm0.03$\\
         PhoIndex ($\Gamma$) &$1.175\pm0.004$\\
         N\textsubscript{PL}$^\ddagger$ &$0.444\pm0.003$\\
         kT$_\mathrm{bbody}$ &$0.096\pm0.001$\\
         N$_\mathrm{bbody}^\S$ &$0.64_{-0.08}^{+0.09}$\\
         Iron emission lines complex -- -- &\\
         E$_\mathrm{Gauss,1}$ &$6.44_{-0.01}^{+0.01}$\\
         $\sigma_\mathrm{Gauss,1}$ &$0.01^\star$\\
         N$_\mathrm{Gauss,1}^\P$ &$9.74^{+0.62}_{-0.64}\times10^{-4}$\\
         E$_\mathrm{Gauss,2}$ &$6.65\pm0.01$\\
         $\sigma_\mathrm{Gauss,2}$ &$0.22_{-0.01}^{+0.01}$\\
          N$_\mathrm{Gauss,2}^\P$ &$(4.16\pm0.02)\times10^{-3}$\\
         E$_\mathrm{Gauss,3}$ &$6.65\pm0.01$\\
         $\sigma_\mathrm{Gauss,3}$ &$1.21_{-0.08}^{+0.10}$\\
         N$_\mathrm{Gauss,3}^\P$ &$0.009\pm0.001$\\
         Other emission lines -- -- &\\
         E$_\mathrm{Gauss,4}$ &$1.018\pm0.004$\\
         $\sigma_\mathrm{Gauss,4}$ &$0.019\pm0.002$\\
         N$_\mathrm{Gauss,4}^\P$ &$0.22_{-0.01}^{+0.01}$\\
         E$_\mathrm{Gauss,5}$ &$1.39_{-0.01}^{+0.01}$\\
         $\sigma_\mathrm{Gauss,5}$ &$0.16_{-0.01}^{+0.01}$\\
         N$_\mathrm{Gauss,5}^\P$ &$0.019\pm0.003$\\
         E$_\mathrm{Gauss,6}$ &$1.89_{-0.01}^{+0.01}$\\
         $\sigma_\mathrm{Gauss,6}$ &$0.10_{-0.01}^{+0.01}$\\
         N$_\mathrm{Gauss,6}^\P$ &$3.43^{+0.80}_{-0.61}\times10^{-3}$\\
         E$_\mathrm{Gauss,7}$ &$2.254\pm0.003$\\
         $\sigma_\mathrm{Gauss,7}$ &$0.1^\star$\\
         N$_\mathrm{Gauss,7}^\P$ &$1.38^{+0.15}_{-0.13}\times10^{-3}$\\
         E$_\mathrm{Gauss,8}$ &$2.68_{-0.01}^{+0.01}$\\
         $\sigma_\mathrm{Gauss,8}$ &$0.10_{-0.01}^{+0.01}$\\
         N$_\mathrm{Gauss,8}^\P$ &$1.68_{-0.19}^{+0.25}\times10^{-3}$\\
         E$_\mathrm{Gauss,9}$ &$3.29_{-0.02}^{+0.02}$\\
         $\sigma_\mathrm{Gauss,9}$ &$0.18_{-0.02}^{+0.02}$\\
         N$_\mathrm{Gauss,9}^\P$ &$0.16_{-0.01}^{+0.01}$\\
         E$_\mathrm{Gauss,10}$ &$3.90_{-0.04}^{+0.03}$\\
         $\sigma_\mathrm{Gauss,10}$ &$0.30_{-0.04}^{+0.05}$\\
         N$_\mathrm{Gauss,10}^\P$ &($2.20\pm0.03)\times10^{-3}$\\
        Flux\textsubscript{2--20 keV} ($10^{-10}$ erg s\textsuperscript{-1} cm\textsuperscript{-2}) &$67.12\pm0.02$\\
         \\
         $\chi^2$ (dof) &2496.02/1813\\
         $\chi^2$\textsubscript{red} &1.38\\
         \hline
    \end{tabular}
    \begin{tablenotes}
        \item $^\ddagger$ Normalization in units of photons s$^{-1}$ cm$^{-2}$ keV$^{-1}$ at 1 keV.
        \item {$^\S$ Normalization in units of $10^{37}$ ergs s$^{-1}$ kpc$^{-2}$.}
        \item $^\P$ Total photons s$^{-1}$ cm$^{-2}$ in the gaussian line.
    \end{tablenotes}
    \label{tab:cenx3_spectra}
\end{table}

\subsection{XTE J1858+034}
XTE J1858+034 is a transient X-ray pulsar spinning at 4.5 mHz. The binary is thought to be a Be-type HMXB due to its transient nature \citep{1858_takeshima_Be-suggestion} and nature of the optical companion \citep{1858_reig_Optical}. However, a recent study by \cite{1858_tsygankov_revised-optical} proposed that the system {could be} a Symbiotic binary hosting a K/M type stellar companion. QPO was first reported in the source from an \textit{RXTE}/PCA observation by \cite{1858_Paul_QPO_rxte-discovery} at 110 mHz. In subsequent outbursts, \cite{mukherjee2006variablextej1854} found the centroid frequency of the QPO being variable from 140 to 185 mHz and that rms amplitude of QPO has a strong correlation with photon energy. QPO was again reported in the \textit{NuSTAR} observation at 196 mHz by \cite{1858_Manoj_QPO_nustar} {during an outburst of the source in 2019}. 

\begin{figure}
    \centering
    \includegraphics[width=\columnwidth]{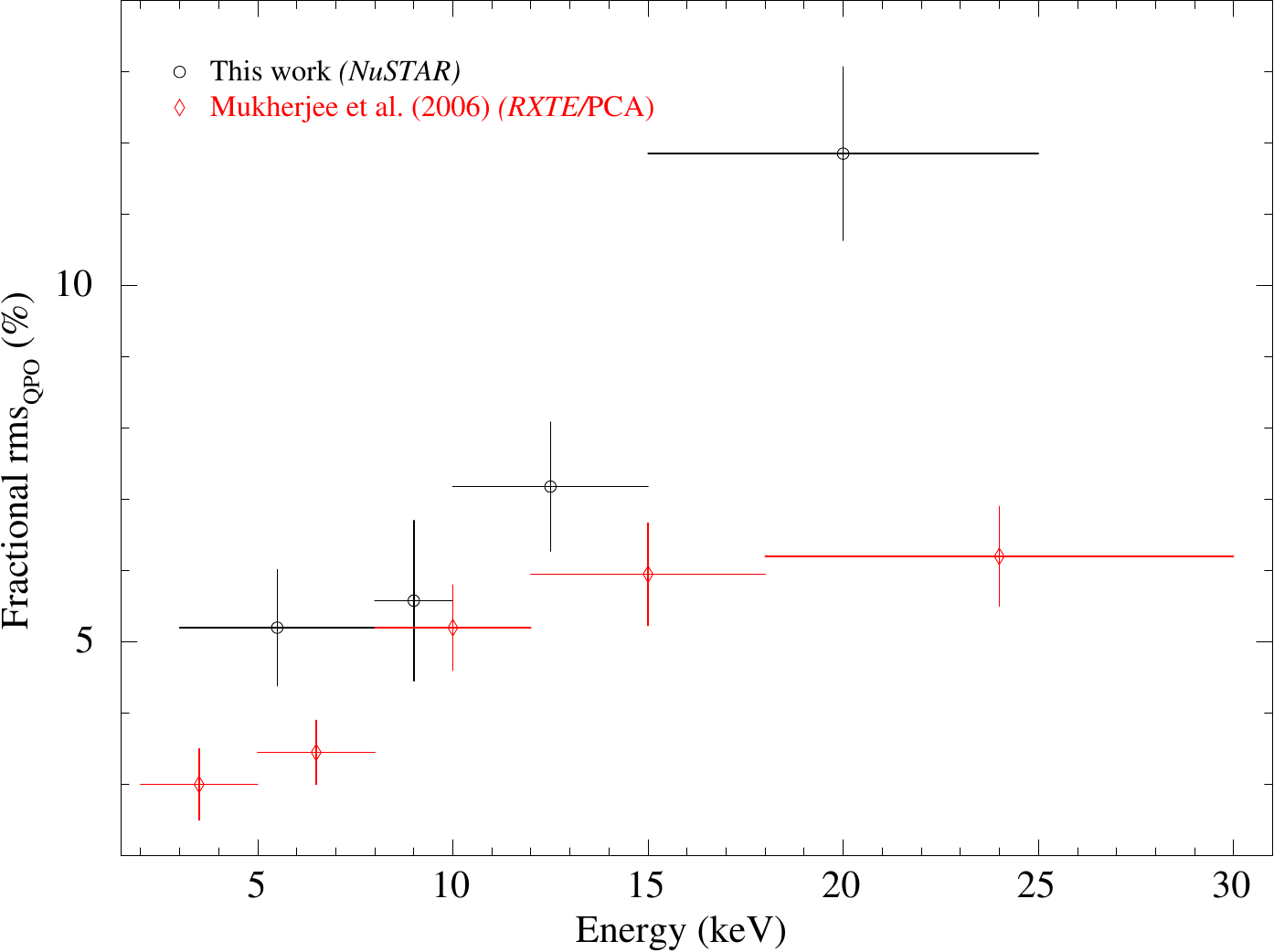}
    \includegraphics[width=\columnwidth]{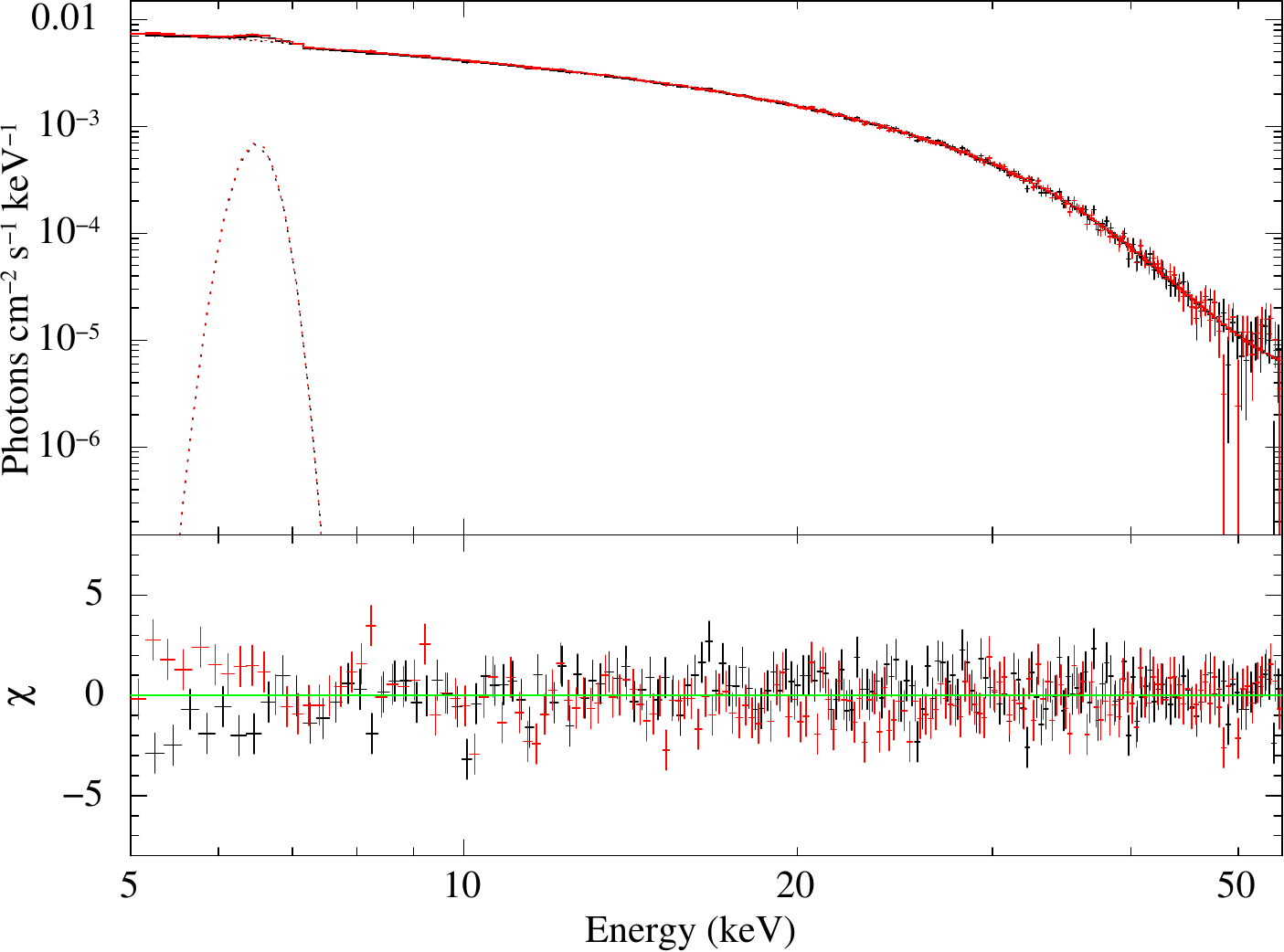}
    \caption{ Top: XTE J1858+034 QPO rms energy dependence from \textit{NuSTAR} (black) and \textit{RXTE}/PCA (red; {\protect \citealt{mukherjee2006variablextej1854}}). Bottom: The 3--60 keV unfolded spectrum and residuals to the best fit spectral model (\texttt{tbabs * (compTT * gabs + gaussian)}) on the \textit{NuSTAR} observation with OID 90501348002.}
    \label{fig:xtej1858-results}
\end{figure}

We generated the energy-dependent variation of {fractional rms amplitude of the} QPO detected at 196 mHz (Fig.~\ref{fig:xtej1858_qpofit}), the same one reported by \cite{1858_Manoj_QPO_nustar}. The QPO is detected only till 25 keV. {The \textit{NuSTAR} PDS in $5-1000$ mHz was fitted with the combination of a \texttt{powerlaw} for the band-limited noise and a \texttt{Lorentzian} for the QPO (Fig.~\ref{fig:igrj19294_qpofit}).} The QPO rms increases from about {5}\% in 3--8 keV to about {11}\% in 15--25 keV (Fig.~\ref{fig:xtej1858-results}).

{We adopted the spectral model of the source used by \cite{malacaria2021x}. The 5--55 keV \textit{NuSTAR} spectrum could be fitted with a \texttt{compTT} continuum, modified by the CRSF at $~50$ keV with a \texttt{gabs}, and an iron emission line at $6.4$ keV  with a \texttt{gaussian} (Fig.~\ref{fig:xtej1858-results}). The best fitting spectral model parameters are given in Table.~\ref{tab:1858-spectrum}.}

\begin{table}
    \centering
    \caption{The results of spectral fit performed on the \textit{NuSTAR} observation of XTE J1858+034 having QPO. The best fit model parameter values and their errors are given. The errors quoted on all the parameters are their 90\% confidence ranges.}
    \begin{tabular}{cc}
         \hline
         Obs. ID &90501348002\\
         \hline
         nH$^\dagger$ &$8.49\pm2.29$\\
         kT$_0$ &$0.98\pm0.05$\\
         kT$_\mathrm{e}$ &$5.80\pm0.15$\\
         $\tau$ &$6.60\pm0.30$\\
         N\textsubscript{norm} &$0.020\pm0.001$\\
         E\textsubscript{CRSF} &$51.09\pm2.06$\\
         $\sigma$\textsubscript{CRSF} &$10.74\pm1.81$\\
         $\tau$\textsubscript{CRSF} &$0.25\pm0.06$\\
         E\textsubscript{Gauss,Fe} &$6.48\pm0.03$\\
         $\sigma$\textsubscript{Gauss,Fe} &$0.25\pm0.05$\\
         N\textsubscript{Gauss,Fe}$^\P$ &$(5\pm1)\times10^{-4}$\\
         Flux\textsubscript{2--20 keV} ($10^{-10}$ erg s\textsuperscript{-1} cm\textsuperscript{-2}) &$10.35\pm0.02$\\
         const. &$1.018\pm0.003$\\
         &\\
         $\chi^2$(dof) &411.50/337\\
         $\chi^2$\textsubscript{red} &1.22\\
         \hline
    \end{tabular}
    \begin{tablenotes}
        \item $^\dagger$ in units of 10$^{22}$ atoms cm$^{-2}$.
        \item $^\P$ Total photons s$^{-1}$ cm$^{-2}$ in the gaussian line.
    \end{tablenotes}
    \label{tab:1858-spectrum}
\end{table}

\onecolumn
\begin{small}
\begin{longtable}{llccccrl}
	\caption{Summary of the QPO fits. The continuum is fitted with either or a combination of \textit{powerlaw}, \textit{lorentzian} and the QPO is fitted with a \textit{lorentzian}. The centre and width of the \textit{lorentzian} are considered the centroid and width of the QPO respectively. Errors assigned to the centroid and width of the QPO are their 90\% confidence intervals, while error assigned to the rms powers are their 68\% confidence intervals.}\\ \toprule
	\label{tab:qpofit}
	\endfirsthead
        \caption* {\textbf{Table \ref{tab:qpofit} Continued:} Summary of QPO fits}\\\toprule
    \endhead
    \endfoot
    \bottomrule
    \endlastfoot
		Source & Obs. ID & Energy range (keV) & Avg. count rate \footnote[1]{Across all PCUs (\textit{RXTE}) or FPMs (\textit{NuSTAR}).} {(cts s$^{-1}$)} &$\nu_\mathrm{QPO}$ (mHz) &{width}\textsubscript{QPO} (mHz) &Power\textsubscript{rms} (\%) &$Q$-factor\\
    \midrule
    4U 1626--67 & 0111070201 &$0.5-1.0$ &$5.53\pm0.03$ &$48\pm1$ &$12\pm3$ &$18.34\pm1.06$ &4.0\\
		    & & $1.0-3.0$ &$13.35\pm0.04$ &$49\pm1$ &$13\pm3$ &$20.74\pm0.96$ &3.8\\
		    & & $3.0-5.7$ &$6.19\pm0.03$ &$49\pm2$ &$14\pm7$ &$15.69\pm1.41$ &3.5\\
		    & & $5.7-10.0$ &$4.08\pm0.02$ &$49\pm1$ &$11\pm4$ &$18.54\pm1.44$ &4.5\\
		\\
		  & 0152620101 &$0.5-1.0$ &$4.63\pm0.01$ &$48\pm1$ &$11\pm2$ &$17.65\pm0.51$ &4.3\\
		    & & $1.0-3.0$ &$10.96\pm0.02$ &$48\pm1$ &$11\pm1$ &$20.61\pm0.48$ &4.2\\
		    & & $3.0-5.7$ &$5.24\pm0.01$ &$47\pm1$ &$11\pm2$ &$17.11\pm0.60$ &4.3\\
		    & & $5.7-10.0$ &$3.53\pm0.10$ &$48\pm1$ &$10\pm2$ &$16.86\pm0.73$ &4.5\\
	    \\
		&P10101 &$2.02-6.70$ &$61.23\pm0.02$ &$48.2\pm0.3$ &$11\pm1$ &$16.83\pm0.44$ &4.4\\
		    & &$6.7-8.5$ &$31.75\pm0.02$ &$48.3\pm0.3$ &$10\pm1$ &$16.65\pm0.39$ &4.8\\
		    & &$8.5-11.1$ &$35.33\pm0.02$ &$48.4\pm0.3$ &$11\pm1$ &$17.45\pm0.35$ &4.4\\
		    & &$11.1-13.0$ &$19.40\pm0.01$ &$48.2\pm0.3$ &$10\pm1$ &$18.23\pm0.41$ &4.8\\
		    & &$13.0-15.4$ &$16.31\pm0.01$ &$48.3\pm0.3$ &$11\pm1$ &$19.83\pm0.36$ &4.4\\
		    & &$15.4-20.2$ &$19.67\pm0.01$ &$48.4\pm0.3$ &$11\pm1$ &$21.45\pm0.35$ &4.4\\
		    & &$20.2-60.0$ &$11.38\pm0.02$ &$48.7\pm0.5$ &$9\pm1$ &$28.32\pm0.98$ &5.4\\
		    \\
            &\cellcolor{orange!0}62038001 &\cellcolor{orange!0}$0.5-0.9$ &\cellcolor{orange!0}$5.89\pm0.03$ &\cellcolor{orange!0}$47\pm1$ &\cellcolor{orange!0}$12\pm3$ &\cellcolor{orange!0}$13.59\pm0.69$ &\cellcolor{orange!0}3.9\\
                & &\cellcolor{orange!0}$0.9-1.3$ &\cellcolor{orange!0}$5.74\pm0.03$ &\cellcolor{orange!0}$47\pm1$ &\cellcolor{orange!0}$12\pm3$ &\cellcolor{orange!0}$17.26\pm0.69$ &\cellcolor{orange!0}3.6\\
                & &\cellcolor{orange!0}$1.3-1.9$ &\cellcolor{orange!0}$5.91\pm0.03$ &\cellcolor{orange!0}$48\pm1$ &\cellcolor{orange!0}$15\pm3$ &\cellcolor{orange!0}$19.59\pm0.83$ &\cellcolor{orange!0}3.2\\
                & &\cellcolor{orange!0}$1.9-3.5$ &\cellcolor{orange!0}$6.44\pm0.03$ &\cellcolor{orange!0}$49\pm2$ &\cellcolor{orange!0}$17\pm5$ &\cellcolor{orange!0}$19.10\pm1.10$ &\cellcolor{orange!0}2.9\\
        \\    
        IGR J19294+1816  & 0841190101 &0.5-3 &$4.38\pm0.02$ &$29\pm1$ &$8\pm5$ &$8.07\pm1.27$ &7.8\\
		& & $3.0-5.7$ &$8.95\pm0.02$ &$31\pm1$ &$9\pm3$ &$9.84\pm0.83$ &3.6\\
		& & $5.7-10.0$ &$7.70\pm0.03$ &$31\pm1$ &$10\pm5$ &$11.00\pm1.03$ &3.0\\
		\\
		V 0332+53  & 0763470301 &0.5-3 &$123.77\pm0.11$ &$38\pm3$ &$38\pm8$ &$9.62\pm0.74$ &$1.0$\\
		    & & $3.0-5.7$ &$153.56\pm0.18$ &$39\pm2$ &$33\pm7$ &$10.67\pm0.76$ &$1.2$\\
		    & & $5.7-10.0$ &$147.28\pm0.12$ &$38\pm2$ &$38\pm7$ &$10.74\pm0.72$ &$1.8$\\
		    \\
      &80102002004 &$3.0-6.0$ &$123.09\pm0.13$ &$40\pm1$ &$37\pm8$ &$9.67\pm0.50$ &1.1\\
            & &$6.0-8.5$ &$139.10\pm0.13$ &$38\pm2$ &$24\pm5$ &$8.81\pm0.69$ &1.6\\
            & &$8.5-11.5$ &$147.92\pm0.13$ &$39\pm2$ &$32\pm8$ &$9.57\pm0.77$ &1.2\\
            & &$11.5-60.0$ &$163.67\pm0.14$ &$40\pm2$ &$36\pm10$ &$9.39\pm0.85$ &1.1\\
            \\
            
		 & 0763470401 &0.5-3 &$84.54\pm0.07$ &$41\pm1$ &$22\pm6$ &$11.38\pm0.66$ &1.9\\
		    & & $3.0-5.7$ &$100.66\pm0.13$ &$42\pm1$ &$21\pm5$ &$12.16\pm0.73$ &2.0\\
		    & & $5.7-10.0$ &$95.23\pm0.07$ &$42\pm1$ &$22\pm5$ &$12.94\pm0.90$ &2.0\\
		    \\
      
      &80102002006 &$3.0-6.0$ &$79.40\pm0.08$ &$41\pm1$ &$21\pm4$ &$11.71\pm0.49$ &2.0\\
        & &$6.0-8.5$ &$89.54\pm0.09$ &$41\pm1$ &$20\pm4$ &$12.24\pm0.54$ &2.1\\
        & &$8.5-11.5$ &$99.43\pm0.09$ &$41\pm1$ &$19\pm3$ &$12.20\pm0.55$ &2.1\\
        & &$11.5-60.0$ &$112.81\pm0.09$ &$41\pm1$ &$20\pm3$ &$11.65\pm0.49$ &2.1\\
        \\
        
		 &90202031002 &3-8 &$14.69\pm0.03$ &$17.1\pm0.4$ &$2.7\pm1.2$ &$9.33\pm1.16$ &6.3\\
		    & &8-10 &$6.00\pm0.02$ &$18.3\pm0.7$ &$3.3\pm1.2$ &$9.80\pm1.37$ &5.5\\
		    & &10-15 &$9.15\pm0.03$ &$17.4\pm0.3$ &$2.0\pm0.9$ &$10.52\pm1.11$ &8.7\\
		    & &15-60 &$6.29\pm0.02$ &$18.0\pm0.5$ &$2.8\pm1.1$ &$9.86\pm1.25$ &6.4\\
		    \\
                & &3-8 &$14.69\pm0.03$ &$2.48\pm0.05$ &$0.27\pm0.13$ &$10.91\pm1.38$ &9.2\\
                & &8-10 &$6.00\pm0.02$ &$2.51\pm0.05$ &$0.11\pm0.11$ &$9.61\pm1.43$ &$\gtrsim$$11.4$\\\
                & &10-15 &$9.15\pm0.03$ &$2.49\pm0.05$ &$0.23\pm0.11$ &$11.95\pm1.43$ &10.8\\
                & &15-60 &$6.29\pm0.02$ &$2.49\pm0.05$ &$0.29\pm0.13$ &$11.92\pm1.44$ &8.6\\
                \\
		 &90202031004 &3-8 &$12.05\pm0.03$ &$17.8\pm0.9$ &$4.4\pm1.9$ &$12.56\pm1.26$ &4.1\\
		    & &8-10 &$4.88\pm0.02$ &$17.5\pm0.8$ &$4.9\pm1.8$ &$15.20\pm1.30$ &3.6\\
		    & &10-15 &$7.46\pm0.02$ &$17.8\pm0.8$ &$5.5\pm2.4$ &$14.56\pm1.38$ &3.2\\
		    & &15-60 &$5.14\pm0.02$ &$18.6\pm1.2$ &$4.7\pm2.2$ &$12.77\pm1.52$ &4.0\\
		    \\
                & &3-8 &$12.05\pm0.03$ &$2.33\pm0.08$ &$0.36\pm0.26$ &$12.55\pm1.33$ &6.5\\
                & &8-10 &$4.88\pm0.02$ &$2.33\pm0.08$ &$0.40\pm0.27$ &$13.64\pm1.41$ &5.8\\
                & &10-15 &$7.46\pm0.02$ &$2.35\pm0.05$ &$0.18\pm0.15$ &$12.36\pm1.26$ &$\gtrsim$$10.2$\\
                & &15-60 &$5.14\pm0.02$ &$2.30\pm0.12$ &$0.40\pm0.36$ &$11.73\pm1.49$ &5.8\\
                \\
    
	    Cen X--3 &040550201 &0.5-3 &$187.88\pm0.06$ &$30\pm1$ &$26\pm4$ &$4.74\pm0.16$ &1.2\\
	        & &$3.0-5.7$ &$186.94\pm0.06$ &$31\pm1$ &$23\pm4$ &$5.98\pm0.19$ &1.4\\
	        & &$5.7-10.0$ &$118.63\pm10.05$ &$30\pm1$ &$23\pm3$ &$6.12\pm0.20$ &1.3\\
                \\
        XTE J1858+034 &90501348002 &3-8 &$23.03\pm0.03$ & $178\pm9$ &$70\pm30$ &$5.20\pm0.69$ &2.5\\
            & &8-10 &$9.08\pm0.02$ &$185\pm8$ &$37\pm22$ &$6.02\pm0.87$ &5.0\\
            & &10-15 &$12.45\pm0.02$ &$183\pm7$ &$56\pm20$ &$7.07\pm0.69$ &3.3\\
            & &15-25 &$3.69\pm0.01$ &$186\pm7$ &$84\pm27$ &$11.73\pm1.03$ &2.2\\

\end{longtable}
\end{small}
\twocolumn

\section{Discussions}\label{sec:DAI}
Our search for QPOs in the archival \textit{XMM-Newton} and \textit{NuSTAR} observations resulted in the detection of QPO in {eleven} observations of five sources viz., 4U 1626--67 (\textit{XMM-Newton}) (Fig.\ref{fig:4u1626_qpofit_xmmn}), IGR J19294+1816 (\textit{XMM-Newton}) (Fig.~\ref{fig:igrj19294_qpofit}), V 0332+53 (\textit{XMM-Newton}, \textit{NuSTAR}) (Figs.~\ref{fig:v0332_qpofit-XMM_2015},~\ref{fig:v0332_qpofit-NuSTAR_2015},~\ref{fig:v0332_qpofit-NuSTAR_2016}), Cen X--3 (\textit{XMM-Newton}) (Fig.~\ref{fig:cenx3_qpofit}) and XTE J1858+034 (\textit{NuSTAR}) (Fig.~\ref{fig:xtej1858_qpofit}). {A summary of the QPO parameters in different energy bands of each observation is given in Table~\ref{tab:qpofit}.  The highest QPO rms is observed in the 48 mHz QPO in 4U 1626$-$67 ($\gtrapprox15\%$), followed by V 0332+53 ($\gtrapprox10\%$). The QPO rms is $<10\%$ in all the other cases. In Cen X--3 and XTE J1858+034, the QPO rms is low at $\lessapprox6$\% below 10 keV. We discuss the observed QPO characteristics in
various contexts below. 

\subsection{Transient nature of QPOs}

The transient nature of QPOs in XRPs is evident from detection and non-detection in different observations of the same source (Tables~\ref{tab:qpoobscatalogue} and~\ref{tab:catalogue}). Some observations could however be made regarding the appearance/disappearance of QPOs in certain observations of a source analysed in this work. 

In the four observations of 4U 1626$-$67 analysed in this work, QPO has been detected in three observations and not detected in one observation. 4U 1626$-$67 was in the spin-down torque-state of the pulsar between 1990 and 2008, and since 2023 \citep{4u1626_SD_sharma2023}. The three observations (Table~\ref{tab:qpoobscatalogue}) having the 48 mHz QPO were taken when the pulsar was in the spin-down state. The source was observed in the spin-up torque state of the pulsar in the observation without QPO (\textit{XMM-Newton} OID 0764860101 in 2015). The appearance of the 48 mHz QPO only during the spin-down state of the pulsar has been previously reported by \cite{jain2010} and \cite{4u1626_SD_sharma2023}. 4U 1626$-$67 is the only XRP where the QPO is persistently present during a specific torque state of the pulsar.

Among the three observations of IGR J19294+1816, QPO was only detected in one observation. The \textit{XMM-Newton} observation in which the QPO at $\sim30$ mHz is present, was taken at the luminosity rising phase during a Type-I outburst of Be-XRP in October 2019. \cite{graman2021astrosat} has reported the presence of QPO at around the same frequency during the luminosity decline phase of the same outburst with \textit{Astrosat}/LAXPC. QPO was not observed in the two \textit{NuSTAR} observations, of which OID 90401306002 was done when the source was in a low-luminosity state in March 2018 and OID 90401306004 was taken immediately following a Type-I outburst in the same month \citep{19294_nustar_tsygankov2019study}.

In the three observations of Cen X--3, a QPO at 30 mHz is present in only one observation (Table~\ref{tab:qpoobscatalogue}) and it covers the out-of-eclipse orbital phase of the source \citep{aftab2019x}, and not present in the two eclipse-egress orbital phase observations (\textit{XMM-Newton} OID 0111010101 that covers orbital phase $0.00-0.37$; \citealt{cenx3_xmm011_eclegress_sanjurjo2021},  and \textit{NuSTAR} OID 30101055002 that covers orbital phase $0.20-0.41$; \citealt{cenx3_nustar_tamba2023orbital}). 30 mHz is the lowest QPO frequency reported in Cen X--3. \cite{harsha_cenx3_2008ApJ...685.1109R} had found QPO frequency clustering around 40 mHz and 90 mHz from \textit{RXTE}/PCA observations during 1996-1998. \cite{liu_Cenx3_40mhZqpo_insight} observed that the QPO frequency varies as a function of the orbital phase from about 33 mHz in the eclipse egress phase ($0.1-0.3$) to about 40 mHz in the pre-ingress phase ($0.8$) from \textit{Insight-HXMT} observations in 2022.

Among the nine observations of V 0332+53, QPO is present in six observations. Seven out of the nine observations were taken during a Type-II outburst of the source in 2015. This includes two \textit{XMM-Newton} and five \textit{NuSTAR} observations. Two quasi-simultaneous \textit{NuSTAR} and \textit{XMM-Newton} observations taken mid-way during the decline of Type-II outburst \citep{v0332_nustar_qpo_noqpo_doroshenko2017luminosity,vybornov_v0332_nustar_crsf} have the presence of QPO at 40 mHz. However, two other \textit{NuSTAR} observations taken towards the end of the decline of the 2015 outburst \citep{v0332_nustar_qpo_noqpo_doroshenko2017luminosity,vybornov_v0332_nustar_crsf} does not show QPO. The two \textit{NuSTAR} observations during a Type-I outburst in 2016 have the presence twin-QPOs at $\sim$2.5 and $\sim$18 mHz (See discussion in Section~\ref{sec:v0332-discussion}). The \textit{XMM-Newton} observation taken during the quiescent state of the source in 2008 \citep{v0332_xmmn_noqpo_elshamouty2016soft, v0332_xmmn_noqpo_tsygankov2017x} does not have a QPO.

QPOs in Be-XRPs are generally associated with the formation of accretion disks during the outbursts \citep{bexrp_reig2013patterns}. In the two Be-XRPs analysed in this work (IGR J19294+1816 and V 0332+53), QPOs were detected in all the observations during high luminosity phases of the Type-II outbursts and were detected in some Type-I outbursts while not in some other Type-I outbursts. However, no QPOs were detected in the observations that were carried out during their low-luminosity states.

\subsection{Energy-dependence of QPOs}

The energy spectra of all five sources had a smooth continuum shape absorbed by ISM at low energies, with atomic emission lines mainly of iron. Cen X$-$3, which has a companion that releases strong stellar wind had the presence of multiple emission lines. CRSF was present in all of them having \textit{NuSTAR} observations. A soft excess was present in 4U 1626$-$67, Cen X$-$3, and V 0332+53 (Section~\ref{disc:1626_qpo_softexcess}).

The overall shape of the broadband noise in PDS does not vary with energy in all the sources (Appendix~\ref{app:qpofit}). However, the strengths of QPO and the pulsations vary with energy. We do not notice any strong trends in the QPO parameters (centre, width, \textit{Q}-factor) except the fractional rms amplitude, as a function of energy in any of the sources (Table~\ref{tab:qpofit}). There is also no secular trend in the QPO rms vs energy exhibited by all the sources. A correlation of fractional rms amplitude with energy is seen in 4U 1626$-$67 in $5-60$ keV, IGR J19294+1816 in $0.5-30$ keV, Cen X$-$3 in $0.5-10$ keV, and XTE J1858+034 in $2-30$ keV. No variation of fractional rms amplitude with energy is visible in V 0332+53. However, differences are noticed in the energy dependence of fractional rms between different observations (having the same QPO frequency) of the same source.

In XTE J1858+034, the fractional rms amplitude of the QPO during the November 2019 outburst (\textit{this work}) is overall higher than during the outburst in May 2004 with \textit{RXTE} observations \citep{mukherjee2006variablextej1854}. The X-ray flux is similar ($\sim10^{-9}$ ergs s$^{-1}$ cm$^{-2}$) for the two observations. The authors also observed that the QPO frequency changes from 150 to 180 mHz while the X-ray flux varies from $2.5-5.5\times10^{-9}$ ergs s$^{-1}$ cm$^{-2}$. The highest 180 mHz QPO frequency was observed at a flux of about $4\times10^{-9}$ ergs s$^{-1}$ cm$^{-2}$. In the \textit{NuSTAR} observation we analysed, the QPO frequency is slightly higher than 180 mHz and the flux is about $1.4\times10^{-9}$ ergs s$^{-1}$ cm$^{-2}$. Even though a correlation of rms amplitude with energy is seen in both the outbursts, the trend is seen to deviate above 10 keV, where it saturates for the 2004 observations while the correlation continues in the 2019 observation. The absence of QPO above 30 keV is in agreement with \cite{mukherjee2006variablextej1854}.

In Cen X$-$3, the rms amplitude of 30 mHz QPO shows a correlation with energy as opposed to no correlation of the 40 mHz QPO in \cite{harsha_cenx3_2008ApJ...685.1109R} and a clear anti-correlation of the 30$-$40 mHz QPO seen in \cite{liu_Cenx3_40mhZqpo_insight}. While \cite{harsha_cenx3_2008ApJ...685.1109R} reported the presence of QPO till 35 keV, \cite{liu_Cenx3_40mhZqpo_insight} observed no QPO above 20 keV.

The overall energy dependence of the rms amplitude of the 30 mHz QPO in IGR J19294+1816 during the onset of Type-I outburst in 2019 from this work is in agreement with \cite{graman2021astrosat} during the decline of the same outburst.

Neither the twin-QPOs at 2.5 and 18 mHz in 2016 (0.5--60 keV) nor the QPO at 40 mHz in 2015 (3--60 keV) detected in V 0332+53 exhibit any energy dependence. This is different from the previous reports of \cite{qu2005discovery_v0332} where the rms of the twin QPOs at 50 mHz and 220 mHz detected during the 2015 Type-II outburst drops beyond 10 keV.

The 48 mHz QPO in 4U $1626-67$ preserves the value of fractional rms amplitude and the correlation of QPO rms with energy across several observations spanning over two decades. During this time, the source has also exhibited multiple torque-state switches \citep{4u1626_SD_sharma2023}. A peculiarity of 4U 1626$-$67 is that it is the only Roche lobe overflown (RLO) LMXB in the list (Table~\ref{tab:qpoobscatalogue}), in which the accretion only proceeds via an accretion disk. Even though accretion in the compact HMXB Cen X$-3$ is also partially driven by RLO, there is also a significant contribution expected from the strong stellar wind. Cen X$-$3 in the long-term has shown a variety of QPO frequencies varying from 30 mHz to 90 mHz and the energy trend of these QPOs are also different \citep[][and this work]{harsha_cenx3_2008ApJ...685.1109R,liu_Cenx3_40mhZqpo_insight}. It is also worth noting that the 48 mHz QPO in 4U 1626$-$67 has exhibited the highest rms amongst the list ($\gtrapprox15\%$), while Cen X--3 has the lowest $(\lessapprox6\%)$.}

 \subsection{Twin QPOs in V 0332+53}\label{sec:v0332-discussion}

The QPOs we detected in V 0332+53 at 2.5 mHz, 18 mHz {(2016 Type-I outburst)} and 40 mHz {(2015 Type-II outburst decay)} in the \textit{XMM-Newton} and \textit{NuSTAR} observations have different frequencies compared to the ones reported earlier {(50 mHz and 220 mHz)}. We also noticed that the PSDs of the observations {during 2015 outburst} have a different overall shape when compared to the PSDs {during the 2016 Type-I outburst (\textit{this work})} and the previously reported PSDs {during two different Type-II outbursts} from \textit{Ginga} \citep{V0332_Ginga_takeshima_QPO-dicovery} and \textit{RXTE} (\citealt{qu2005discovery_v0332,v0332_psd_garcia}). This variable nature of the PSD directly indicates that the factors contributing towards aperiodic variability in the source vary over time.

{The 2015 \textit{XMM-Newton} and \textit{NuSTAR}}  observations show a QPO at 40 mHz (Figs.~\ref{fig:v0332_qpofit-XMM_2015},\ref{fig:v0332_qpofit-NuSTAR_2015}) while the {2016} \textit{NuSTAR} observations show twin QPOs at 2.5 mHz and 18 mHz (Fig.~\ref{fig:v0332_qpofit-NuSTAR_2016}). Twin QPOs have been reported in {V 0332+53 \citep{qu2005discovery_v0332} and in} GX 304-1 \citep{devasia2011timing}. {However, the twin QPOs in V 0332+53 reported before have centroid frequencies of 50 mHz and 220 mHz, where 220 mHz is the spin period of the NS. \cite{qu2005discovery_v0332} proposed that inhomogeneous accretion flow at the polar caps results in coupling between the spin-variability and noise-variability in the PDS, which could lead to the appearance of QPO at NS spin frequency. In GX 304$-$1,} the second QPO was a harmonic of the first, which is not the case here.

Spectra from all the observations could be modelled with a soft-black body component and a power-law-based comptonization component (Table~\ref{tab:v0332_spectra}). The 2015 observations were performed when the source was $\ge5$ time brighter than during the 2016 observations. The higher QPO frequency during a high X-ray flux state favours the inner accretion disk origin of QPO, which might move closer to the neutron star, resulting in an enhanced Keplerian orbital frequency. Following the relations $r\propto\dot M^{-2/7}$ and Keplerian relation of $\nu\propto r^{-3/2}$, we get $\nu\propto\dot M^{3/7}$. Under this argument, an increase in QPO frequency to a factor of $\sim$2 (from 18 mHz to 40 mHz) requires $\dot M$ (L\textsubscript{X}) to scale up to a factor of $\sim5$, which {close to} the measured X-ray flux from the spectral analysis. However, a similar argument for an increase in QPO frequency from 2.5 mHz to 40 mHz would require a factor of $\ge600$ increase in $\dot M$. Thus, the 40 mHz QPO observed {during the 2015 outburst} and 18 mHz QPO observed {during the 2016 outburst} are likely of similar origin, while the 2.5 mHz QPO is likely of a different origin. The \textit{Q}-factor of 18 mHz QPO present at low flux is also about a factor of $\gtrsim2$ higher that of the 40 mHz QPO present at a high flux (Table~\ref{tab:qpofit}). A similar increment in the \textit{Q}-factor with decreasing flux has also been observed for the 220 mHz QPO by \cite{qu2005discovery_v0332}. 

{A careful consideration of the estimate of Alfven radius from \cite{lamb1989model_xrp} in Section~\ref{subsec:kfm_bfm_applicability} however shows that the 40 mHz QPO could be explained with BFM, and a shift in QPO frequency to 18 mHz is unlikely to be due to change in luminosity. Moreover}, this interpretation {of the 18 mHz and 40 mHz having a common origin} contrasts with the observation of \cite{qu2005discovery_v0332}, where it was shown that the QPO {centroid} frequency does not evolve with source flux.

\subsection{Association of QPO with the soft excess}\label{disc:1626_qpo_softexcess}

XRPs sufficiently away from the galactic plane are known to exhibit soft excess in their energy spectrum and is usually associated with the X-ray emission from the NS reprocessed by the inner accretion disk (\citealt{Paul_2002_softexcess}, \citealt{Hickox_2004}). It is usually modelled with a low temperature ($kT\sim100-200$ eV) blackbody component. The idea that both soft excess and QPOs in XRPs are considered to originate from the inner accretion disk motivated us to look for patterns in the QPO strength in the soft-excess energy band. \textit{XMM-Newton}/PN is the most suitable detector for such a study, as it has low energy coverage till 0.5 keV to detect the soft excess and a relatively good effective area for high significance detection of QPO. However, sources that exhibit both QPO in the light curve and soft excess in the spectrum were needed to perform such a study.

Out of the five sources in which we detected \textit{XMM-Newton} observations, only 4U 1626--67 satisfies this criterion. {Despite soft-excess being also detected in V 0332$+$53 and Cen X$-$3, the blackbody dominant spectral band is outside the sensitive instrument energy band to perform QPO rms estimation.} We modelled the soft excess in 4U 1626--67 with a blackbody component of $kT_\textrm{BB}$$\sim$0.3 keV. The black body contribution is expected to peak around 0.8 keV ($\sim2.8$ $kT_\textrm{BB}$). Incidentally, QPO also shows high rms values of about 20\% in the 0.5--3 keV spectral band in 4U 1626--67 (Fig.~\ref{fig:4u1626-results}). An alternate interpretation is that the QPO rms is consistently high in the 0.5--10 keV with an abrupt drop in 3--5.7 keV, which seems unlikely when looking at it together with QPO rms from other wavelengths (See Section~\ref{sec:4u1626}) and the \textit{RXTE}/PCA observation ({Fig.}~\ref{fig:4u1626_qpofit_rxtepca}). 

\begin{figure}
    \centering
    \includegraphics[width=\columnwidth]{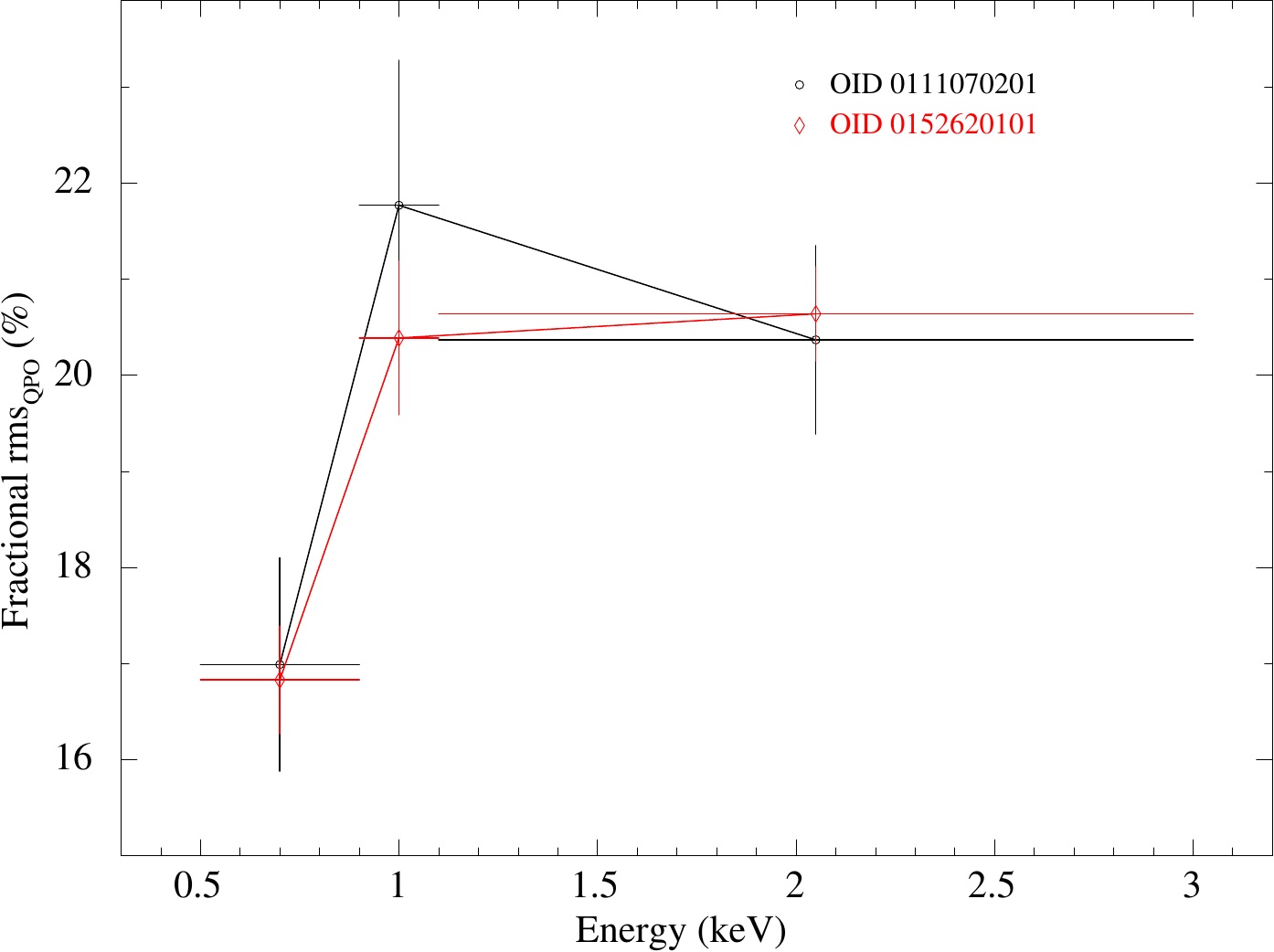}
    \caption{QPO rms in three different energy ranges of two \textit{XMM-Newton} observations of 4U 1626--67. {The QPO rms shows a rise throughout the 0.9--3.0 keV band in both observations.}}
    \label{fig:1626-qpo-SE}
\end{figure}

However, apart from the soft excess, 1--3 keV also contains the Ne emission complex (See Fig.~\ref{tab:4u1626_spectra}), the origin of which is believed to be the O/Ne rich accretion disk of 4U 1626--67 \citep{schulz_1626_chandrahetg_ONeaccdisk}. To check if the QPO rms show any anomaly in the energy range corresponding to the Neon emission complex, we calculated the QPO rms in 0.5--0.9 keV, 0.9--1.1 keV and 1.1--3 keV energy bands, {and there seems to be no significant rise in the QPO rms peculiar to the Neon spectral range in the two \textit{XMM-Newton} observations (Fig.~\ref{fig:1626-qpo-SE}). The QPO rms as a function of energy from combined multiple \textit{NICER} observations also show that the QPO rms is maximum in $1.3-1.9$ keV, i.e, neither at the soft-excess energy band nor at the Neon line energy band (Fig.~\ref{fig:1626-nicer-results}). Therefore, we found no evidence to associate the origins of soft-excess and QPO.}

\subsection{Applicability of KFM and BFM}\label{subsec:kfm_bfm_applicability}
The KFM treats QPO as the NS emission modulated by the inhomogeneous matter orbiting at Keplerian orbit in the inner accretion disk. KFM imposes the condition $\nu_{\textrm{NS}}<\nu_{\textrm{QPO}}$, and based on this the applicability of KFM is invalid on four of the five QPOs we detected (See Table~\ref{tab:qpo-kfm_bfm-applicability}). It could be applicable in XTE J1858+034, which gives the inner accretion disk orbital frequency as 196 mHz. But considering that soft X-rays are vulnerable to absorption from cold matter compared to hard X-rays, the QPO rms is expected to peak at lower energies, i.e., exhibiting anti-correlation with energy. However, a strong positive correlation is exhibited by XTE J1858+034 between QPO rms and photon energy in both the \textit{NuSTAR} (this work) and \textit{RXTE}/PCA \citep{mukherjee2006variablextej1854}, with the QPO rms reaching about 10\% in 15--25 keV in the \textit{NuSTAR} observation. 

{KFM} is therefore {not} apt to explain the QPOs we observed. From the observed QPO frequency and the pulsar spin frequency, we estimated the orbital frequency of the inner accretion disk in each case (Table~\ref{tab:qpo-kfm_bfm-applicability}). Both KFM and BFM also predict the QPO centroid frequency to vary with the X-ray luminosity, which is not generally observed in XRPs \citep{FINGER1998_qpo-review, harsha_cenx3_2008ApJ...685.1109R}. Exceptions are A 0535+262 \citep{finger1996quasi, FINGER1998_qpo-review, ma_1a0535_mHzqpo_Erelation-distinct} and XTE J1858+034 \citep{mukherjee2006variablextej1854}, where, a positive correlation of the QPO frequency with X-ray flux was observed. We found such a variation in V 0332+53, where the QPO centroid frequency is almost doubled when the X-ray flux increased $\sim$5 times (See the discussion in Section~\ref{sec:v0332-discussion}).

{
The inner accretion disk is expected to terminate at the Alfven radius ($\mathrm{r_M}$), and $\mathrm{r_M}$ \citep{lamb1989model_xrp,becker2012spectral_GL} is given by

\begin{align}\label{eq:GL}
    \mathrm{r_M} = 2.73\times10^{7} \textrm{cm} &\left(\frac{\Lambda}{0.1}\right) \left(\frac{M_\star}{1.4 M_\odot}\right)^{1/7} \left(\frac{R_\star}{10\ \textrm{km}}\right)^{10/7} \nonumber\\ &\times \left(\frac{B_\star}{10^{12}\  \textrm{G}}\right)^{4/7} \left(\frac{L_x}{10^{37} \textrm{ergs\ s}^{-1}}\right)^{-2/7}
\end{align}

$\Lambda$ is a constant, and $\Lambda=1$ for spherical accretion, and $\Lambda\approx0.22\alpha^{18/69}$ for disk accretion, where $\alpha$ is the alpha-disk parameter. $\Lambda=0.1$ is a fair approximation for typical values of $\alpha$ between 0.01--0.1 \citep{becker2012spectral_GL}. $M_\star$ is the mass of the NS, $R_\star$ is the radius of NS, $B_\star$ is the magnetic field at the surface of the NS, and $L_x$ is the X-ray luminosity of the source. We checked the difference in $\mathrm{r_M}$ predicted by equation~\ref{eq:GL} from the inner accretion radius ($\mathrm{r_{BFM}}$) predicted by BFM, given by

\begin{align}
    \mathrm{r_{BFM}}=\left(\frac{GM_\mathrm{NS}}{4\pi^2(\nu_\mathrm{QPO}+\nu_\mathrm{NS})^2}\right)^{1/3}
\end{align}

The results are summarized in Table~\ref{tab:qpo-kfm_bfm-applicability}. The BFM predicted inner accretion radius and the Alfven radius are within a factor of 1.1 in IGR J19294+1816, XTE J1858+034, and the 40 mHz QPO in V 0332+53. However, they are off by a factor of 1.5 or less in the 2.5 mHz and 18 mHz QPOs in V 0332+53, and by a factor of 1.5 or more in 4U 1626--67 and Cen X--3. BFM could therefore explain the QPOs only in IGR J19294+1816, V 0332+53 (40 mHz), and XTE J1858+034. The estimate of Alfven radius from equation~\ref{eq:GL} is also, however, bound to have uncertainties from various variables.

Beyond KFM and BFM, we also acknowledge the existence of other QPO models, for instance, the magnetically driven disk precession model \citep{Shirakawa_2002}. However, this model does not predict the energy dependence of QPO rms.

}

\begin{table*}
    \centering
    \caption{Characterisitcs of the observed QPOs and applicability of Keplerian and Beat frequency models.}
    \scalebox{0.9}{
    \begin{tabular}{c|cccccccccc}
        \hline
         Source &Type &$\nu$\textsubscript{QPO} {in mHz} &E-relation &\multicolumn{2}{c}{$\nu$\textsubscript{orb} (inner acc. disk) {in mHz}} &{Unabsorbed Flux$_\mathrm{2-20keV}$} &\multicolumn{3}{c}{inner acc. disk radius in km} &{References$^\top$}\\
         &&&&KFM$^\dagger$ &BFM$^\S$ &{in $10^{-10}$ ergs s$^{-1}$ cm$^{-2}$} &{r$_\textrm{KFM}$} &{r$_\textrm{BFM}$ } &{r$_\textrm{M}$ }\\
         \hline
         4U 1626--67& Persistent &48 &+ve &N.A &178 &{2.8} &-- &{5305} &{1055} &\cite{staubert2019cyclotron}\\
         IGR J19294+1816 &Transient &30 &+ve &N.A &{113} &{4.0} &-- &{7182} &{6616} &\cite{staubert2019cyclotron}\\
         V 0332+53 &Transient &{2.5} &{nil} &{N.A} &{230} &{6.1--7.5} &-- &{4478} &{5386--5713} &\cite{staubert2019cyclotron}\\
         & &{18} &{nil} &{N.A} &{245} &{6.1--7.5} &-- &{4287} &{5386--5713} &\cite{staubert2019cyclotron}\\
         & &{40} &{nil} &{N.A} &{267}  &{29.1--124.8}  &-- &{4049} &{2411-3656} &\cite{staubert2019cyclotron}\\
         Cen X--3 & Persistent &{30} &+ve &N.A &{238} &{67.1} &-- &{4371} &{2775} &\cite{staubert2019cyclotron}\\
         XTE J1858+034 &Transient &{185} &+ve &{185} &{190} &{10.4} &{5170} &{5088} &{5363} &\cite{malacaria2021x}\\
         \hline
    \end{tabular}
    }
    \begin{tablenotes}
        \item $^\dagger$ $\nu$\textsubscript{orb} = $\nu_\textrm{QPO}$. KFM is not applicable if $\nu_\textrm{QPO}<$$\nu_\textrm{NS}$ (denoted by N.A).
        \item $^\S$ $\nu$\textsubscript{orb} = $\nu_\textrm{QPO}+$$\nu_\textrm{NS}$.
        \item $^\top$ References for distance to the source, and magnetic field strength.
    \end{tablenotes}
    \label{tab:qpo-kfm_bfm-applicability}
\end{table*}

\subsection{QPOs in other accretion-powered X-ray sources}
Besides XRPs, black hole binaries (BHBs) and low magnetic field ($10^8$--$10^9$ G) neutron stars in Low mass X-ray binaries (NSBs) are two other classes of accretion-powered X-ray binaries containing primary compact stellar objects with a mass of the order of M$_\odot$, that exhibits QPOs which has been well studied. The QPOs exhibited by BHBs are broadly classified into two types based on their centroid frequency, namely the low-frequency {QPOs (LFQPOs)} (0.1$-$30 Hz) and the high-frequency {QPOs (HFQPOs)} ($>30$ Hz). HFQPOs in BHBs like the QPOs exhibited by XRP are also a transient phenomena \citep{HFQPO_BHB_belloni_review}. NSBs exhibit QPOs at kHz frequencies, known as kHz QPOs (they sometimes appear in pairs, and are then known as twin kHz QPOs). Due to the centroid frequency being close to the Keplerian frequency of their inner accretion disks, the kHz QPOs in NSBs and HFQPOs in BHBs are generally associated with the accretion disk \citep{BHreview_remillard_2006}. Moreover, BHBs and {NSBs} usually exhibit different spectral states like high-soft state (high luminosity and soft spectrum), low-hard state and intermediate state. The kHz QPOs in NSBs and HFQPOs in BHBs are usually observed during the soft states \citep{link_NH-BH-QPO}, which are generally associated with the accretion disk. Considering their connection with the accretion disk, the mHz QPOs in XRPs, kHz QPOs in {NSBs} and HFQPOs in BHBs could be discussed in the same context.

The twin kHz QPOs in NSBs are usually explained by two models; the sonic point beat-frequency model \citep{miller_sonicpoint-kHzqpo} and the relativistic precession model \citep{stella_lensthir-precession_kHzqpo}. The relativistic precession model interprets the HF kHz QPO from the inner accretion disk and LF kHz QPO as relativistic precession modes at that orbit. The sonic point beat-frequency model interprets the high-frequency kHz QPO to be related to the clumps in the innermost accretion disk and the LF kHz QPO as the beat frequency between neutron star spin and the HF kHz QPO. The sonic point model is a combination of the KFM and BFM employed in XRPs. If the sonic point model is employed to explain the twin QPOs (2 {and} 18 mHz) observed in two \textit{NuSTAR} observations of V 0332+53, the model predicts a spin frequency of NS around 20 mHz, while the true value stands at 220 mHz.

In general, QPOs in XRPs (\textit{this work}), HFQPOs (\citealt{Morgan_1997_grs1915_khzqpodiscover_qporms-E-correlation}, \citealt{klis_kHzQPO-millisecondpulsar_2000}) and kHz QPOs (\citealt{kHzQPO-rmsphotonE-correlate_Strohmayer_1996}, \citealt{Zhang_1996_lowB-NS}, \citealt{kHzQPO-rmsphotonE-correlate_Berger_1996}, \citealt{Wijnands_1997_lowB-NS_GX17p2}) exhibit a positive correlation of QPO rms with photon energy. A key aspect that sets the QPOs in XRPs apart is their QPO rms that regularly goes over 10\% up to 30\% (this work) when compared to $<20\%$ in kHz QPOs (\citealt{Zhang_1996_lowB-NS}, \citealt{Wijnands_1997_lowB-NS_GX17p2}) and even lower values for the rarely detected transient HFQPOs \citep{HFQPO_BHB_belloni_review}.

\section{Conclusions}
We analysed {99} \textit{XMM-Newton} and \textit{NuSTAR} observations of 29 accreting X-ray pulsars and searched for the presence of mHz Quasi-periodic oscillations in them. We detected QPOs in six \textit{XMM-Newton} observations and three \textit{NuSTAR} observations and estimated the variation of QPO properties with photon energy. {The Magnetospheric beat frequency model (BFM) is favourable over the Keplerian frequency model (KFM) in IGR J19294+1816, V 0332+53 (40 mHz) and XTE J1858+034. Twin QPOs were detected in the \textit{NuSTAR} observation of V 0332+53.} 

\section*{Acknowledgements}

This research has made use of data and/or software provided by the High Energy Astrophysics Science Archive Research Center (HEASARC), which is a service of the Astrophysics Science Division at NASA/GSFC. 
This publication makes use of the observations obtained with XMM-Newton, an ESA science mission with instruments and contributions directly funded by ESA Member States and NASA. This research has made use of the NuSTAR Data Analysis Software (NuSTARDAS) jointly developed by the ASI Science Data Center (Italy) and the California Institute of Technology (USA). 
HM thank Ketan Rikame and Kinjal Roy for the lightcurves of Her X--1 and Cen X--3, respectively.
{We thank the anonymous reviewer for the useful suggestions that helped us improve the manuscript.}

\section*{Data Availability}

All the data underlying this research work is publicly available in NASA's HEASARC Data archive. 



\bibliographystyle{mnras}
\bibliography{mnras_template} 




\appendix

\section{QPO fits}\label{app:qpofit}

\begin{figure}
    \centering
    \includegraphics[width=\columnwidth]{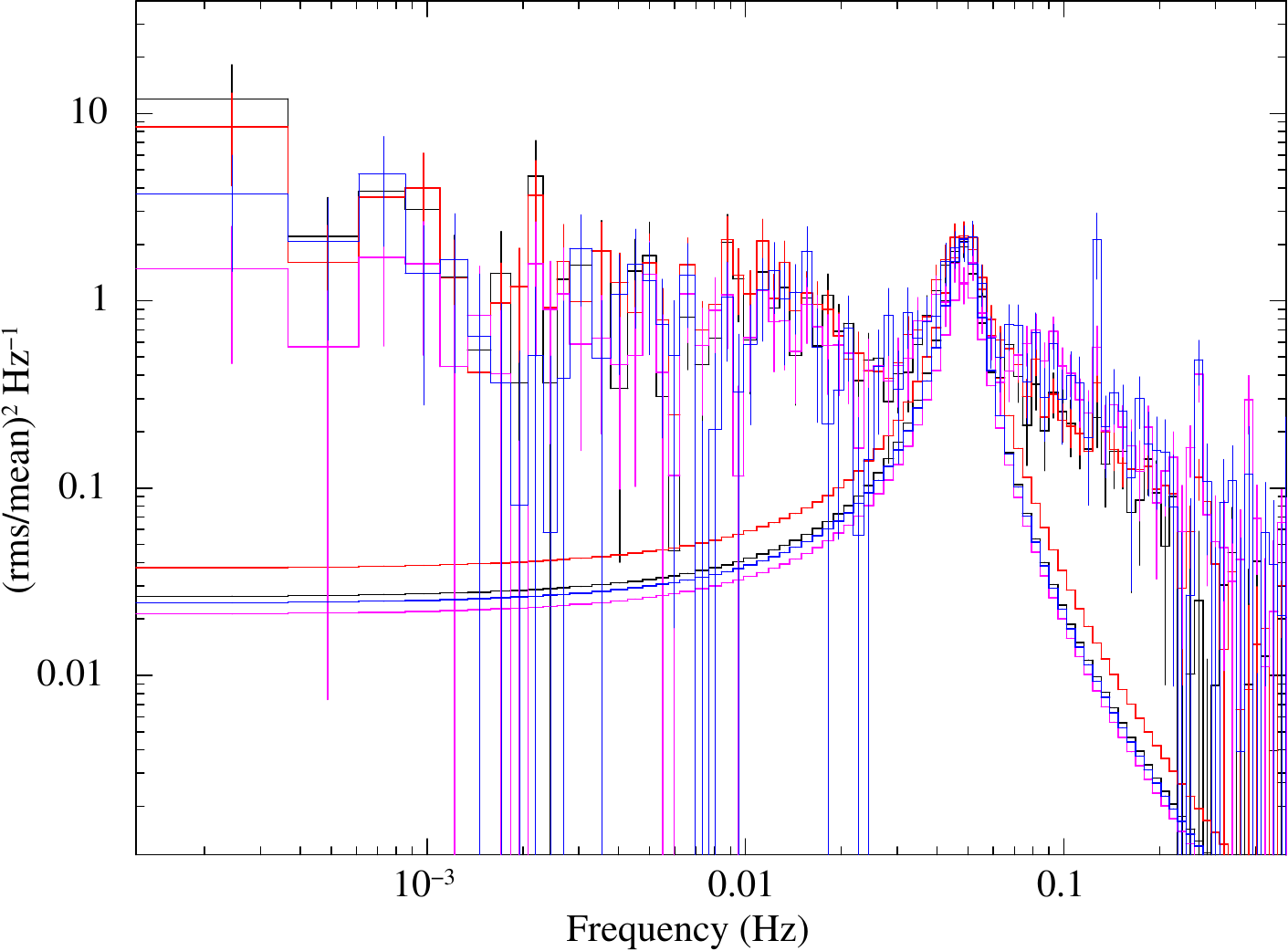}
    \includegraphics[width=\columnwidth]{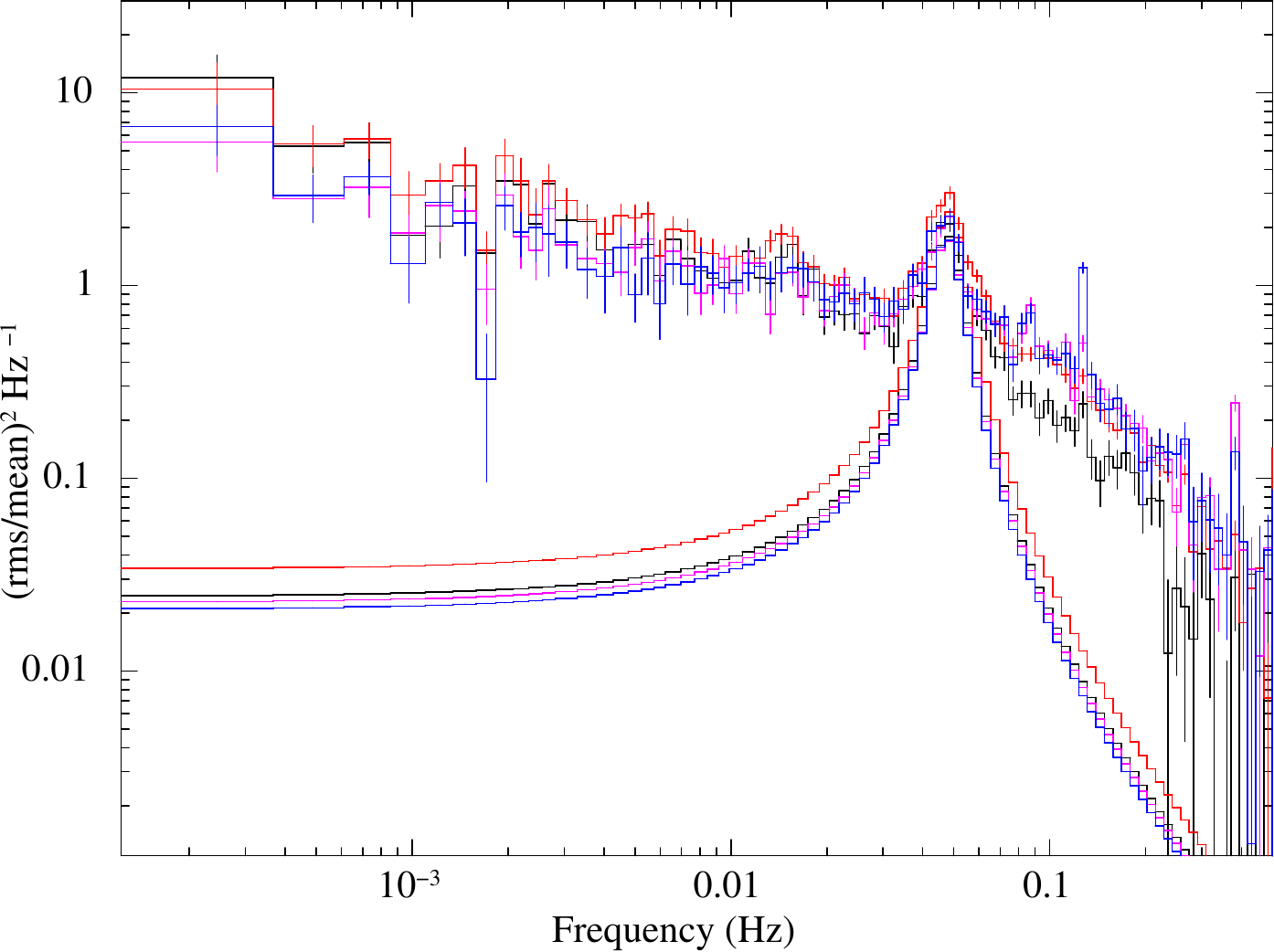}
    \caption{The \textit{XMM-Newton}/PN PSD of 4U 1626--67 in different energy bands (0.5--1 keV in black, 1--3 keV in red, 3--5.7 keV in magenta, $5.7-10.0$ keV in blue) of OID 0152620101 (Top) and OID 111070201 (Bottom). QPO at 40 mHz is fitted with lorentzian profile.}
    \label{fig:4u1626_qpofit_xmmn}
\end{figure}

\begin{figure}
    \centering
    \includegraphics[width=\columnwidth]{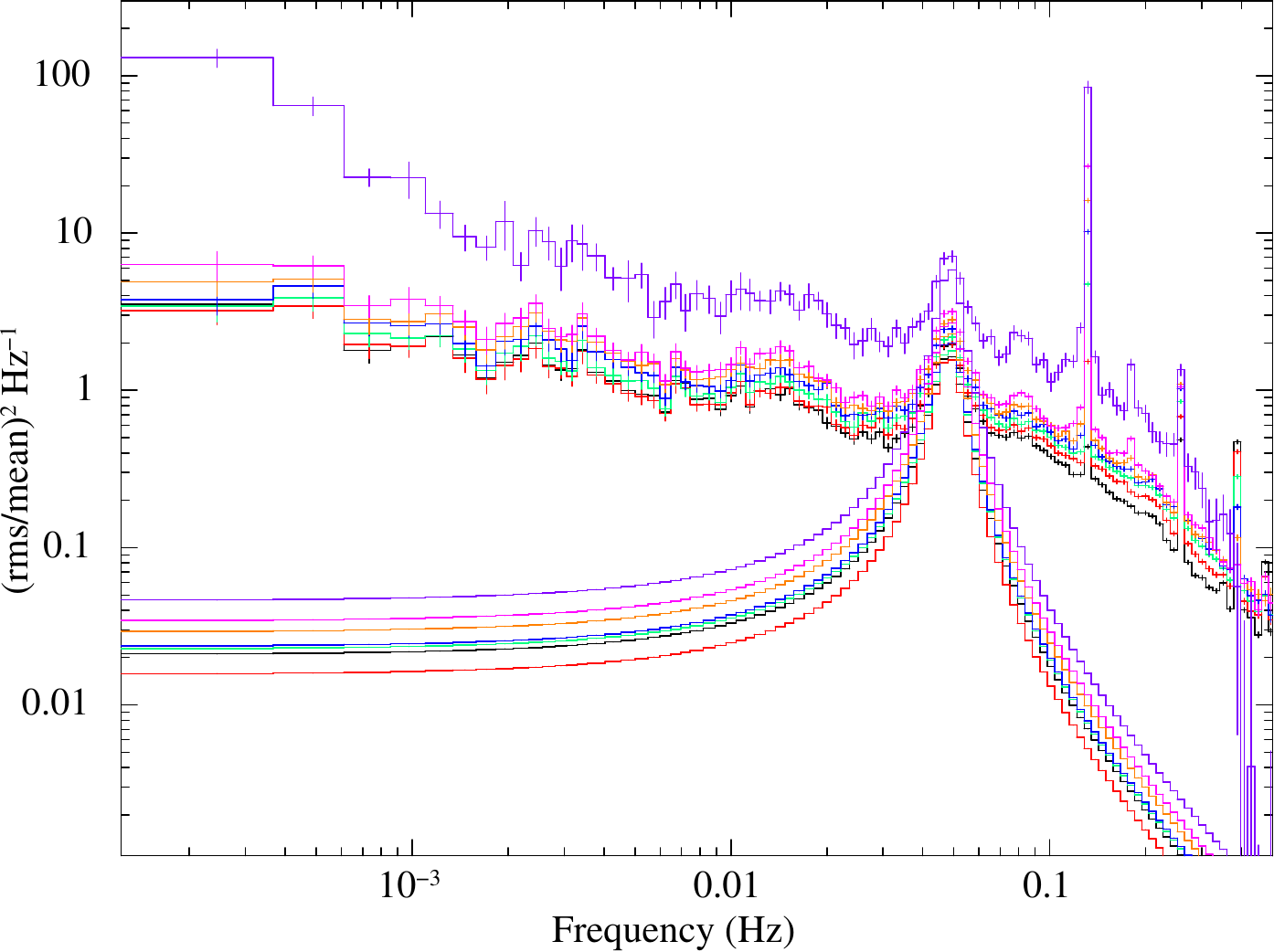}
\caption{The \textit{RXTE}/PCA PSD of 4U 1626--67 in different energy bands ($2.02-6.70$ keV in black, $6.7-8.5$ keV in red, $8.5-11.1$ keV in green, $11.1-13.0$ keV in blue, $13.0-15.4$ keV in orange, $15.4-20.2$ keV in magenta and $20.2-60.0$ keV in purple). The QPO at 48 mHz is fitted with a \texttt{Lorentzian} profile.}
    \label{fig:4u1626_qpofit_rxtepca}
\end{figure}

\begin{figure}
    \centering
    \includegraphics[width=\columnwidth]{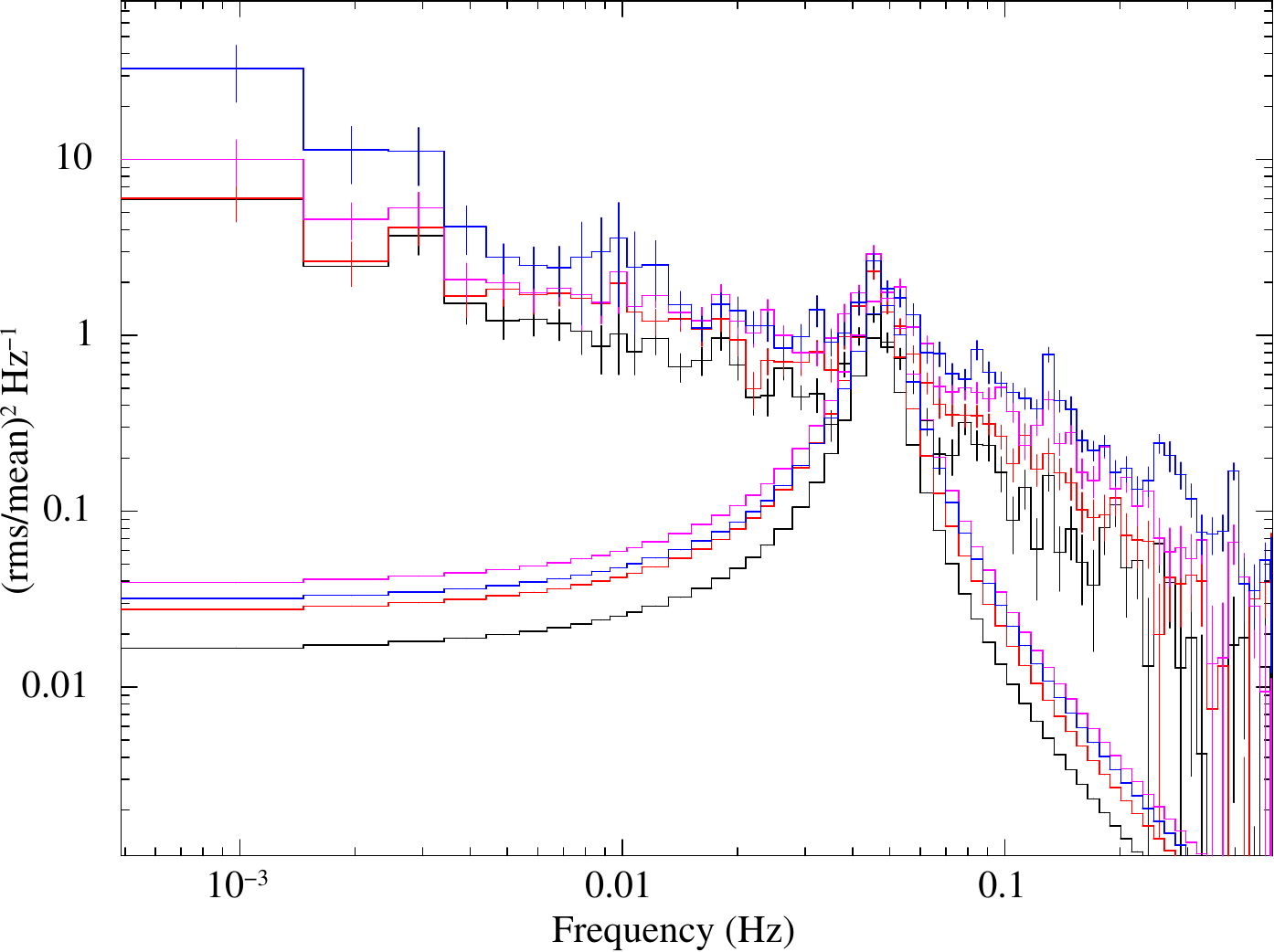}
\caption{The \textit{NICER} PSD of 4U 1626--67 in different energy bands ($0.5-0.9$ keV in black, $0.9-1.3$ keV in red, $1.3-1.9$ keV in magenta, $1.9-3.5$ keV in blue). The QPO at 48 mHz is fitted with a \texttt{Lorentzian} profile.}
    \label{fig:4u1626_qpofit_nicer}
\end{figure}

\begin{figure}
    \centering
    \includegraphics[width=\columnwidth]{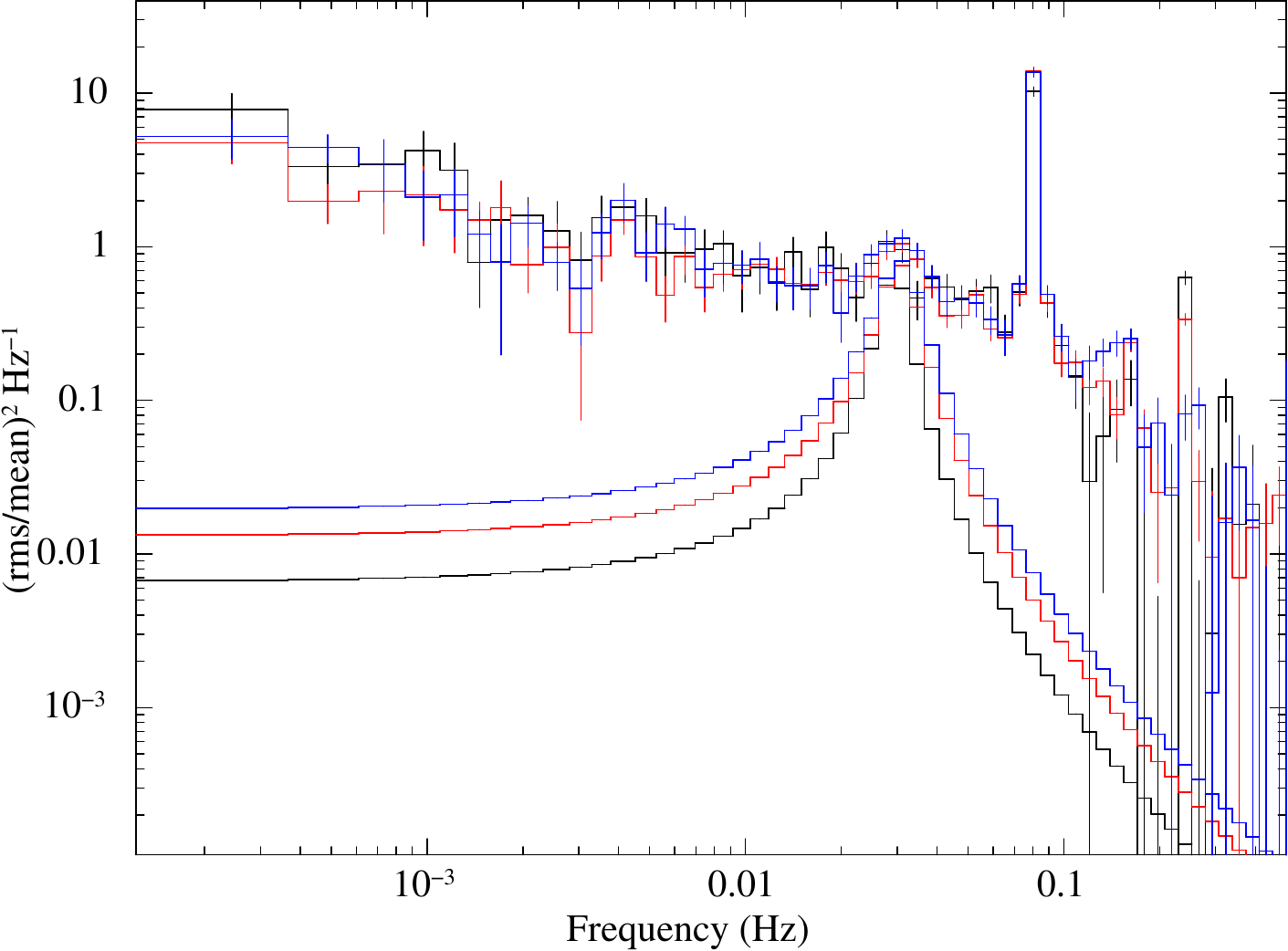}
    \caption{The \textit{XMM-Newton}/PN PSD of IGR J19296+1816 in different energy bands (0.5--3 keV in black, 3--5.7 keV in red, 5.7--10 keV in blue). The QPO at 30 mHz is fitted with a \texttt{Lorentzian} profile.}
    \label{fig:igrj19294_qpofit}
\end{figure}

\begin{figure}
    \centering
    \includegraphics[width=\columnwidth]{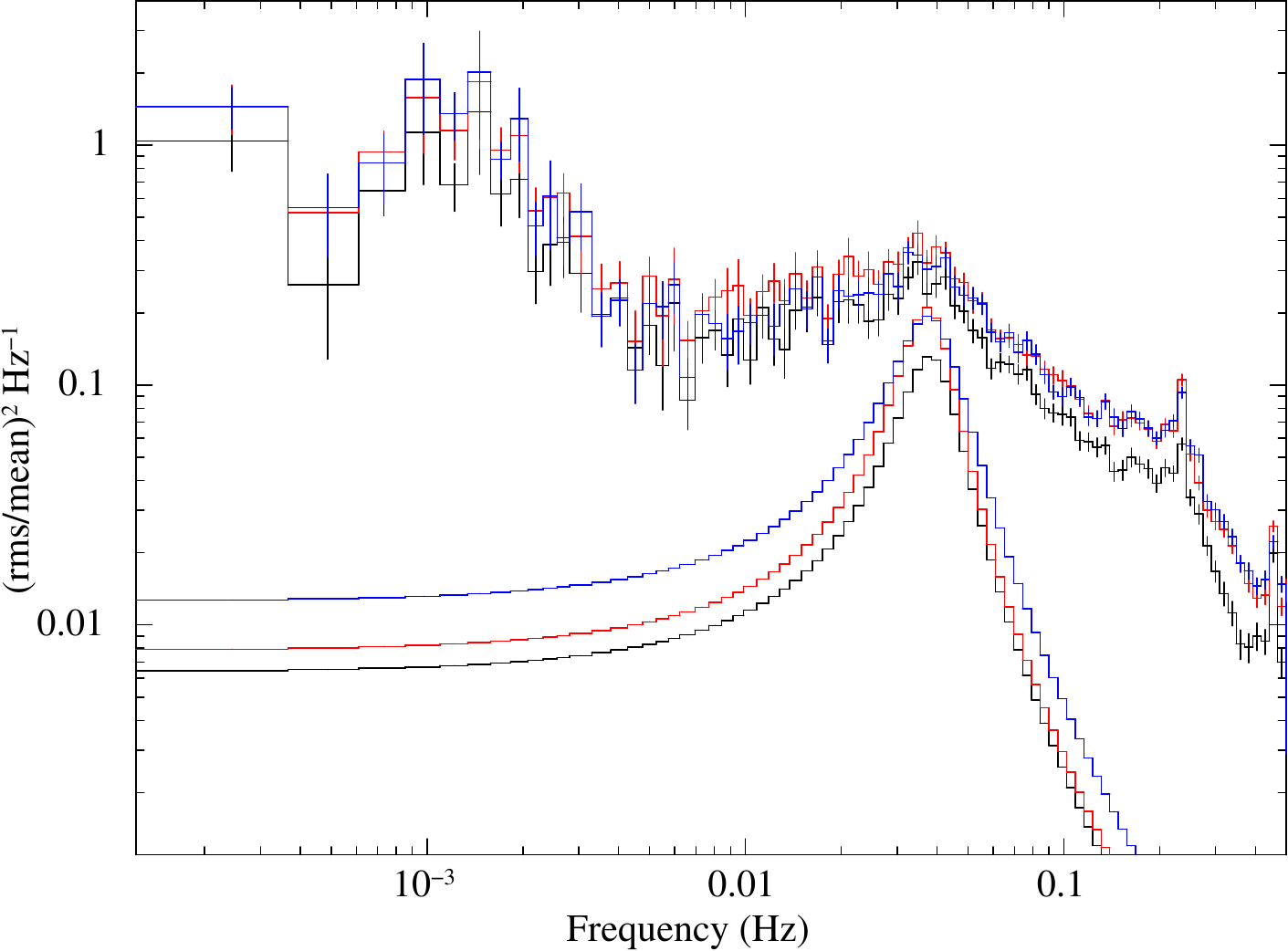}
    \includegraphics[width=\columnwidth]{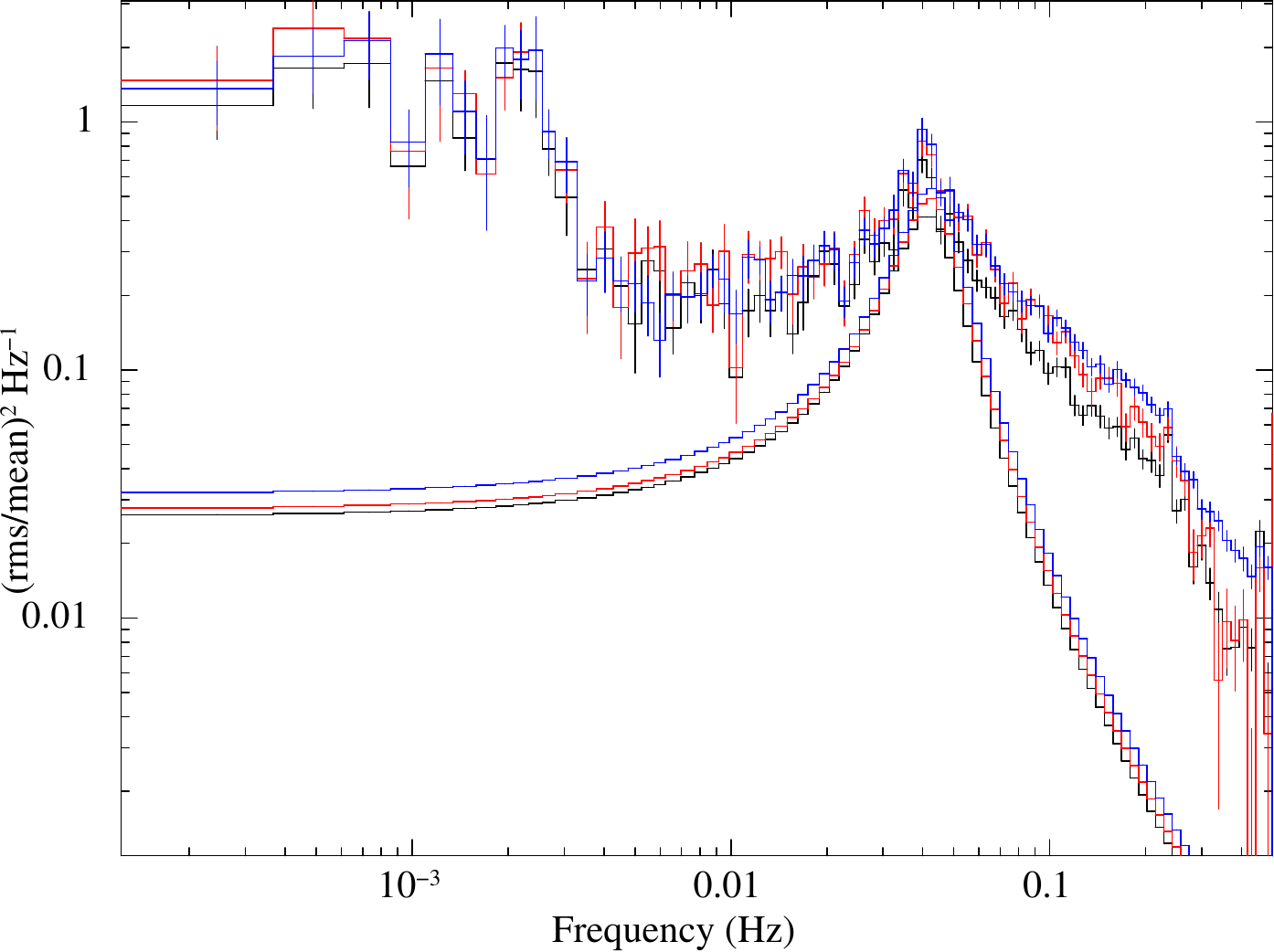}
    \caption{The \textit{XMM-Newton}/PN PSD of V 0332+53 in different energy bands (0.5--3 keV in black, 3--5.7 keV in red, 5.7--10 keV in blue) of OID 0763470301 (Top) and OID 0763470401 (Bottom). QPO at 40 mHz is fitted with a \texttt{Lorentzian} profile.}
    \label{fig:v0332_qpofit-XMM_2015}
\end{figure}

\begin{figure}
    \centering
    \includegraphics[width=\columnwidth]{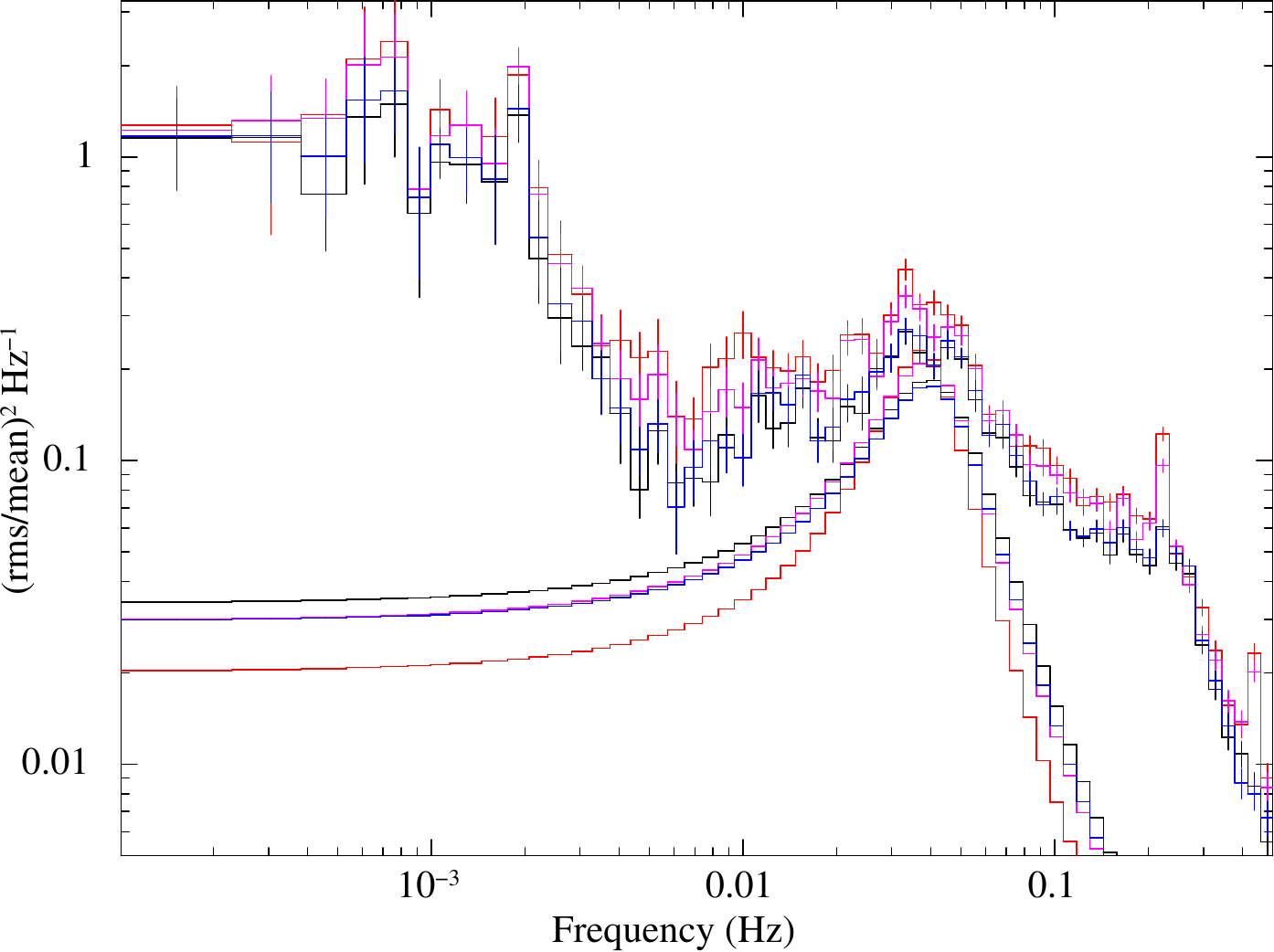}
    \includegraphics[width=\columnwidth]{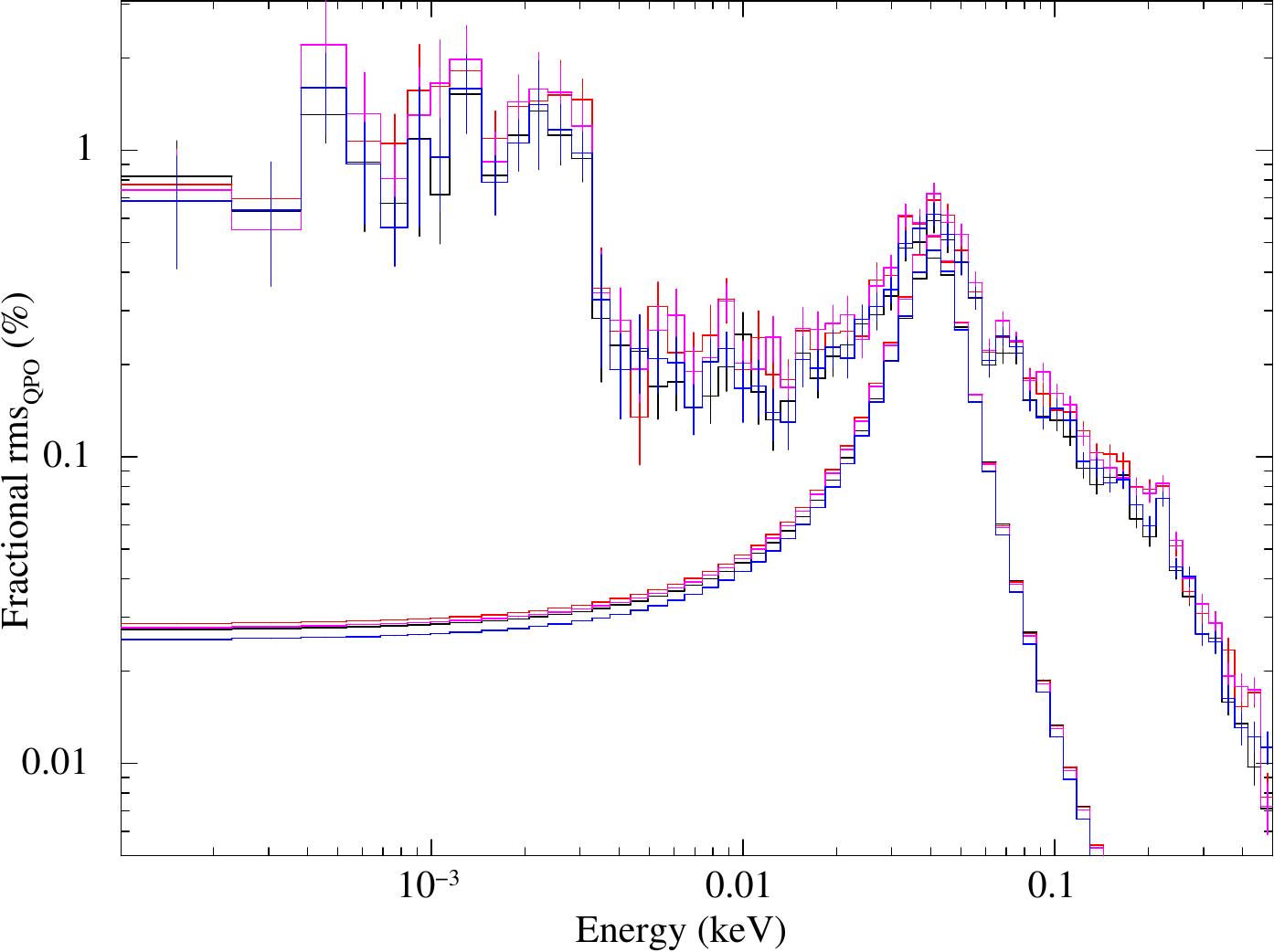}
    \caption{The \textit{NuSTAR} PSD of V 0332+53 in different energy bands (3--6 keV in black, 6--8.5 keV in red, 8.5--11.5 keV in magenta, and 11.5--60 keV in blue) of OID 80102002004 (Top) and OID 80102002006 (Bottom). QPO at 40 mHz is fitted with a \texttt{Lorentzian} profile.}
    \label{fig:v0332_qpofit-NuSTAR_2015}
\end{figure}

\begin{figure}
    \centering
    \includegraphics[width=\columnwidth]{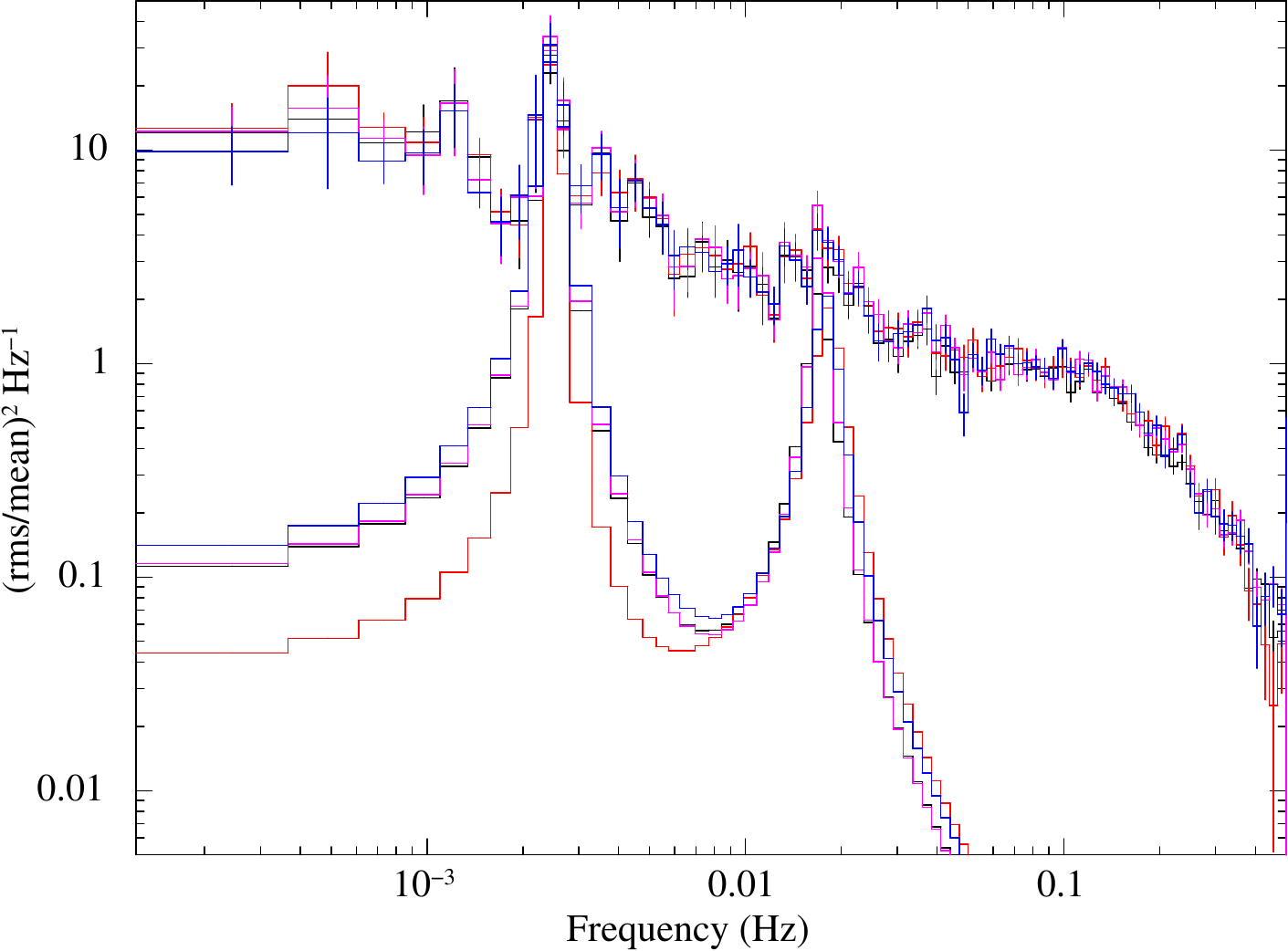}
    \includegraphics[width=\columnwidth]{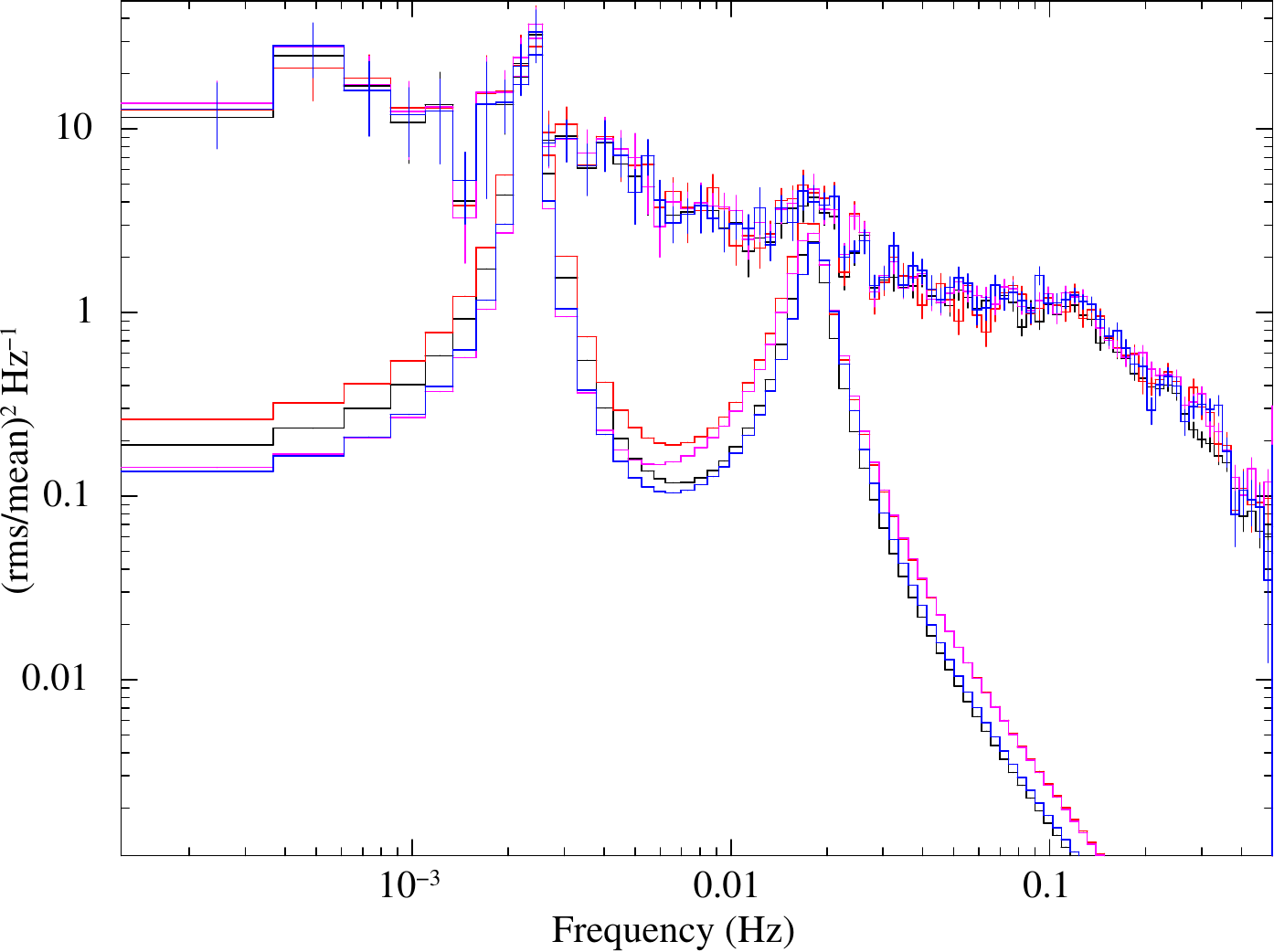}
    \caption{The \textit{NuSTAR} PSD of V 0332+53 in different energy bands (3--8 keV in black, 8--10 keV in red, 10--15 keV in magenta, 15--25 keV in blue). Twin QPOs at 2.5 mHz and 18 mHz were fitted with two \texttt{Lorentzian} profiles.}
    \label{fig:v0332_qpofit-NuSTAR_2016}
\end{figure}

\begin{figure}
    \centering
    \includegraphics[width=\linewidth]{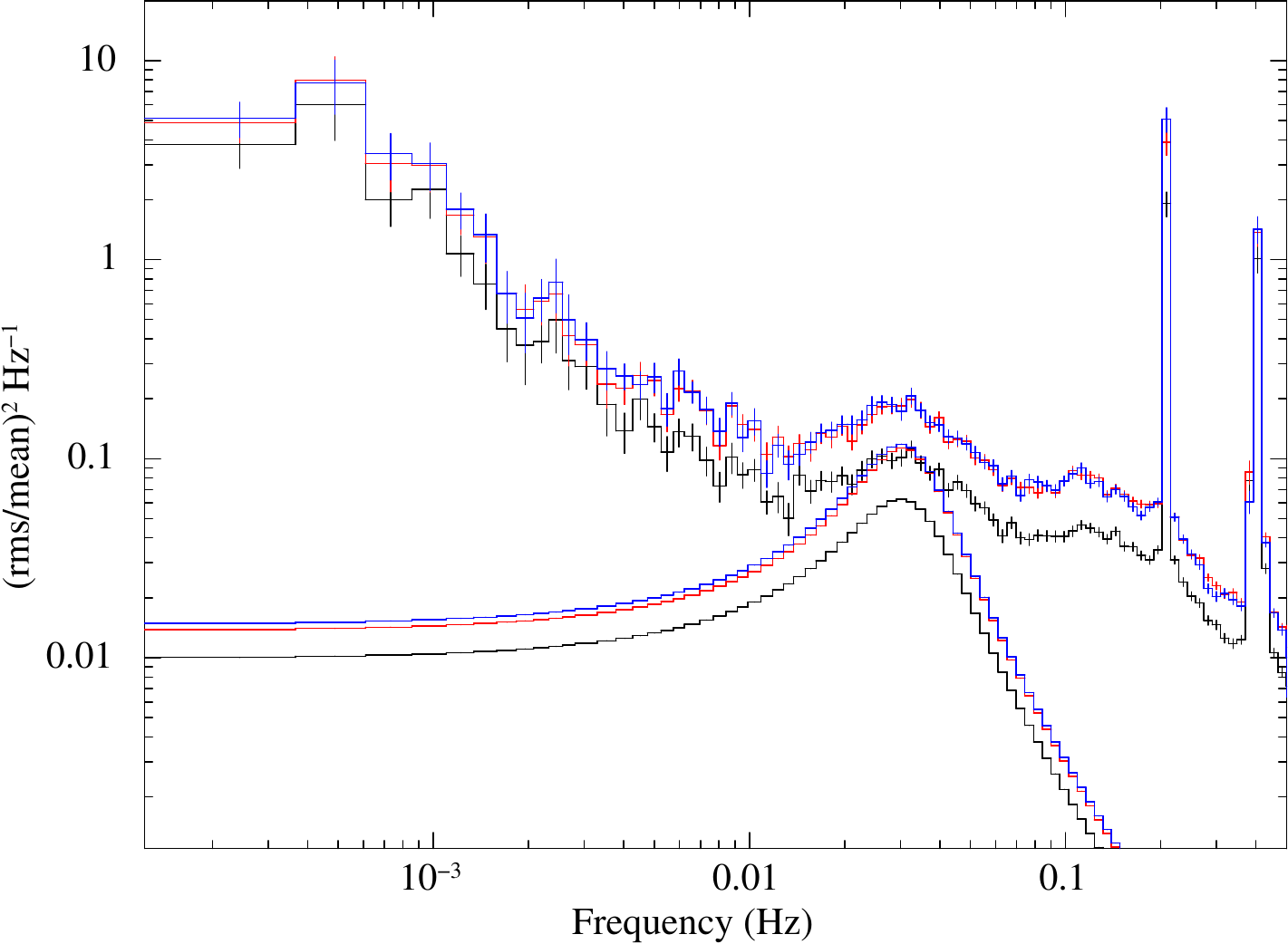}
    \caption{The \textit{XMM-Newton}/PN PSD of Cen X--3 in three different energy bands (0.5--3 keV in black, 3--5.7 keV in red, 5.7--10 keV in blue). QPO at 30 mHz is fitted with \texttt{Lorentzian} profile.}
    \label{fig:cenx3_qpofit}
\end{figure}

\begin{figure}
    \centering
    \includegraphics[width=\columnwidth]{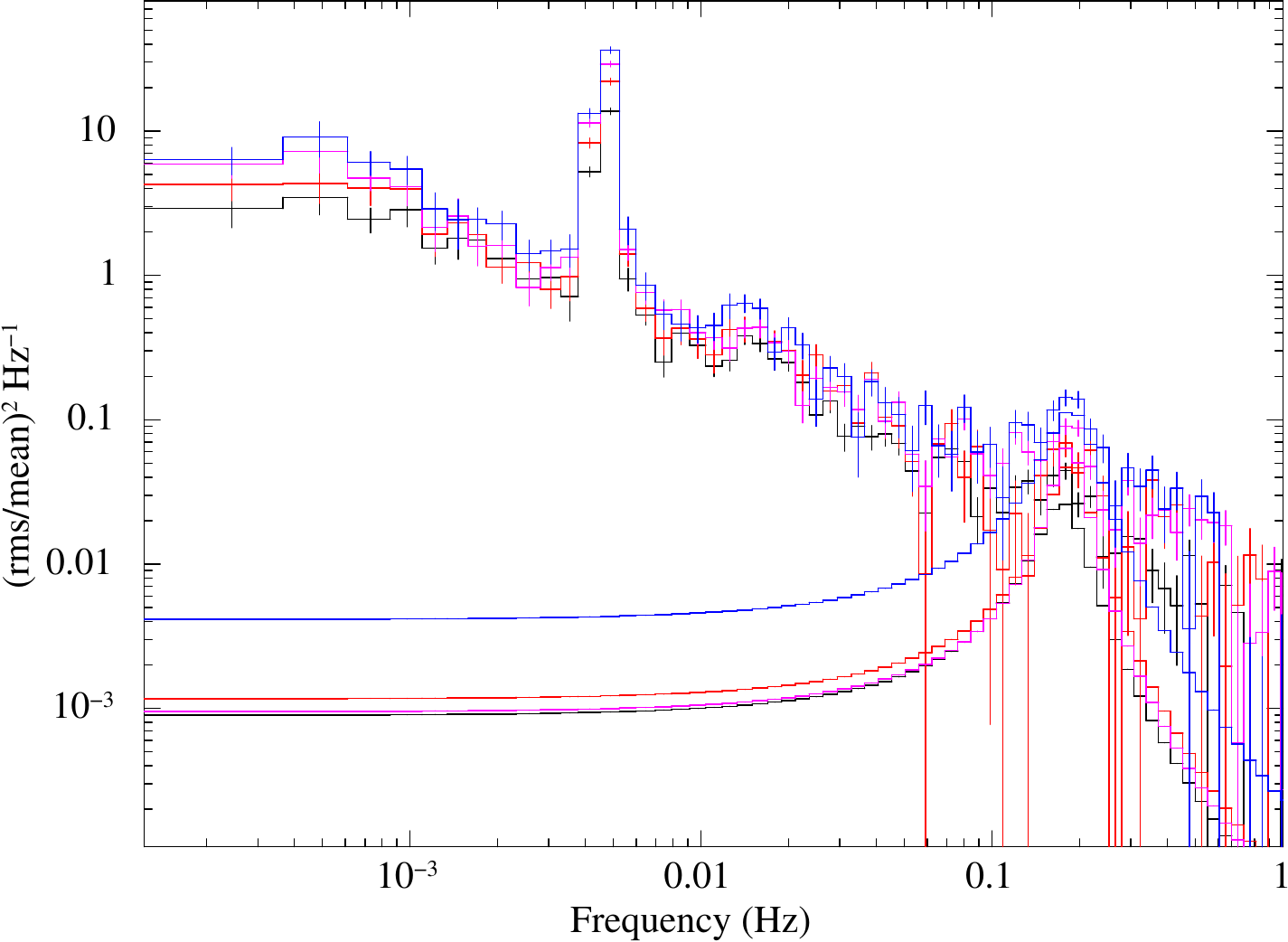}
    \caption{The \textit{NuSTAR} PSD of XTE J1858+034 in four different energy bands (3--8 keV in black, 8--10 keV in red, 10--15 keV in magenta, 15--25 keV in blue). QPO at $\sim185$ mHz was fitted with a \texttt{Lorentzian} profile.}
    \label{fig:xtej1858_qpofit}
\end{figure}

\section{Observations}\label{app:obslog}
\onecolumn
\begin{longtable}{lllccc}
    \caption{Observations {log}}\label{tab:catalogue}\\\toprule
    \endfirsthead
    \caption* {\textbf{Table \ref{tab:obscat1} Continued:} Observations {log}}\\\toprule
    \endhead
    \endfoot
    \bottomrule
    \endlastfoot
         Sl no. &Source & Observatory/Instrument & Obs. ID & Observation mode & Observation duration (ks)\\
         \midrule
         1 &1A 0535+262 & \textit{XMM-Newton}/PN &0674180101 &PrimeFullWindow &58 \\
         2 && \textit{NuSTAR}/FPMA,B &80001016002 &-&43\\
         3 && \textit{NuSTAR}/FPMA,B &80001016004 &-&56\\
         4 && \textit{NuSTAR}/FPMA,B &90401370001 &-&118\\
         5 &2S 1553--542& \textit{NuSTAR}/FPMA,B &90101002002 &-&50 \\
         6 &4U 0115+63& \textit{NuSTAR}/FPMA,B &90102016002 &-&38 \\
         7 && \textit{NuSTAR}/FPMA,B &90102016004 &-&41 \\
         8 &4U 1538--522& \textit{XMM-Newton}/PN &0152780201 &PrimeFullWindow &79 \\
         9 && \textit{NuSTAR}/FPMA,B &30201028002 &-&85\\
         10 &4U 1626--67& \textit{XMM-Newton}/PN &0111070201 &PrimeSmallWindow &16 \\
         11 && \textit{XMM-Newton}/PN &0152620101 &PrimeSmallWindow &84 \\
         12 && \textit{XMM-Newton}/PN &0764860101 &FastTiming &54 \\
         13 && \textit{NuSTAR}/FPMA,B &30101029002 &-&114\\
         14 && \textit{RXTE}/PCA &P10101 &Good Xenon &395\\
         15 &4U 1700--37& \textit{XMM-Newton}/PN &0600950101 &PrimeFullWindow &50 \\
         16 && \textit{NuSTAR}/FPMA,B &30101027002 &-&74\\
         17 &4U 1901+03& \textit{NuSTAR}/FPMA,B &90501305001 &-&44 \\
         18 && \textit{NuSTAR}/FPMA,B &90501324002 &-&102 \\
         19&& \textit{NuSTAR}/FPMA,B &90502307002 &-&38 \\
         20 && \textit{NuSTAR}/FPMA,B &90502307004 &-&55 \\
         21 &4U 1907+09& \textit{XMM-Newton}/PN &0555410101 &FastTiming &21 \\
         22 && \textit{NuSTAR}/FPMA,B &30401018002 &-&154\\
         23 &4U 2206+54& \textit{XMM-Newton}/PN &0650640101 &PrimeLargeWindow &75 \\
         24 && \textit{NuSTAR}/FPMA,B &30201015002 &-&108\\
         25 &Cen X--3& \textit{XMM-Newton}/PN &0111010101 &PrimeSmallWindow &67 \\
         26 && \textit{XMM-Newton}/PN &0400550201 &FastTiming &80\\
         27 && \textit{NuSTAR}/FPMA,B &30101055002 &-&39\\
         28 &Cep X--4& \textit{NuSTAR}/FPMA,B &80002016002 &-&79\\
         29 && \textit{NuSTAR}/FPMA,B &80002016004 &-&76\\
         30 &EXO 2030+375& \textit{XMM-Newton}/PN &0745240201 &FastTiming &31 \\
         31 && \textit{NuSTAR}/FPMA,B &90201029002 &-&117\\
         32 && \textit{NuSTAR}/FPMA,B &90701336002 &-&50\\
         33 &GRO J1008--57& \textit{NuSTAR}/FPMA,B &80001001002 &-&32\\
         34 &GRO J1744--28& \textit{XMM-Newton}/PN &0506291201 &FastTiming &38 \\
         35 && \textit{XMM-Newton}/PN &0729560401 &FastTiming &82 \\
         36 && \textit{NuSTAR}/FPMA,B &80002017002 &-&67\\
         37 && \textit{NuSTAR}/FPMA,B &80202027002 &-&56\\
         38 &GX 301--2& \textit{XMM-Newton}/PN &0555200301 &FastTiming &59 \\
         39 && \textit{XMM-Newton}/PN &0555200401 &FastTiming &47 \\
         40 && \textit{NuSTAR}/FPMA,B &30001041002 &-&51\\
         41 && \textit{NuSTAR}/FPMA,B &30101042002 &-&53\\
         42 &GX 304--1& \textit{NuSTAR}/FPMA,B &90401326002 &-&108\\
         43 &IGR J16393-4643& \textit{XMM-Newton}/PN &0206380201 &PrimeLargeWindow &9 \\
         44 && \textit{XMM-Newton}/PN &0604520201 &PrimeSmallWindow &19 \\
         45 && \textit{NuSTAR}/FPMA,B &30001008002 &-&96\\
         46 &IGR J17329--2731& \textit{XMM-Newton}/PN &0795711701 &FastTiming &37 \\
         47 && \textit{NuSTAR}/FPMA,B &90301012002 &-&38\\
         48 &IGR J17544--2619& \textit{XMM-Newton}/PN &0679810401 &PrimeSmallWindow &15 \\
         49 && \textit{XMM-Newton}/PN &0679810501 &PrimeSmallWindow &15 \\
         50 && \textit{XMM-Newton}/PN &0744600101 &PrimeFullWindow &135 \\
         51 && \textit{NuSTAR}/FPMA,B &30002003003 &-&50\\
         52 &IGR J18027--2016& \textit{XMM-Newton}/PN &0206380601 &PrimeLargeWindow &10 \\
         53 && \textit{XMM-Newton}/PN &0745060401 &PrimeFullWindow &43 \\
         54 && \textit{XMM-Newton}/PN &0745060501 &PrimeFullWindow &16 \\
         55 && \textit{XMM-Newton}/PN &0745060601 &PrimeFullWindow &17 \\
         56 && \textit{XMM-Newton}/PN &0745060701 &PrimeFullWindow &14 \\
         57 && \textit{XMM-Newton}/PN &0745060801 &PrimeFullWindow &17 \\
         58 && \textit{NuSTAR}/FPMA,B &30101049002 &-&85\\
         59 &IGR J19294+1816& \textit{XMM-Newton}/PN &0841190101 &PrimeFullWindow &67 \\
         60 && \textit{NuSTAR}/FPMA,B &90401306002 &-&79\\
         61 && \textit{NuSTAR}/FPMA,B &90401306004 &-&79\\
         62 &KS1947+300& \textit{XMM-Newton}/PN &0727961201 &FastTiming &12 \\
         63 && \textit{NuSTAR}/FPMA,B &80002015002 &-&38\\
         64 && \textit{NuSTAR}/FPMA,B &80002015004 &-&42\\
         65 && \textit{NuSTAR}/FPMA,B &80002015006 &-&56\\
         66 &RX J0520.5--6932& \textit{XMM-Newton}/PN &0701990101 &PrimeFullWindow &20 \\
         67 && \textit{XMM-Newton}/PN &0729560201 &FastTiming &2 \\
         68 && \textit{XMM-Newton}/PN &0729560301 &FastTiming &10 \\
         69 && \textit{NuSTAR}/FPMA,B &80001002002 &-&54\\
         70 && \textit{NuSTAR}/FPMA,B &80001002004 &-&66\\
         71 &SMC X--1& \textit{XMM-Newton}/PN &0784570201 &FastTiming &19 \\
         72 && \textit{XMM-Newton}/PN &0784570301 &FastTiming &19 \\
         73 && \textit{XMM-Newton}/PN &0784570401 &FastTiming &21 \\
         74 && \textit{XMM-Newton}/PN &0784570501 &FastTiming &19 \\
         75 && \textit{XMM-Newton}/PN &0893400101 &PrimeSmallWindow &21 \\
         76 && \textit{XMM-Newton}/PN &0893400301 &PrimeSmallWindow &22 \\
         77 && \textit{NuSTAR}/FPMA,B &30202004002 &-&42\\
         78 && \textit{NuSTAR}/FPMA,B &30202004004 &-&42\\
         79 && \textit{NuSTAR}/FPMA,B &30202004006 &-&38\\
         80 && \textit{NuSTAR}/FPMA,B &30202004008 &-&43\\
         81 &SMC X--2& \textit{XMM-Newton}/PN &0770580701 &FastTiming &8 \\
         82 && \textit{NuSTAR}/FPMA,B &90101017002 &-&48\\
         83 && \textit{NuSTAR}/FPMA,B &90102014002 &-&48\\
         84 && \textit{NuSTAR}/FPMA,B &90102014004 &-&50\\
         85 && \textit{XMM-Newton}/PN &0770580901 &PrimeSmallWindow &31 \\
         86 &V 0332+53& \textit{XMM-Newton}/PN &0506190101 &PrimeFullWindow &36 \\
         87 && \textit{XMM-Newton}/PN &0763470301 &FastTiming &32 \\
         88 && \textit{XMM-Newton}/PN &0763470401 &FastTiming &31 \\
         {89} && {\textit{NuSTAR}/FPMA,B} &{80102002004} &- &{41}\\
         {90} && {\textit{NuSTAR}/FPMA,B} &{80102002006} &- &{38}\\
         {91} && {\textit{NuSTAR}/FPMA,B} &{80102002008} &- &{38}\\
         92 && \textit{NuSTAR}/FPMA,B &80102002010 &-&44\\
         93 && \textit{NuSTAR}/FPMA,B &90202031002 &-&44\\
         94 && \textit{NuSTAR}/FPMA,B &90202031004 &-&44\\
         95 &X Persei& \textit{XMM-Newton}/PN &0151380101 &PrimeFullWindow &30 \\
         96 && \textit{XMM-Newton}/PN &0600980101 &PrimeFullWindow &124 \\
         97 && \textit{NuSTAR}/FPMA,B &30401033002 &- &12\\
         98 &XTE J1829--098& \textit{XMM-Newton}/PN &0135746701 &PrimeFullWindow &1 \\
         99 && \textit{NuSTAR}/FPMA,B &90401332002 &-&55\\
         100 &XTE J1858+034 &\textit{NuSTAR}/FPMA,B &90501348002 &- &90\\
         
    \label{tab:obscat1}
\end{longtable}

\bsp	
\label{lastpage}
\end{document}